\documentclass[twocolumn,longauthor]{aastex63}

\usepackage{hyperref}
\usepackage{color}
\usepackage{diagbox} 
\usepackage{amsmath}

\newcommand{\chandra}{{\sl Chandra}}
\newcommand{\xmm}{XMM-{\sl Newton}}
\newcommand{\swift}{{\sl Swift}}
\newcommand{\egp}{HD\,179949}
\newcommand{\ciao}{{\tt CIAO}}
\newcommand{\behr}{{\tt BEHR}}
\newcommand{\cstat}{{\tt cstat}}
\newcommand{\phot}{{\tt phot}}
\newcommand{\coro}{{\tt cor}}
\newcommand{\angr}{{\tt angr}}
\newcommand{\aspl}{{\tt aspl}}
\newcommand{\grsa}{{\tt grsa}}
\newcommand{\testa}{{\tt DT05}}
\newcommand{\luck}{{\tt Lu18}}
\newcommand{\ecu}{{\tt E+04}}
\newcommand{\brwr}{{\tt B+16}}

\newcommand{\updatebf}[1]{{#1}}
\newcommand{\updatetwo}[1]{{ #1}}

\makeatletter
\newcommand\footnoteref[1]{\protected@xdef\@thefnmark{\ref{#1}}\@footnotemark}
\makeatother

\received{}
\revised{\today}
\accepted{}
\submitjournal{ApJ}
\shorttitle{X-ray Activity on \egp}
\shortauthors{Acharya et al.}
\graphicspath{{./}{figures/}}

\begin{document}

\title{X-ray Activity Variations and Coronal Abundances of the Star-Planet Interaction candidate HD\,179949}

\correspondingauthor{Anshuman Acharya}
\email{anshuman@mpa-garching.mpg.de}

\author[0000-0003-3401-4884]{Anshuman Acharya}
\affiliation{Department of Physical Sciences, Indian Institute of Science Education and Research (IISER) - Mohali, \\
SAS Nagar, Punjab - 140306, India.}
\affiliation{Max Planck Institut für Astrophysik, 85748 - Garching, Germany.}
\author[0000-0002-3869-7996]{Vinay L.\ Kashyap}
\affiliation{Center for Astrophysics $|$ Harvard \& Smithsonian \\
60 Garden St.\ \\
Cambridge, MA 02138, USA}

\author[0000-0001-7032-8480]{Steven H. Saar}
\affiliation{Center for Astrophysics $|$ Harvard \& Smithsonian \\
60 Garden St.\ \\
Cambridge, MA 02138, USA}

\author[0000-0001-6952-3887]{Kulinder Pal Singh}
\affiliation{Department of Physical Sciences, Indian Institute of Science Education and Research (IISER) - Mohali, \\
SAS Nagar, Punjab - 140306, India.}

\author[0000-0002-8883-2930]{Manfred Cuntz}
\affiliation{Dept.\ of Physics, University of Texas at Arlington \\
502 Yates St.\ \\
Arlington, TX 76019, USA}

\begin{abstract}
We carry out detailed spectral and timing analyses of the \chandra\ X-ray data of \egp, a prototypical example of a star with a close-in giant planet with possible star-planet interaction (SPI) effects.
\updatetwo{We find a low coronal abundance $A({\rm Fe})/A{\rm H}){\approx}0.2$ relative to the solar photospheric baseline of \citet{Anders_1989},} \updatebf{and significantly lower than the stellar photosphere as well.  We further find low abundances of high First Ionization Potential (FIP)} elements 
\updatetwo{$A({\rm O})/A({\rm Fe}){\lesssim}1$, $A({\rm Ne})/A({\rm Fe}){\lesssim}0.1$, but with indications of higher abundances of $A({\rm N})/A({\rm Fe}){\gg}1, A({\rm Al})/A({\rm Fe}){\lesssim}10$.} 
\updatebf{We estimate a FIP bias for this star in the range $\approx{-0.3}$ to ${-0.1}$, larger than the ${\lesssim}-$0.5 expected for stars of this type, but similar to stars hosting close-in hot Jupiters.} We detect significant intensity variability over time scales ranging from 100~s - 10~ks, and also evidence for spectral variability over time scales of 1-10~ks.  We combine the \chandra\ flux measurements with \swift\ and \xmm\ measurements to detect periodicities, and determine that the dominant signal is tied to the stellar polar rotational period, consistent with expectations that the corona is rotational-pole dominated.  We also find evidence for periodicity at both the planetary orbital frequency and at its beat frequency with the stellar polar rotational period, suggesting the presence of a magnetic connection between the planet and the stellar pole. \updatebf{If these periodicities represent an SPI signal, it is likely driven by a quasi-continuous form of heating (e.g., magnetic field stretching) rather than sporadic, hot, impulsive flare-like reconnections.}
\end{abstract}


\section{Introduction} \label{sec:intro}

The first confirmed \updatebf{detections of exoplanets started occurring around the late 1980s \citep{Wolszczan_1992,Hatzes_2003}}, and since then, the number of exoplanet detections has been increasing rapidly, reaching $\approx$5000 confirmed detections by October 2022. Among them, giant planets are commonly detected, with $\approx$450 such planets having semi-major axis,  $a_p < 0.15$~AU and mass of the planet $M_p$ in the range $0.5~M_{Jupiter} < M_p < 15~M_{Jupiter}$\footnote{\url{http://exoplanet.eu/catalog/}}.

Magnetic and tidal interactions between the planet and the host star in such systems can affect stellar activity \citep{Rubenstein_2000,Cuntz_2000}. Such star-planet interactions (SPI) could have measurable effects \citep{Shkolnik_2018book, Strugarek_2021} on both chromospheric and X-ray output. For example, \citet{Shkolnik_2003} identified chromospheric variability in \egp\ tied to planetary phase and \citep{Kashyap_2008} claimed an average of 4$\times$ enhancement of X-ray luminosity in stars with close-in planets (though other studies \updatebf{like \citet{Scharf_2010,Poppenhaeger_2010,Miller_2015,Viswanath_2020}} have not confirmed this). \citet{Poppenhaeger_2010} and \citet{Pillitteri_2014a, Pillitteri_2014b, Pillitteri_2022} have demonstrated that stellar activity in HD\,189733 appears to be affected by the planetary phase and that a coeval star in the same system without a close-in giant planet is considerably weaker in activity. 

\egp\ is the prototypical example of a stellar system with a close-in giant planet which could be affected by SPI.  It has a hot Jupiter orbiting it at a distance of \updatebf{$\approx$0.04~AU} with a period of \updatebf{$\approx$3.1 days} (Table~\ref{table:baseparams}). The star itself has an equatorial rotational period of $7.62 \pm 0.05$ days and a polar rotational period of $10.30 \pm 0.80$ days \citep{Fares_2012}. \cite{Cauley_2019} found the star's magnetic field strength to be 3.2$\pm$0.3~G. 

\citet{Shkolnik_2003} \updatetwo{first} presented evidence for planet-related emission variability in \egp. This effect lasted for over a year and was correlated with the orbital phase of the planet, \updatebf{indicating} a magnetic interaction. \citet{Shkolnik_2003} suggested that these interactions could produce a chromospheric hot spot which rotates in phase with the planet's orbit, and is thus modulated by the orbital period. However, only intermittent variability rather than regular periodicity has been observed. Nevertheless, this intermittent variability is expected for some cases of magnetospheric SPI due to the patchy nature of stellar surface fields \citep[e.g.,][]{Cranmer_2007}. Indeed, in an early survey of SPI detections, \cite{Saar_2004} found that magnetic interactions were the most likely mechanism driving SPI.  They argued that a comparison of relative SPI energy levels with the physical properties of the various detected systems suggested flare-like reconnection was the most likely source \updatebf{(see also \citealt{Lanza_2009,Lanza_2012}).  Below, we suggest field line stretching \citep{Lanza_2013} also as a plausible mechanism.}

Further, \citet{Shkolnik_2008} observed synchronicity of the Ca\,II~H\&K emission with the orbit in four out of six epochs, while rotational modulation with a period of \updatebf{$\approx$7} days is apparent in the other two seasons. They then claim that if there are activity cycles, then that may be a possible explanation for the on/off nature of SPI, where there are periods of no SPI, where they detected a period close to the stellar equatorial rotation period. \citet{Scandariato_2013} again carried out observations of Ca\,II~H\&K emission, and suggested that there may be intermittent variability in the chromosphere, due to obscuring by other signatures. They further noted variability with period $\approx 11$ days, and thus concluded variability tied to the stellar \textit{polar} rotational period instead, and did not find planet induced stellar variability.
\updatetwo{Thus, while \citealt{Shkolnik_2008} found evidence for planet induced emission (PIE) in Ca\,II~H\&K, it was not always observable.  The PIE effect in \egp\ was detected by \citet{Gurdemir_2012} in Ca\,II~K, but not in Ca\,II~H emission; and \citet{Scandariato_2013} found only a periodicity consistent with stellar rotation.}

To follow up with X-ray observations of the stellar corona, data was collected in 2005 using the \chandra\ X-ray telescope. A preliminary analysis of this data, which presented a tentative detection of a SPI, was presented by \cite{Saar_2008}. \updatebf{Here we carry out a comprehensive reanalysis of the data: we construct count rate light curves to determine the level and scale of temporal variability; compute hardness ratio light curves to determine {\sl spectral} variability; obtain robust spectral fits to measure coronal abundances; compare with photospheric abundances to characterize abundance anomalies; and combine with \xmm\ and \swift\ measured fluxes obtained from the literature to carry out an exhaustive periodicity analysis to establish the character of SPI on \egp.}

We describe the \chandra\ data of \egp\ and subsequent processing in Section~\ref{sec:data}. We describe our analysis in detail in Section~\ref{sec:analysis}; spectral fitting in Section~\ref{sec:spec}, and the detection of temporal variability in light curves in Section~\ref{sec:lc}, and in hardness ratios in Section~\ref{sec:HR}.  Detection of periodicities via Lomb-Scargle periodograms are described in Section~\ref{sec:LS}.  We discuss our results and their implications in Section~\ref{sec:discuss}, and summarize in Section~\ref{sec:summary}.

\begin{deluxetable*}{lll}
\tabletypesize{\footnotesize}
\tablecolumns{3}
\tablewidth{0pt}
\tablecaption{Stellar and Planetary parameters \updatetwo{and notation} \label{table:baseparams}}
\tablehead{\colhead{Property} & \colhead{Value} & \colhead{Reference}}
\startdata
\multicolumn{3}{c}{\egp: HIP/HIC 94645, HR 7291, GJ 749, Gumala} \\
\hline
Spectral Type & F8V & \cite{Houk_1988} \\
Distance, $d_*$ & $27.5 \pm 0.6$ pc & \cite{GaiaCollaboration_2018} \\
Age, $\tau_*$ & $1.20 \pm 0.60$ Gyr & \cite{Bonfanti_2016} \\
Mass, $M_*$ & $1.23 \pm 0.01$ $M_{\odot}$ & \cite{Bonfanti_2016} \\
Radius, $R_*$ & $1.20 \pm 0.01$ $R_{\odot}$ & \cite{Bonfanti_2016} \\
m$_V$, $B-V$ & 6.237$^m$, 0.534$^m$ & \cite{Hog_2000} \\
Proper motion (RA,Dec) & (+118.52,-102.235) mas/yr & \cite{GaiaCollaboration_2018} \\
Equatorial rotational period, $P_{\rm Equatorial}$ & $7.62 \pm 0.05$ days & \cite{Fares_2012} \\
Polar rotational period, $P_{\rm Polar}$ & $10.3 \pm 0.8$ days & \cite{Fares_2012} \\
Magnetic field strength, $B_*$ & 3.2 $\pm$ 0.3~G & \cite{Cauley_2019} \\
Bolometric Luminosity, $L_{bol}$ & 1.95$\pm$0.01 L$_{\odot}$ & \cite{Bonfanti_2016} \\
Photospheric ${\rm [Fe/H]}_{\phot|\angr}$ & 0.056$\pm$0.055$~^\dag$ & \cite{Gonzalez_2007} \\
Effective Temperature T$_{eff}$ & 6220 $\pm$ 28 K & \cite{Bonfanti_2016} \\
\hline
\multicolumn{3}{c}{Hot Jupiter \egp{b}} \\
\hline
Mass, $M_p$ & $0.980 \pm 0.004$ M$_{Jupiter}$ & \cite{Butler_2006} \\
Radius, $R_p$ & 1.22 $\pm$ 0.18 R$_{Jupiter}$ & \cite{Cauley_2019} \\
\updatebf{Magnetic field strength}, $B_p$ & $86 \pm 29$~G & \citet[][]{Cauley_2019}~$^\ddag$ \\
Orbital semi-major axis, $a_p$ & $0.044 \pm 0.003$ AU & \cite{Butler_2006} \\
Orbital period, $P_{\rm Orbit}$ & $3.09285 \pm 0.00056$ days & \cite{Shkolnik_2003} \\
Ephemeris ($\phi=0$) & $2452479.823 \pm 0.093$ days & \cite{Shkolnik_2003} \\
\enddata
\tablenotetext{$\dag$}{\updatebf{Corrected from a baseline of $\grsa$ (${\rm [Fe]}=7.5$; \citealt{1998SSRv...85..161G}) to $\angr$ (${\rm [Fe]}=7.67$; \citealt{Anders_1989}); see Appendix~\ref{sec:appendix_abun}.}}
\tablenotetext{$\ddag$}{\updatebf{For an assumed fraction of the total energy disssipated in Ca\,II\,K, $\epsilon = E_{\rm Ca\,II\,K}/E_{\rm tot} = 0.2\%$}}
\end{deluxetable*}

\section{Data} \label{sec:data}

We primarily use archival data from the \chandra\ X-ray Observatory (\chandra). The data span five observations of ${\approx}30$~ks each obtained in May 2005 (PI: S.Saar; see observation log in Table~\ref{table:obslog}).  The observations were carried out over a variety of planetary phase ranges spanning three consecutive orbits.  We adopt the ephemeris of \cite{Shkolnik_2003} (see Table~\ref{table:baseparams}) to describe the planetary orbit.  Each observation covers ${\approx}40^{\circ}$ of the orbital phase, and the expected uncertainty in the absolute planetary phase at any given epoch is ${\lesssim}25^{\circ}$.  All the observations were carried out over one stellar polar rotation.

\begin{deluxetable*}{llll}
\tabletypesize{\footnotesize}
\tablecolumns{4}
\tablewidth{0pt}
\tablecaption{\updatebf{\chandra } Observation Log \label{table:obslog}}
\tablehead{
\colhead{Observation ID} & \colhead{Observation} & \colhead{Exposure} & \colhead{Planetary phase range$^*$} \vspace{-0.2cm}\\
\colhead{(ObsID)} & \colhead{Start Time [UT]} & \colhead{[ksec]} & \colhead{[deg]} \vspace{-0.4cm} \\}
\startdata
5427 & 2005-05-21~18:12:41 & 29.581 & [292.945 - 332.795] $\pm$ 24.303 \\
6119 & 2005-05-22~11:37:35 & 29.648 & [17.490 - 57.431] $\pm$ 24.316 \\
6120 & 2005-05-29~16:51:03 & 29.648 & [137.526 - 177.467] $\pm$ 24.453 \\
6121 & 2005-05-30~12:00:30 & 29.644 & [203.933 - 243.870] $\pm$ 24.464 \\
6122 & 2005-05-31~10:50:19 & 31.763 & [318.626 - (360+)1.416] $\pm$ 24.482 \\
\vspace{-1em} \enddata 
\vspace{-0.2cm}
\tablenotetext{*}{Phase $\phi=0$ is when \egp{b} is in conjunction, i.e., between \egp\ and the observer along the line of sight. Error bars are computed by propagating the uncertainty in the phase to the observation epoch.}
\end{deluxetable*}

For each observation, we locate the source position by finding the centroid of photons inside a circle of radius $2''$ centered at the proper motion corrected position. We extract the source counts from a circle of radius $3.6''$ and background counts from an annulus with radii $10.8''-28.8''$ corresponding to an area 55$\times$ larger than the source region.  A summary of the observation specific source locations and strengths is in Table~\ref{table:summaryobs}.

\begin{deluxetable*}{lllcccc}
\tabletypesize{\footnotesize}
\tablecolumns{7}
\tablewidth{0pt}
\tablecaption{ Summary of \updatebf{\chandra\ measurements} \label{table:summaryobs}}
\tablehead{
\colhead{Observation ID} & \colhead{Measured Position} & \colhead{Offset from} & \colhead{Counts in} & \colhead{Counts in} & \colhead{Net Counts*$^\ddag$}  & \colhead{Net Rate*$^\ddag$} \vspace{-0.2cm}\\
\colhead{(ObsID)} & \colhead{(RA, Dec)} & \colhead{expected position$^\dag$} & \colhead{source region*} & \colhead{background region*} & \colhead{[ct]} & \colhead{[ct~ksec$^{-1}$]} \vspace{-0.4cm} \\}

\startdata
\vspace{-0.1cm} 5427 & (19:15:33.2990,$-24^{\circ}$ 10' 46.127") & (+0.520",-0.011") & 1597 & 124 & 1590$\pm$41 & 54$\pm$1.4\\
& $\pm$ (0.405",0.351") & \multicolumn{5}{c}{\hfil} \\
\vspace{-0.1cm} 6119 & (19:15:33.2853,$-24^{\circ}$ 10' 46.159") & (+0.305",-0.044") & 1300 & 118 & 1300$\pm$37 & 44$\pm$1.3 \\
& $\pm$ (0.300",0.311") & \multicolumn{5}{c}{\hfil} \\ 
\vspace{-0.1cm} 6120 & (19:15:33.2849,$-24^{\circ}$ 10' 46.084") & (+0.299",+0.033") & 1458 & 117 & 1460$\pm$39 & 49$\pm$1.3\\
& $\pm$ (0.435",0.361") & \multicolumn{5}{c}{\hfil} \\ 
\vspace{-0.1cm} 6121 & (19:15:33.2850,$-24^{\circ}$ 10' 46.061") & (+0.304",+0.056") & 1599 & 116 & 1600$\pm$41 & 54$\pm$1.4\\
& $\pm$ (0.405",0.360") & \multicolumn{5}{c}{\hfil} \\ 
\vspace{-0.1cm}6122 & (19:15:33.2887,$-24^{\circ}$ 10' 46.049") & (+0.358",+0.069") & 1934 & 497 & 1920$\pm$45 & 61$\pm$1.4\\
& $\pm$ (0.390",0.347")& \multicolumn{5}{c}{\hfil} \\ 
\vspace{-1em} \enddata  
\tablenotetext{$\dag$}{Based on positions from \citet{GaiaCollaboration_2018}, corrected for proper-motion.}
\vspace{-0.2cm}
\tablenotetext{$*$}{in broad band 0.5-7.0 keV.}
\vspace{-0.2cm}
\tablenotetext{$\ddag$}{background scaling factor = 1/55.}

\end{deluxetable*}

\begin{figure}[!htbp]
\includegraphics[width=\columnwidth,keepaspectratio]{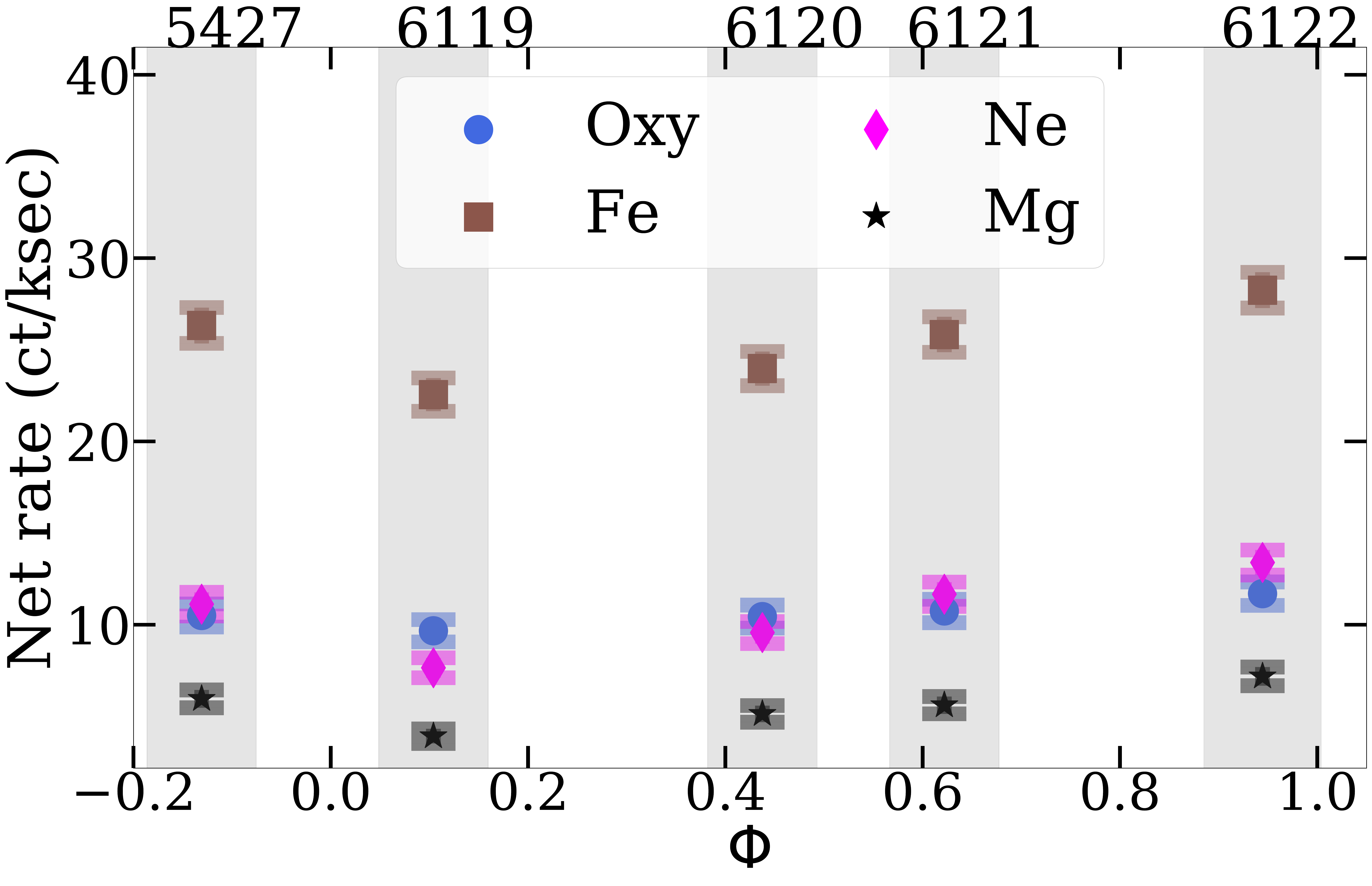}
\includegraphics[width=\columnwidth,keepaspectratio]{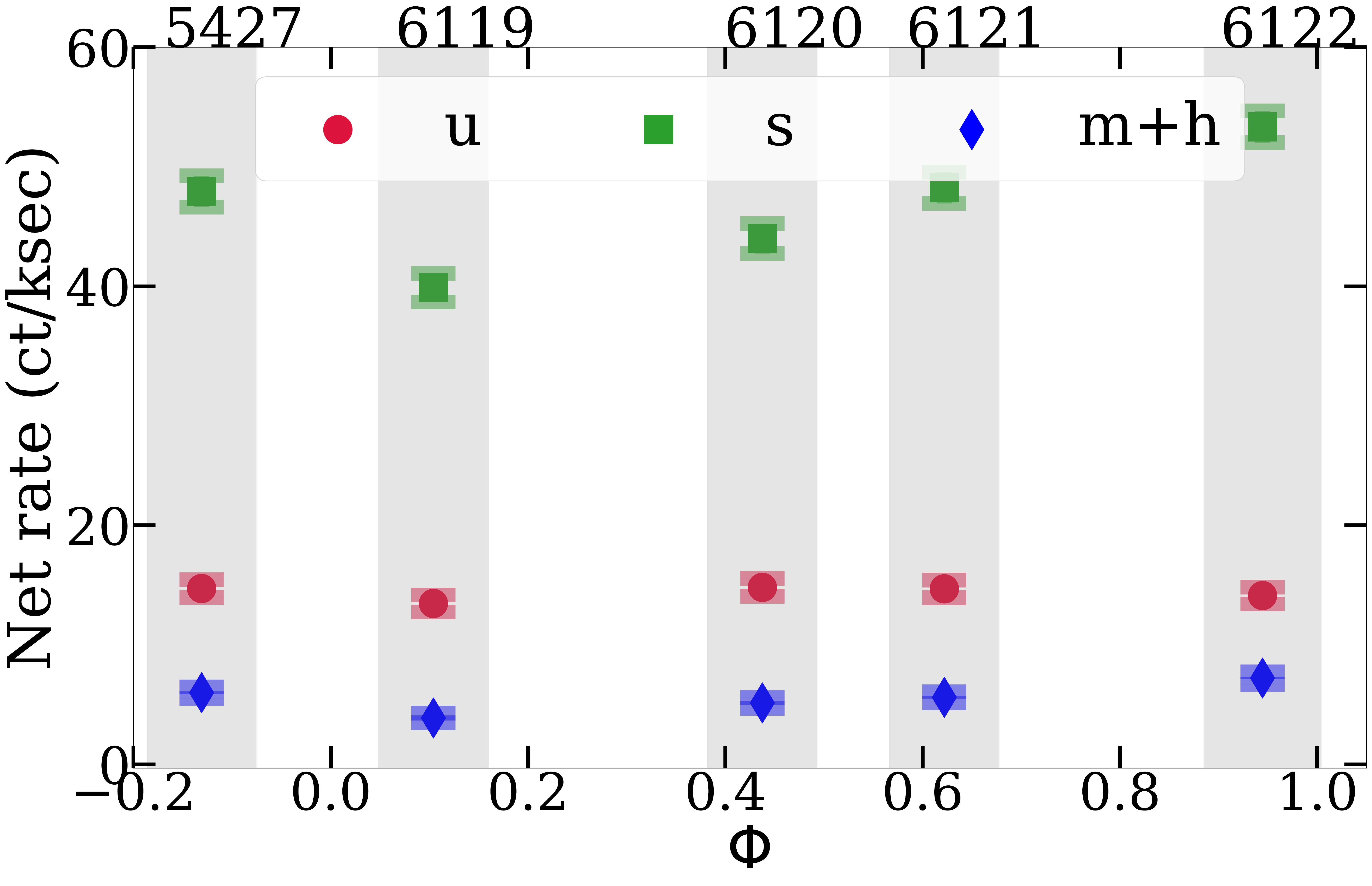}
\caption{\updatebf{Net count rates during \chandra\ observations of \egp\ as a function of planetary orbital phase, \updatetwo{with $\phi=0$ representing conjunction, where the planet between the observer and the star}.  The upper panel shows the count rates in line-centered passbands and the lower panel shows the count rates in broader bands used in the \chandra\ Source Catalog (see Table~\ref{table:passbands}).  The grey shaded regions depict the phase range covered during each observation (ObsID listed along the upper edge of each panel).
}
}  
\label{fig_rates}
\end{figure}

We perform detailed intensity and hardness ratio light curve analyses over several passbands.  We choose both standard broad passbands (such as those defined by the \chandra\ Source Catalog, CSC) to allow comparisons with other analyses, as well as narrow passbands targeted to particular line complexes.  The list of passbands is in Table~\ref{table:passbands}.  We typically merge the {\sl medium} and {\sl hard} bands due to the dearth of counts in the latter.  The narrow bands are centered around strong emission lines in the H- and He-like ions of Oxygen, Neon, and Magnesium, and the Fe\,XVII-XVIII lines that dominate coronal radiation at $\approx$6~MK.  Because the emission from these lines dominate at different temperatures, we expect that relative changes in intensity in the narrow passbands are diagnostic of temperature fluctuations as well as abundance variations.  The count rates in each of the bands are shown in Figure~\ref{fig_rates}.

\begin{deluxetable}{lll}
\tabletypesize{\footnotesize}
\tablecolumns{8}
\tablewidth{0pt}
\tablecaption{\updatebf{Adopted} Passbands \label{table:passbands}}
\tablehead{
\colhead{Band} & \colhead{Energy Range [keV]} & \colhead{Comment}} 
\startdata
\multicolumn{3}{c}{CSC$^\dag$ bands} \\
\hline
u & 0.2 - 0.5 & ultrasoft \\
s & 0.5 - 1.2 & soft \\
m+h & 1.2 - 8.0 & medium + hard\\
b & 0.5 - 7.0 & broad \\
\hline
\multicolumn{3}{c}{Line dominated bands} \\
\hline
Oxy & 0.5 - 0.7 & OVII \& OVIII \\
Fe & 0.7 - 0.9 & FeXVII and FeXVIII \\
Ne & 0.9 - 1.2 & NeX and NeXI \\
Mg & 1.2 - 3.0 & MgXIII and MgXII\\
\vspace{-1em} \enddata  
\tablenotetext{$\dag$}{\updatebf{Based on \chandra\ Source Catalog passbands}}
\end{deluxetable}

\begin{figure*}[!htbp]
\centering
\includegraphics[width=2.0\columnwidth,keepaspectratio]{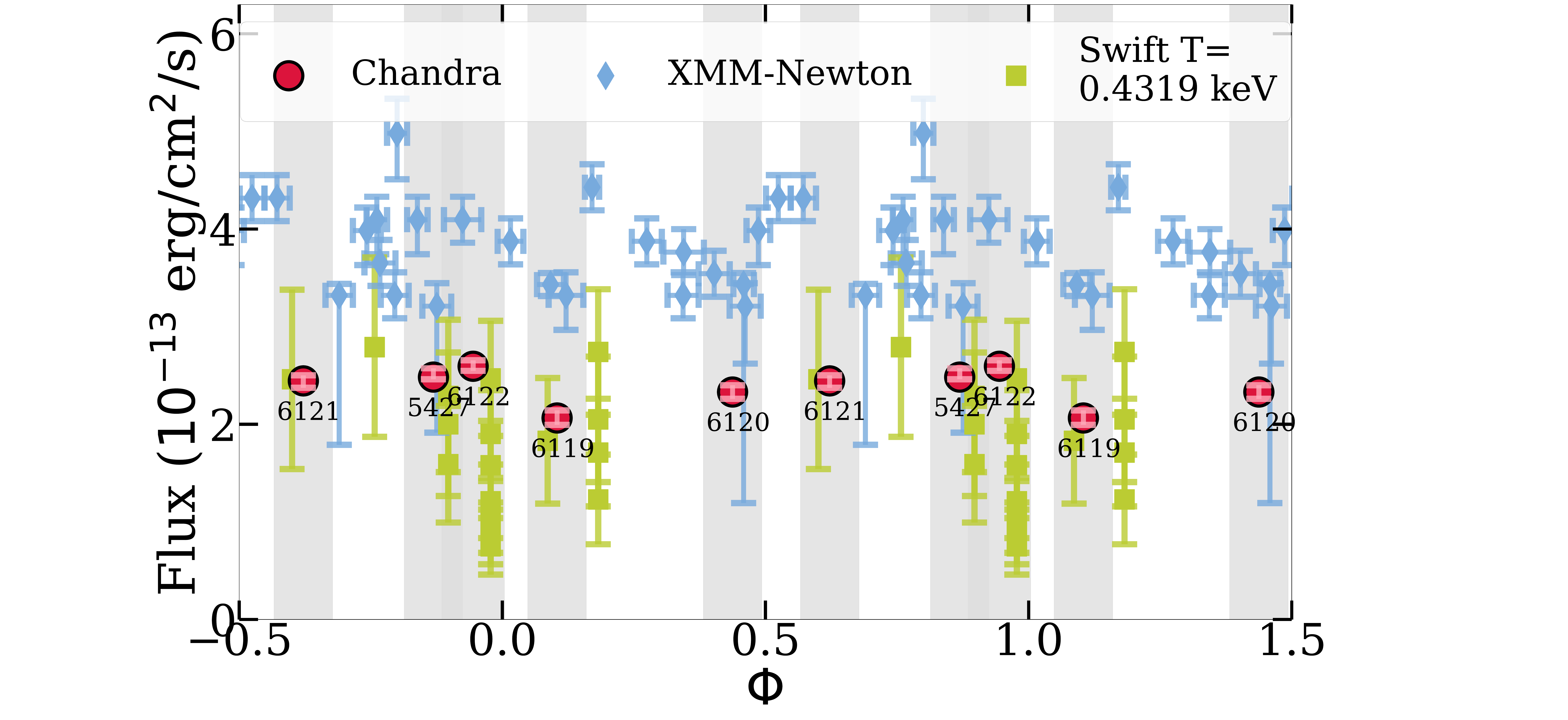}
\includegraphics[width=2.0\columnwidth,keepaspectratio]{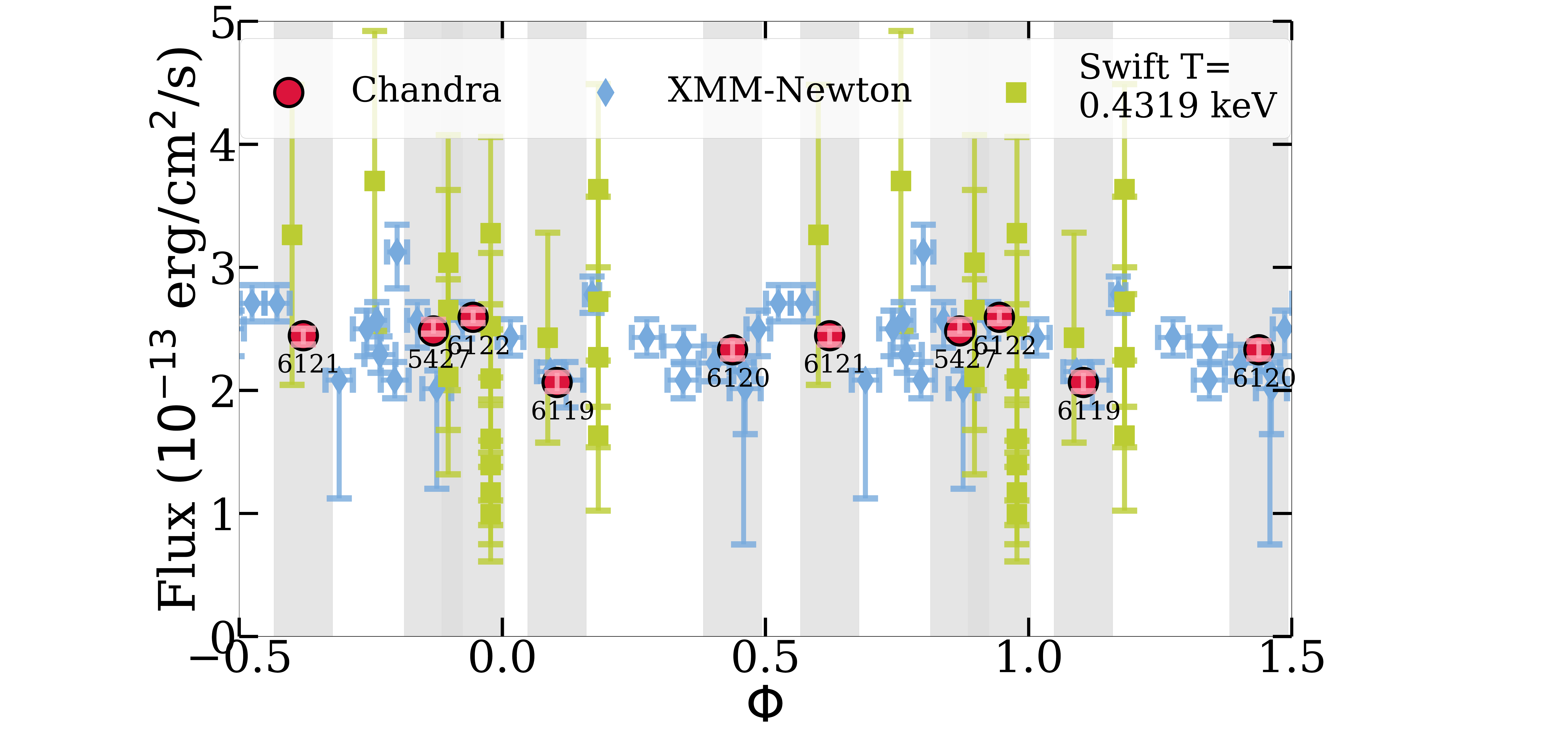}
\caption{Measured X-ray fluxes of \egp\ over phase $\Phi$.  The {\sl top panel} shows raw fluxes obtained from \xmm\ \citep[blue points;][]{Scandariato_2013}, \swift\ \citep[green points;][]{D_Elia_2013}, and the \chandra\ fluxes (red points) measured here (see Section~\ref{sec:spec}).  The {\sl bottom panel} shows the same data, with the \xmm\ and \swift\ fluxes separately normalized \updatebf{to match \chandra\ on average}.  The grey bands show the \chandra\ phase coverage of {\egp}b \updatebf{(see Table~\ref{table:obslog})}.
}  
\label{fig_fluxes}
\end{figure*}

\updatebf{To augment phase coverage and explore periodicities, we supplement the \chandra\ fluxes with those from \xmm\ \citep[from][]{Scandariato_2013} and \swift\ \citep[from][]{D_Elia_2013}.  There is an apparent flare in two of the \swift\ observations with fluxes $>$7$\times$10$^{-13}$~ergs~s$^{-1}$~cm$^{-2}$ which we exclude from further consideration here since isolated flares are detrimental to assessments of periodicties.  These fluxes are shown in Figure~\ref{fig_fluxes} along with the \chandra\ fluxes estimated here (see Section~\ref{sec:spec}).  The upper panel of Figure~\ref{fig_fluxes} presents the fluxes as originally estimated and the lower panel shows the \xmm\ and \swift\ fluxes multiplied by factors $0.63\times$ and $1.3\times$ to match \chandra\ fluxes on average.  These renormalization factors correct for various systematic effects, like calibration offsets (a few percent for \xmm\ and $\approx$15\% for \swift; see \citealt{2021AJ....162..254M}) passband effects\footnote{The \xmm\ fluxes are estimated for a passband of 0.2-2.5~keV and the \swift\ fluxes for a passband of 0.3-3.0~keV.  For an $\approx$5~MK plasma with metallicity of 0.2 (see Table~\ref{table:fitparams}), we compute values of 9\% and 28\% higher in our adopted \chandra\ passband of 0.15-4~keV using WebPimms at \url{https://cxc.cfa.harvard.edu/toolkit/pimms.jsp}.} as well as long-term activity variations across time scales of several years.  Indeed, \citet{Shkolnik_2008} show that there is a slow change or $\approx$10\% in the baseline \updatetwo{Ca\,II~K} band flux over a time scale of $\approx$5 years; by renormalizing the fluxes from each campaign, we also remove the effects of such long-term activity variations in periodicity estimations.
}

\section{Analysis} \label{sec:analysis}

\subsection{Spectral Fits} \label{sec:spec}

We carry out a detailed analysis of the \chandra\ ACIS-S (Advanced CCD Imaging Spectrometer - Spectroscopic array) spectra of \egp\ obtained during each observation using the Sherpa fitting package \citep{brefsdal-proc-scipy-2011} in \chandra\ Interactive Analysis of Observations \citep[{\sl CIAO} v4.13;][]{Fruscione_2006}.  We adopt a collisionally excited thermal emission model for the corona (The Astrophysical Plasma Emission Code [APEC]; \citealt{Brickhouse_2000}, \citealt{Smith_2001}, \citealt{Foster_2016}), and an interstellar absorption model \citep[T\"{u}bingen-Boulder; see][]{Wilms_2000} with the H column density fixed at $N_{\rm H}=10^{19}$~cm$^{-2}$.  In all cases we compute abundances relative to the solar photosphere as listed by \citet{Anders_1989} \updatebf{(see Appendix~\ref{sec:appendix_abun}).} We carry out the modeling in stages of increasing complexity, evaluating the quality of the fit at stage and continuing with more complex models only if necessary.  Our fit results are thus designed to be robust against fluctuations, and describe the most information that may be reliably gleaned from the data.  The models we consider are:
\begin{enumerate}
    \item[{\bf 1m}] An APEC model with one temperature component, fitting normalization, temperature, and metallicity;
    \item[{\bf 2m}] An APEC model with two temperature components, fitting the normalization and temperature for each component separately and the metallicity jointly;
    \item[{\bf 2v}] As in {\bf Model 2m}, but allowing the abundances to vary freely in groups based on similar First Ionization Potential (FIP) values, as
    \{C,O,S\}, \{N,Ar\}, \{Ne\}, \{Mg,Al,Si,Ca,Fe,Ni\}; and
    \item[{\bf 2v/Z}] As in {\bf Model 2v}, but the abundance of a selected element Z fitted separately.  E.g., for model {\bf 2v/Al}, the abundances are fit in groups \{C,O,S\}, \{N,Ar\}, \{Ne\}, \{Mg,Si,Ca,Fe,Ni\}, \{Al\}.  This is designed to explore whether enhancements of specific elements are measurable.
\end{enumerate}

We carry out the fits in all cases by minimizing the c-statistic \citep[\cstat; see][]{1979ApJ...228..939C} over the passband 0.4-7.0~keV.  Background is negligible ($<0.5$\%, see Table~\ref{table:summaryobs}) in all cases and is ignored.
We use the MCMC-based {\tt pyBLoCXS} method \citep[][]{van_Dyk_2001} as implemented in Sherpa.  We compute the observed \cstat$_{obs}$, and the nominal expected \cstat$_{model}$, and the variance $\sigma^2_\cstat$ as described by \citet{Kaastra_2017}, and reject the model as an adequate fit if the difference between the observed and model expected values, $\Delta\cstat=|\cstat_{obs}-\cstat_{model}|$ exceeds $2\sigma_\cstat$.

\begin{figure*}[!htbp]
\centering
\includegraphics[width=1.05\columnwidth,keepaspectratio]{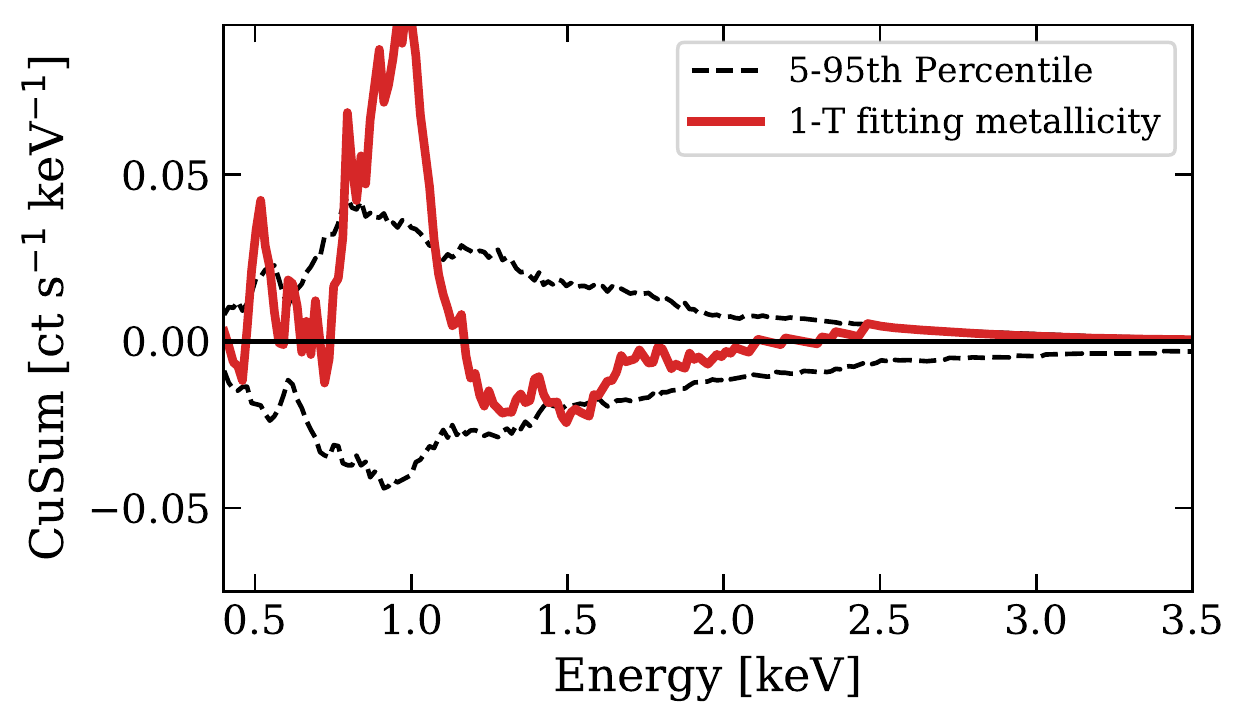}
\includegraphics[width=1.0\columnwidth,keepaspectratio]{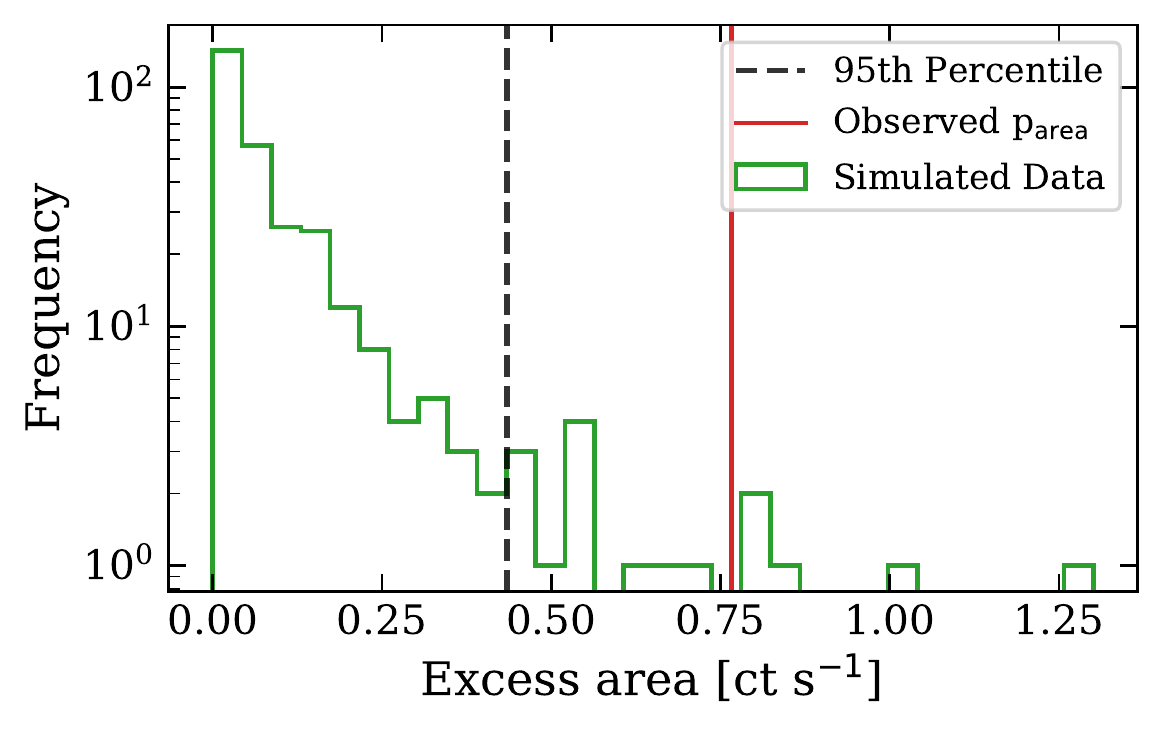} \\
\includegraphics[width=1.05\columnwidth,keepaspectratio]{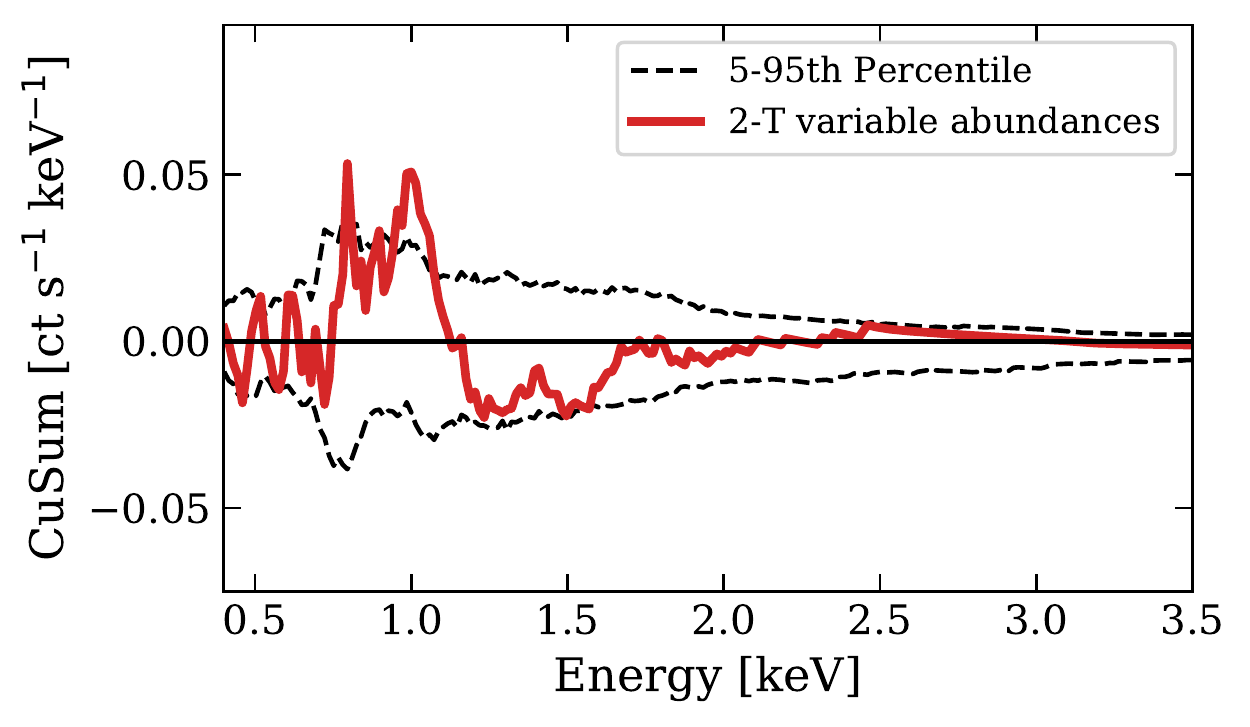}
\includegraphics[width=1.0\columnwidth,keepaspectratio]{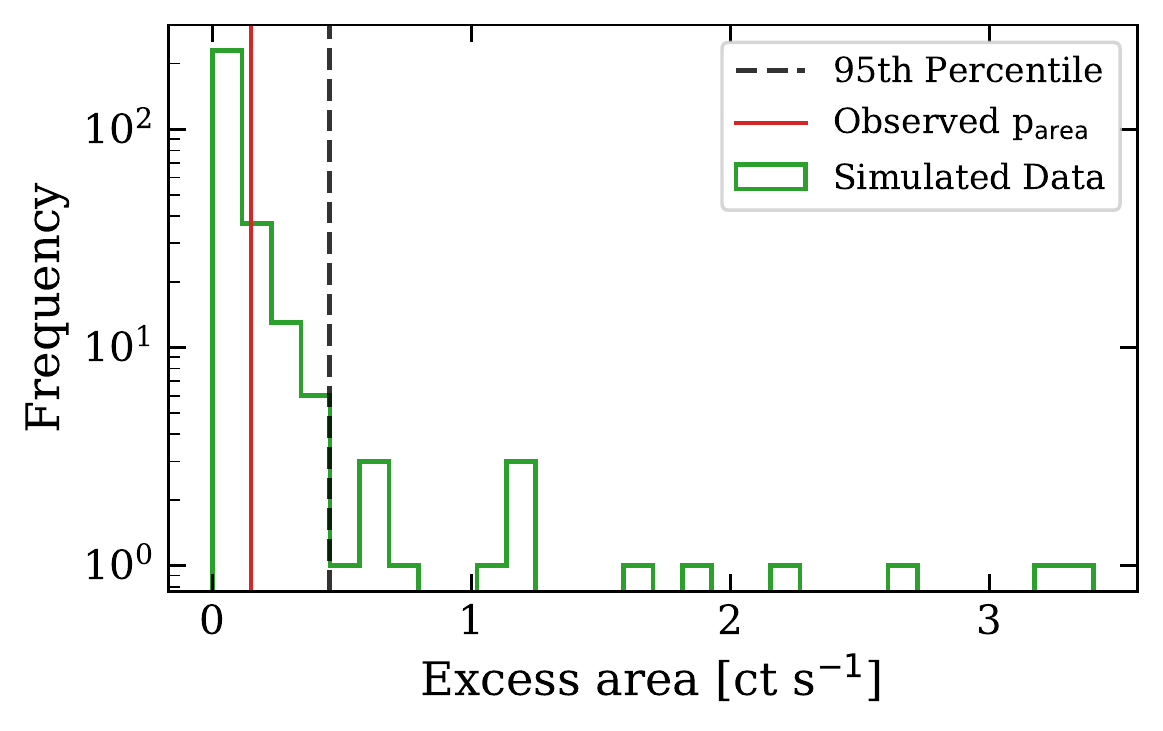}
\caption{Demonstrating improving spectral fit quality by analysis of residuals in ObsID~6119.  The panels at the {\sl top} use model {\bf 2m} and those at the {\sl bottom} use model {\bf 2v}.  The cumulative sum of residuals (CuSum) is shown at {\sl left} for the data (red curve) along with the $5-95$\% bounds determined from simulations (dashed curves), and the magnitudes of the area deviations are shown at {\sl right} as the null distribution derived from the simulations (green histogram), and the observed deviation corresponding to p$_{area}$ (red vertical line), and the threshold we use to stop the exploration of more complex models (i.e., when $p_{area}>0.05$ we desist from rejecting the null; black dashed vertical line; \updatebf{see text}). 
} 
\label{fig_cusumcompare}
\end{figure*}

When $\Delta\cstat$ is deemed acceptable, we further analyze the residuals to the spectral fit, since acceptable fits may still be accompanied by structures in residuals that indicate that some aspect of the model is inadequate over small energy ranges.  We evaluate the quality of the residuals by computing the cumulative sum of the residuals (CuSum) and comparing it against a null distribution generated from the fitted model.  We generate fake spectra based on the best-fit parameter values, and build up a set of CuSum curves.  Then we calculate measures that test both the width and strength of the residual structure by carrying out two tests:
\begin{enumerate}
    \item First, we compute the $\pm$90\% point-wise envelope of CuSum over the energy range 0.5-3.0~keV, and evaluate the percentage of bins, pct$_{CuSum}$, where the observed CuSum exceeds the 90\% bounds.  If this percentage is ${\gtrsim}10$\%, the model is considered inadequate and the next stage of complexity in the model is considered.  If, on the other hand, the percentage is ${\ll}10$\% this is taken as a sign of {\sl overfitting}, and the less-complex model is accepted.  
    \item Second, for each simulated spectrum, we calculate the total area of the CuSum curve that falls outside the 5-95\% bounds from the ensemble of the simulations and construct a null distribution of the excess area beyond the 90\% bounds.  We then compute the same quantity for the observed CuSum curve, and hence the corresponding $p$-value ($p_{area}$) with reference to the null distribution.  This process is similar to the posterior predictive $p$-value calibration procedure described by \citet{2002ApJ...571..545P}.  If this $p_{area}{\ll}0.05$, we consider the corresponding model an inadequate fit.
\end{enumerate}
These tests are illustrated in Figure~\ref{fig_cusumcompare}.  The left panels in Figure~\ref{fig_cusumcompare} show two examples of pct$_{CuSum}$ calculated for two successive models ({\bf 2m} at the top and {\bf 2v} at the bottom) for the same dataset (ObsID~6119).  In both cases, the $\Delta$\cstat\ is acceptable ($+0.3$ for model {\bf 2m} and $-0.86$ for model {\bf 2v});  using just this information, there is no reason to prefer the more complex spectral model.  Considering the fraction of bins that show systematic runs shows that the structure in the residuals is reduced for model {\bf 2v}, with pct$_{CuSum}$ dropping \updatebf{from $8.6$\% to $4.2$\%}.  However, the magnitude of the residuals is reduced significantly for model {\bf 2v}: the right panels show the computation of $p_{area}$ for the same models as in the corresponding left panels, and we find $p_{area}$ changes from an unacceptable 0.01 for model {\bf 2m} \updatebf{($p_{area}\ll{0.05}$ makes it an inadequate fit)}, to an acceptable $0.20$ for model {\bf 2v} \updatebf{($p_{area}>{0.05}$, and the model cannot be rejected)}.  We thus accept model {\bf 2v} as the preferred spectral model over model {\bf 2m}.

\begin{figure}[!htbp]
\centering
\includegraphics[width=0.75\columnwidth,keepaspectratio]{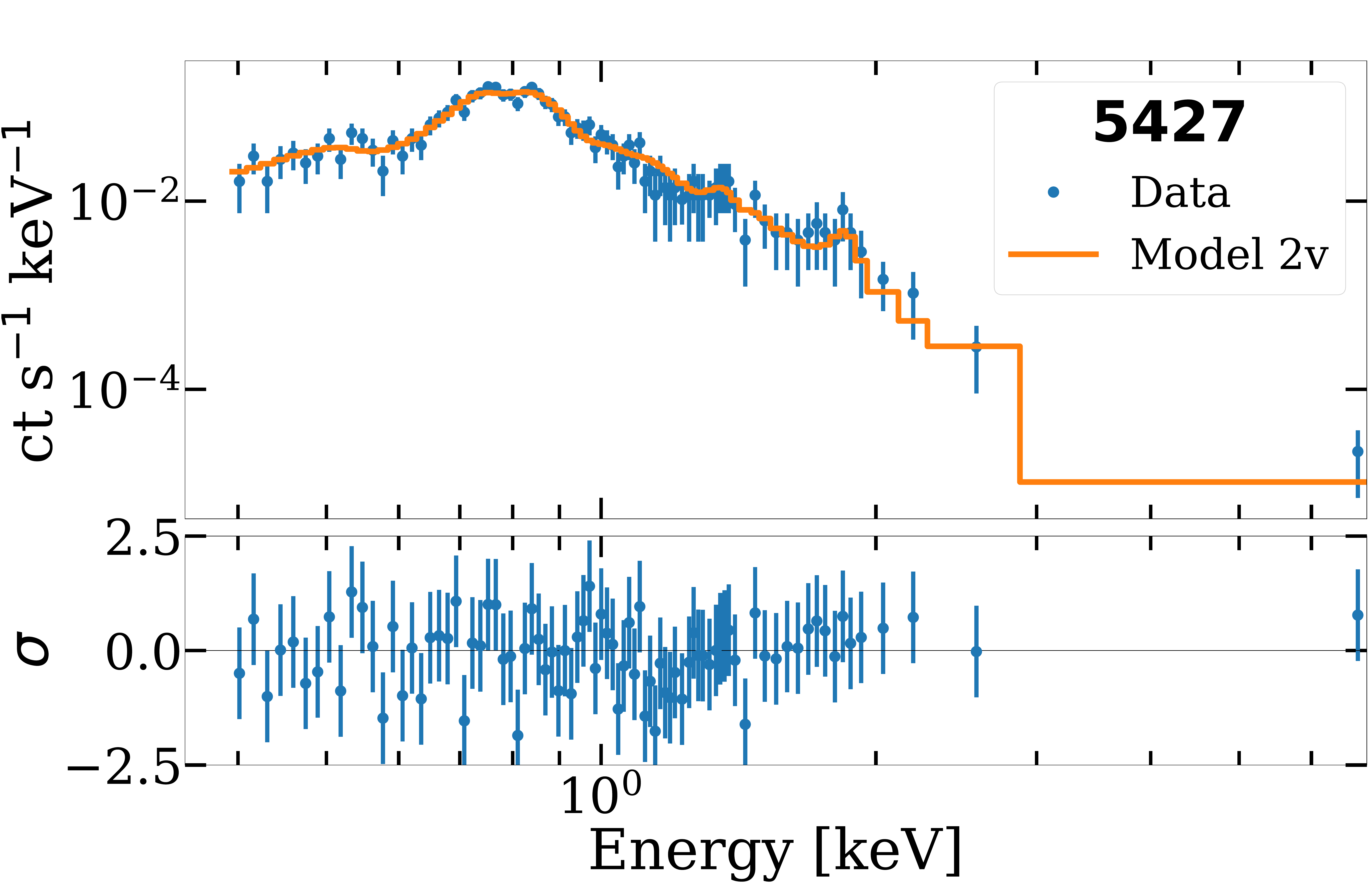} 
\includegraphics[width=0.75\columnwidth,keepaspectratio]{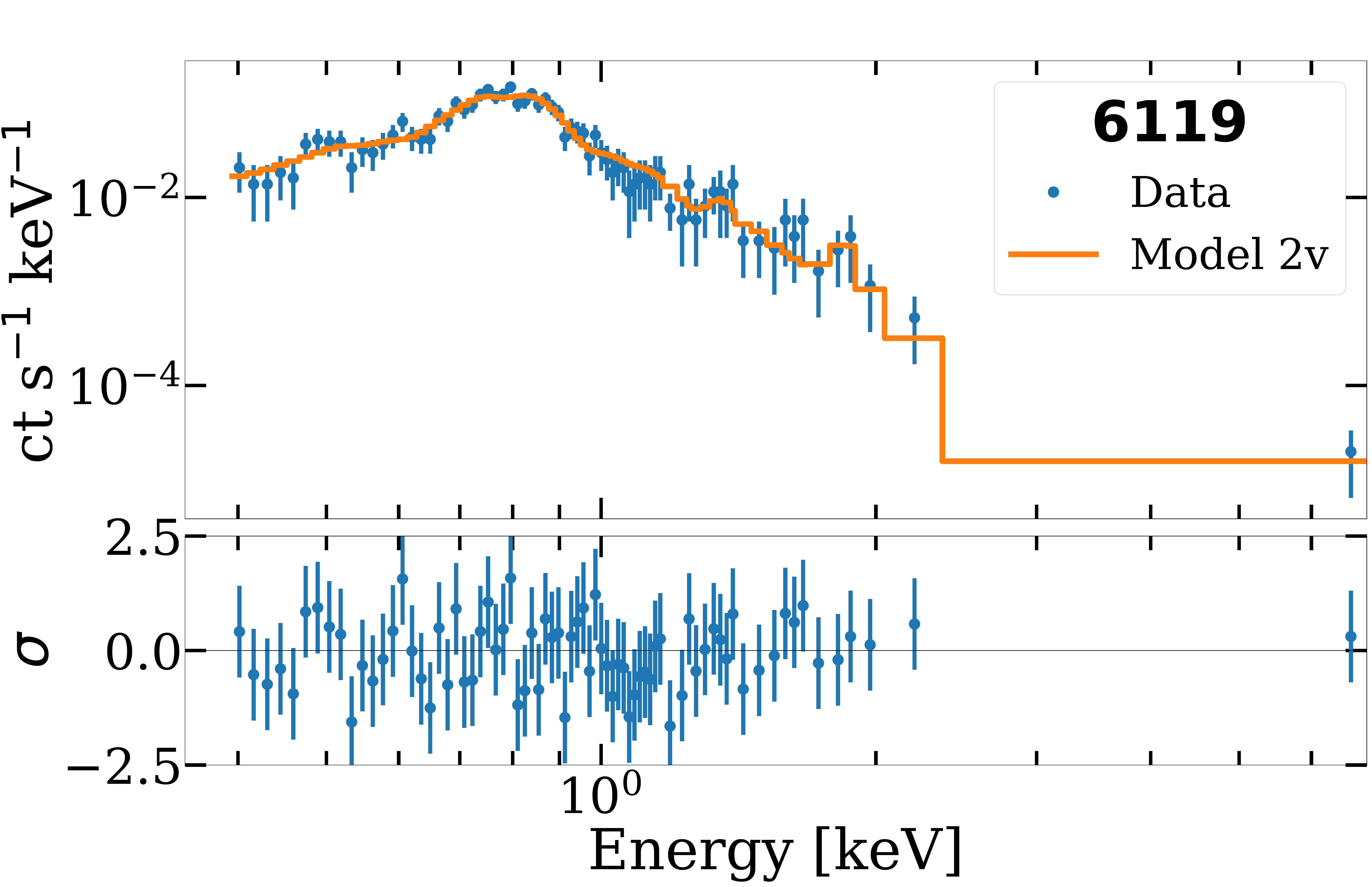} \\
\includegraphics[width=0.75\columnwidth,keepaspectratio]{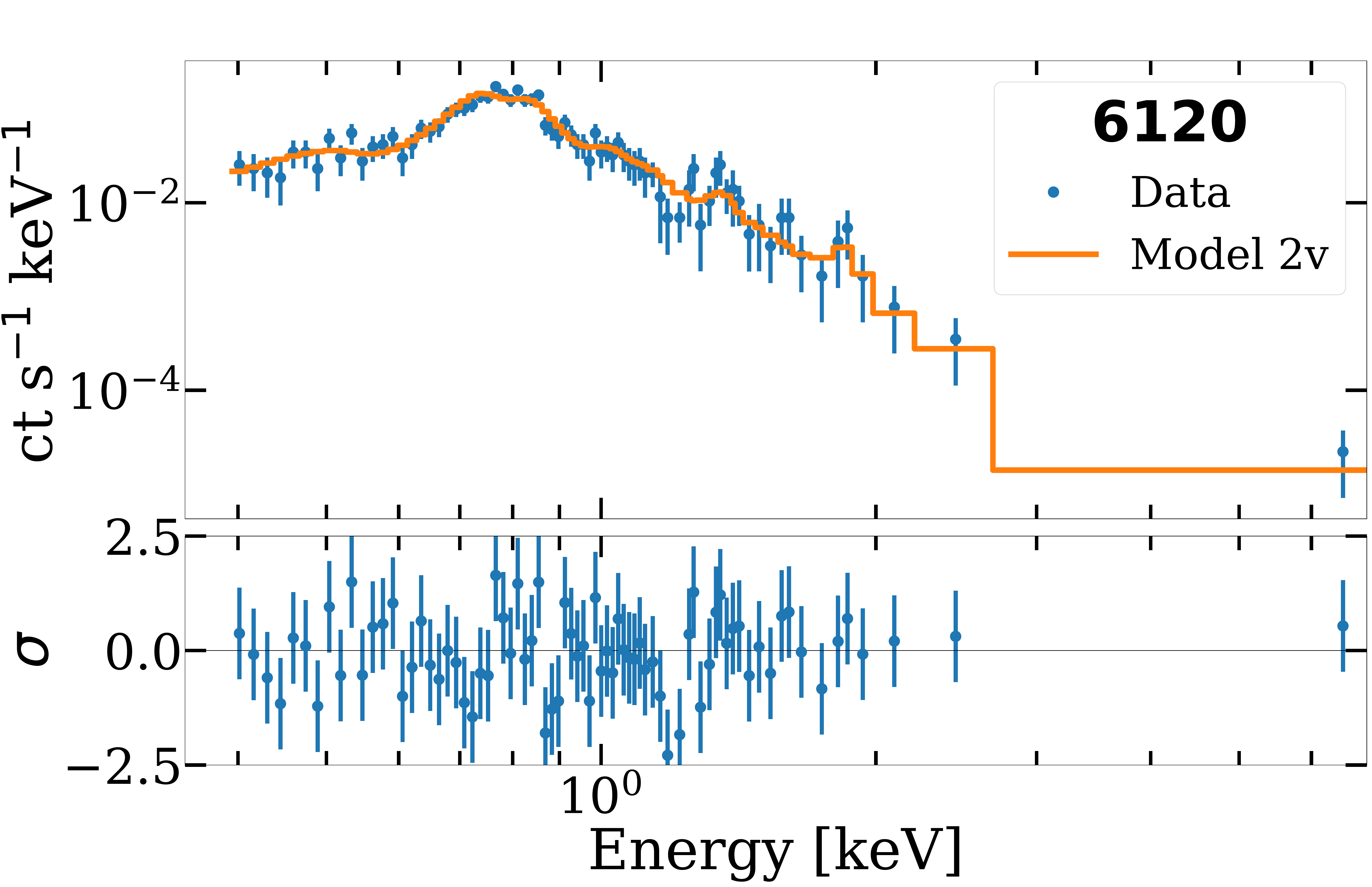}
\includegraphics[width=0.75\columnwidth,keepaspectratio]{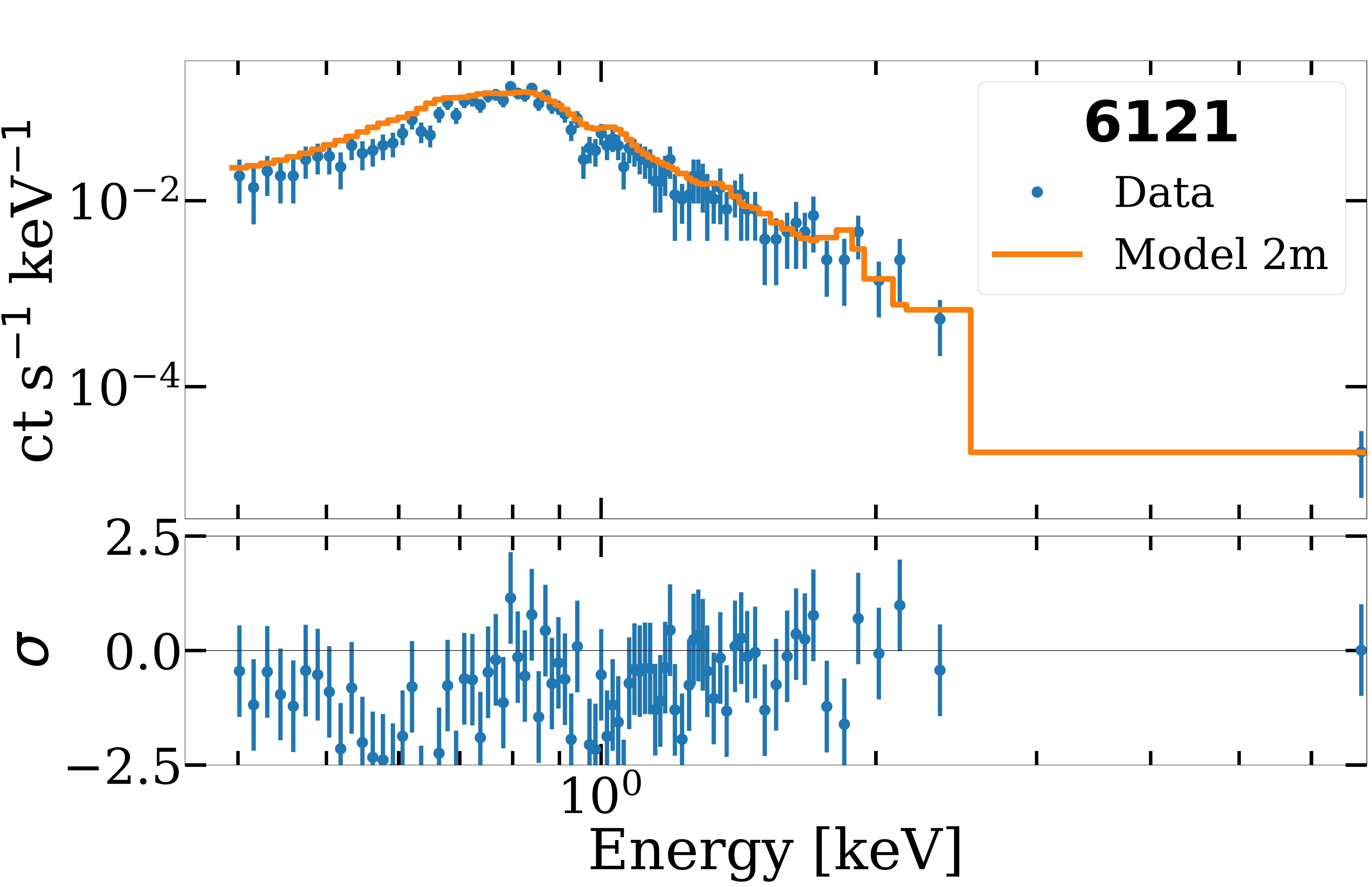}
\includegraphics[width=0.75\columnwidth,keepaspectratio]{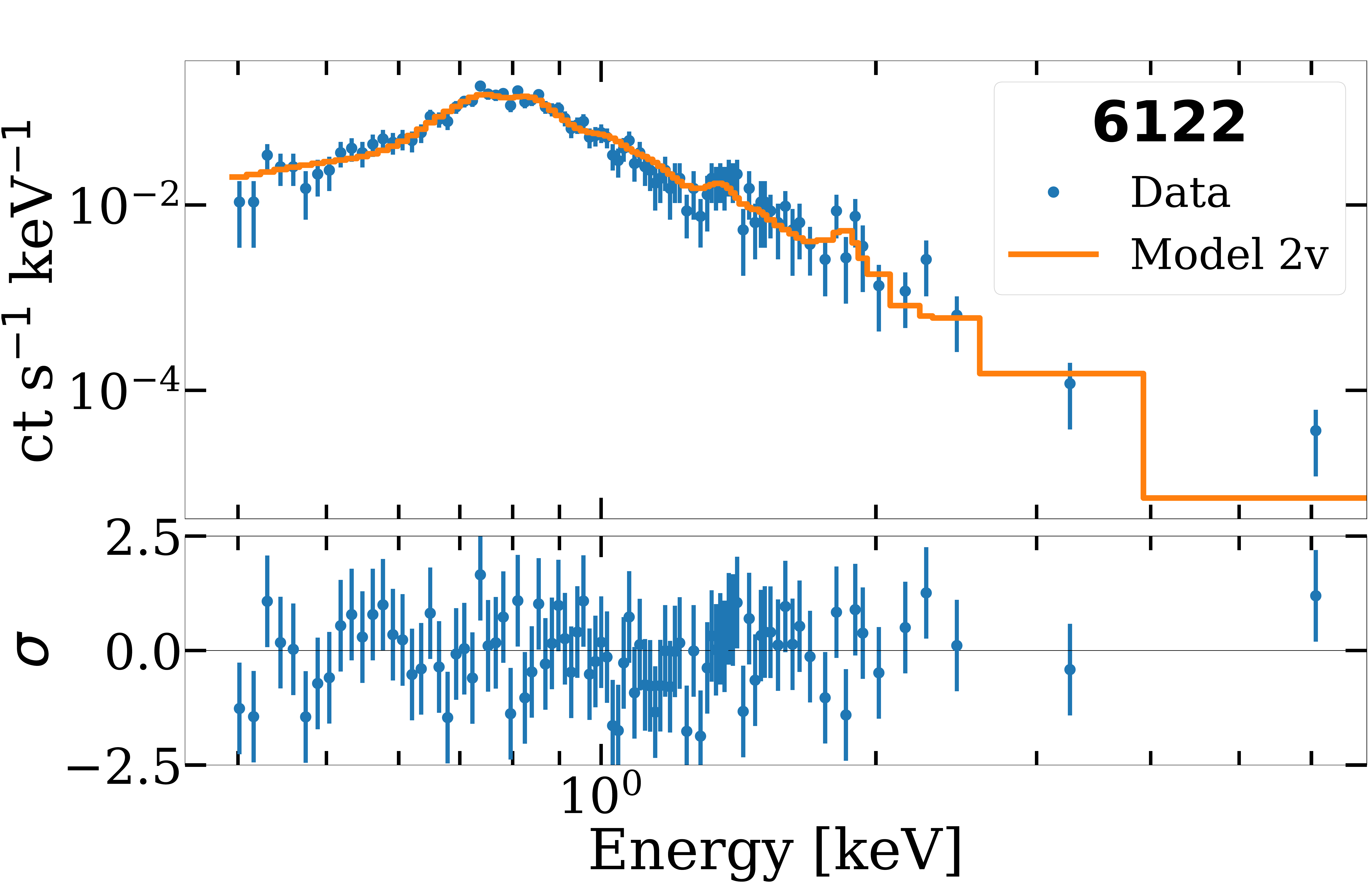}
\caption{Model fits (Table~\ref{table:fitparams}) and residuals for each \chandra\ observation.  The data are shown as the blue points with Gehrels error bars, grouped by 5 counts in each bin for the sake of visualization, and the model is shown as the orange stepped curve.  The residuals relative to the errors are shown in the lower segment of each panel.} 
\label{fig_specfit}
\end{figure}

\begin{deluxetable*}{cccccc}
\tabletypesize{\footnotesize}
\tablecolumns{8}
\tablewidth{0pt}
\tablecaption{Parameters of \updatebf{Fitted} Spectral Models \label{table:fitparams}}
\tablehead{
\colhead{Parameters$^\ddag$} & \multicolumn{5}{c}{Observation IDs (ObsIDs)} \\
\colhead{} & \colhead{5427}& \colhead{6119} & \colhead{6120} & \colhead{6121$^*$}& \colhead{6122}
}
\startdata
\vspace{-0.05cm}T$_{\text{Low}}$ [MK] & 2.00$^{+3.73}_{-2.00}$ & 2.09$^{+0.75}_{-0.75}$ & 4.87$^{+0.67}_{-0.67}$ & 2.55$^{+1.94}_{-1.94}$ & 5.10$^{+1.67}_{-1.67}$ \\
\vspace{0.15cm} & 7.02$[1.86-9.86]$ & 2.06$[1.64-3.24]$ & 5.11$[4.64-5.80]$ & 2.60$[1.28-6.73]$ & 5.80$[4.99-6.73]$ \\
\vspace{-0.05cm}T$_{\text{High}}$ [MK] & 6.61$^{+0.31}_{-0.31}$ & 6.62$^{+0.32}_{-0.32}$ & 11.02$^{+1.64}_{-1.64}$ & 6.73$^{+0.86}_{-0.86}$ & 10.79$^{+3.40}_{-3.40}$ \\
\vspace{0.15cm} & 6.63$[5.10-7.54]$ & 6.73$[6.26-7.31]$ & 11.71$[5.34-19.49]$ & 6.61$[5.22-7.54]$ & 10.97$[9.05-17.86]$ \\
\vspace{-0.05cm}N$^\mathparagraph$ & 1.25$^{+1.31}_{-1.25}$  & 2.10$^{+2.73}_{-2.10}$ & 0.69$^{+0.94}_{-0.69}$ & 0.21 & 0.00$^{+0.60}_{-0.00}$ \\
\vspace{0.15cm} & 1.79$[1.03-4.24]$  & 1.48$[1.62-4.49]$ & 0.76$[0.10-2.22]$ & 0.18 & 0.10$[0.00-0.71]$ \\
\vspace{-0.05cm}O$^\mathparagraph$ & 0.12$^{+0.16}_{-0.12}$ & 0.39$^{+0.38}_{-0.39}$ & 0.06$^{+0.08}_{-0.06}$ & 0.21 & 0.17$^{+0.23}_{-0.23}$ \\
\vspace{0.15cm} & 0.13$[0.05-0.36]$ & 0.50$[0.12-0.57]$ & 0.09$[0.01-0.19]$ & 0.18 & 0.27$[0.11-0.40]$ \\
\vspace{-0.05cm}Ne$^\mathparagraph$ & 0.00$^{+0.09}_{-0.00}$ & 0.00$^{+0.21}_{-0.00}$ & 0.00$^{+0.09}_{-0.00}$ & 0.21 & 0.00$^{+0.08}_{-0.00}$ \\
\vspace{0.15cm} & 0.02$[0.00-0.13]$ & 0.03$[0.00-0.16]$ & 0.04$[0.00-0.14]$ & 0.18 & 0.01$[0.00-0.10]$ \\
\vspace{-0.05cm}Fe$^\mathparagraph$ & 0.20$^{+0.05}_{-0.05}$ & 0.45$^{+0.23}_{-0.23}$ & 0.20$^{+0.08}_{-0.08}$ & 0.21$^{+0.10}_{-0.10}$ & 0.21$^{+0.09}_{-0.09}$ \\
\vspace{0.15cm} & 0.27$[0.18-0.32]$ & 0.48$[0.31-0.73]$ & 0.21$[0.13-0.30]$ & 0.18$[0.14-0.24]$ & 0.23$[0.16-0.29]$ \\
\vspace{-0.1cm}EM$_{\rm Low}^{\dag}$ & 0.26$^{+0.57}_{-0.26}$ & 0.38$^{+0.53}_{-0.38}$ & 2.67$^{+1.02}_{-1.00}$ & 0.52$^{+0.63}_{-0.27}$ & 2.34$^{+0.35}_{-0.71}$ \\
\vspace{0.15cm} & 0.32$[0.07-1.04]$ & 0.16$[0.03-0.69]$ & 2.04$[1.57-2.90]$ & 0.62$[0.18-1.41]$ & 2.22$[1.54-2.81]$ \\
\vspace{-0.05cm}EM$_{\rm High}^{\dag}$ & 2.40$^{+0.55}_{-0.59}$ & 0.95$^{+0.43}_{-0.43}$ & 0.33$^{+0.21}_{-0.22}$ & 2.02$^{+1.43}_{-0.62}$ & 0.75$^{+0.53}_{-0.72}$ \\
\vspace{0.15cm} & 1.66$[0.49-2.36]$ & 0.78$[0.51-1.09]$ & 0.05$[0.00-0.30]$ & 1.85$[1.22-2.84]$ & 0.33$[0.11-0.68]$ \\
Flux (0.15-4~keV) & 2.48$^{+0.09}_{-0.02}$ & 2.06$^{+0.08}_{-0.07}$ & 2.33$^{+0.07}_{-0.07}$ & 2.44$^{+0.06}_{-0.07}$ & 2.59$^{+0.06}_{-0.05}$ \\
{[}$10^{-13}$ erg~s$^{-1}$~cm$^{-2}${]} & & & & & \\
\hline
\multicolumn{6}{c}{Fit characteristics} \\
\hline
model & {\bf 2v} & {\bf 2v} & {\bf 2v} & {\bf 2m}$^*$ & {\bf 2v} \\
\cstat/dof & 164.444/445 & 140.063/445 & 184.616/445 & 201.862/448 & 218.968/445 \\
$\frac{\Delta\cstat}{\sigma_{\cstat}}$ & $-0.30$ & $-0.86$ & $+0.91$ & $+1.82$ & $+1.30$ \\
pct$_{CuSum}$ [\%] & 11.26 & 4.19 & 5.74 & 4.86 & 12.58 \\
$p_{area}$ & 0.09 & 0.20 & 0.10 & 0.23 & 0.07 \\
 \enddata  
\tablenotetext{$\ddag$}{The upper row represents the best-fit value and the \updatebf{95\%} equal-tail bounds on the parameter, while the lower row represents the mode and the 90\% highest-posterior density intervals \updatebf{obtained using MCMC sampling with {\tt pyBLoCXS}}.}
\tablenotetext{*}{The best-fit model for ObsID~6121 is model {\bf 2m}, with all abundances tied together.
}
\tablenotetext{\mathparagraph}
{ Fractional number abundances $\frac{A(X)/A(H)}{(A(X)/A(H))_{\angr}}$ relative to solar photospheric \citep{Anders_1989}. }
\tablenotetext{$\dag$}{Emission Measure at the source, $EM={\rm normalization} \times 4~\pi \times {\rm distance}^2$, in units of [$10^{51}$~cm$^{-3}$].}
\end{deluxetable*}

\begin{deluxetable*}{lcccccc}[!hbt]
\tabletypesize{\footnotesize}
\tablecolumns{6}
\tablewidth{0pt}
\tablecaption{\updatebf{Abundance measurements used for F$_{bias}$ calculation} \label{table:Fbiassummary}}
\tablehead{ 
\colhead{ObsID} &\colhead{$\rm [C/Fe]_{\coro|\angr}$} &\colhead{$\rm [O/Fe]_{\coro|\angr}$} & \colhead{$\rm [N/Fe]_{\coro|\angr}$} & \colhead{$\rm [Ne/Fe]_{\coro|\angr}$} & \colhead{F$_{bias}|_{\luck,\ecu}$} & \colhead{F$_{bias}|_\brwr$}}
\startdata
5427$~^a$ & -0.15 $\pm$ 0.26 & -0.15 $\pm$ 0.26 & 0.96 $\pm$ 0.19 & -0.58 $\pm$ 0.46 & -0.28 $\pm$ 0.33 & -0.04 $\pm$ 0.31 \\
6119$~^a$ & -0.08 $\pm$ 0.19 & -0.08 $\pm$ 0.19 & 0.69 $\pm$ 0.29 & -0.87 $\pm$ 0.46 & -0.39 $\pm$ 0.32 & -0.16 $\pm$ 0.30 \\
6120$~^a$ & -0.37 $\pm$ 0.33 & -0.37 $\pm$ 0.33 & 0.70 $\pm$ 0.29 & -0.56 $\pm$ 0.46 & -0.45 $\pm$ 0.38 & -0.22 $\pm$ 0.36 \\
6121$~^{a,b}$ & 0 $\pm$ 0.1 & 0 $\pm$ 0.1 & 0 $\pm$ 0.1 & 0 $\pm$ 0.1 & -0.30 $\pm$ 0.15 & -0.07 $\pm$ 0.11 \\
6122$~^a$ & 0.04 $\pm$ 0.17 & 0.04 $\pm$ 0.17 & 0.12 $\pm$ 0.49 & -0.78 $\pm$ 0.56 & -0.45 $\pm$ 0.40 & -0.22 $\pm$ 0.39 \\
\hline
\multicolumn{7}{c}{$\langle$F$_{bias}\rangle|_{\luck,\ecu}~^c= -0.33 \pm 0.03~{\rm scatter} \pm 0.32~{\rm statistical}$} \\
\multicolumn{7}{c}{$\langle$F$_{bias}\rangle|_\brwr~^c= -0.09 \pm 0.02~{\rm scatter} \pm 0.29~{\rm statistical}$ } \\
\vspace{-1em} \enddata  
\tablenotetext{$a$}{\updatebf{The mean and standard deviations of the coronal abundances ${\rm [X/Fe]}_{\coro|\angr}$ computed from the MCMC draws for the best-fit spectral model (see Section~\ref{sec:spec}, Table~\ref{table:fitparams}).  The corresponding photospheric abundances are recast into the same baseline ($\angr$; see Appendix~\ref{sec:appendix_abun}).}
}
\tablenotetext{$b$}{\updatebf{Since the best-fit spectral model for ObsID 6121 has all metal abundances tied to Fe, ${\rm [X/Fe]}$ are identically $0$, but we include an error bar due to the uncertainty on ${\rm [Fe/H]}_{\coro|\angr}$.}
}
\tablenotetext{$c$}{\updatebf{Computed as the inverse variance weighted mean, with the first error term signifying the weighted variance and hence the scatter in the estimated values, and the second error term signifying an average measurement error (see Equation~\ref{eq:FIPwERR}).
}
}
\end{deluxetable*}

We report the results of the spectral fits for each Observation ID (ObsID) in Table~\ref{table:fitparams} for the best acceptable model obtained for each case, along with the \cstat, $\Delta$\cstat, pct$_{CuSum}$, and $p_{area}$ for the accepted model.  The corresponding fits and residuals are shown in Figure~\ref{fig_specfit}, and the CuSum plots and distributions are shown in Figure~\ref{fig_cusum} in Appendix~\ref{sec:cusum}.  We note that the data are never fit well with model {\bf 1m}, and never require models more complex than model {\bf 2v}.  ObsID 6121 is fit adequately with model {\bf 2m}; model {\bf 2v} can be rejected due to an abnormally low pct$_{CuSum} = 1.10 \%$.  In no case is model {\bf 2v/Z} or other complex modifications statistically justifiable.  While some exploratory fits indicated that allowing Al to be freely fit produced better fit statistics, there is no reason to choose those fits over those of model {\bf 2v}.  In addition to the plasma temperatures, component normalizations, abundances, and the corresponding fluxes, we also compute the FIP bias (F$_{bias}$; \citealt{Wood_2012}, \citealt{Testa_2015}, \citealt{Wood_2018}). The FIP bias is a {\bf logarithmic} summary of how much the abundances of high-FIP elements like C, N, O, and Ne deviate relative to a low-FIP element like Fe, in the corona relative to the photosphere,
\begin{equation}
    {\rm F}_{bias} = \frac{1}{4} \sum_{X=C,N,O,Ne} \log{\rm [X/Fe]}_{\tt cor} - \log{\rm [X/Fe]_{\phot}} \,.
    \label{eq:fipbias}
\end{equation}
\updatebf{The average value F$_{bias,\rm ObsID}$ and the propagated error bars $\sigma_{bias,\rm ObsID}$ are shown in Table~\ref{table:Fbiassummary} for each ObsID, and these individual measurements are combined to form a weighted estimate
$$\langle {\rm F}_{bias} \rangle = \frac{\sum_{\rm ObsID} {\rm F}_{bias,\rm ObsID} \cdot w_{\rm ObsID}}{\sum_{\rm ObsID} w_{\rm ObsID}} \,,$$
with the weights $w_{\rm ObsID}=\frac{1}{\sigma_{bias,\rm ObsID}^2}$\,.  The weighted variance
$${\rm wtVar}=\sum_{\rm ObsID} \frac{({\rm F}_{bias,\rm ObsID} - \langle {\rm F}_{bias} \rangle)^2\cdot w_{\rm ObsID}^2}{(\sum_{\rm ObsID} w_{\rm ObsID})^2}\,,$$
represents the scatter in the individual estimates \citep[][]{sarndal2003model}.  The average of the individual variances, 
$${\rm avVar}=\frac{1}{5}\sum_{\rm ObsID} \sigma_{bias,\rm ObsID}^2$$
represents an average measurement error.  We find
\begin{equation}
    \langle {\rm F}_{bias} \rangle \pm \sqrt{\rm wtVar} \pm {\rm \sqrt{\rm avVar}} = -0.33 \pm 0.03 \pm 0.32 \,.
    \label{eq:FIPwERR}
\end{equation} on using photospheric abundances of C,O and Fe from \cite{Luck_2018}, and N form \cite{Ecuvillon_2004}, and
\begin{equation}
    \langle {\rm F}_{bias} \rangle \pm \sqrt{\rm wtVar} \pm {\rm \sqrt{\rm avVar}} = -0.09 \pm 0.02 \pm 0.29 \,,
    \label{eq:FIPwERR2}
\end{equation}
using photospheric abundances of C, N, O and Fe from \citet{2016ApJS..225...32B} (see Appendix~\ref{sec:appendix_abun} for values scaled to \citealt{Anders_1989}). Note that in both cases, we adopt ${\rm [Ne/O]}_{\phot|\angr}=+0.48$ from \citet{Drake_2005}. 
\updatetwo{In both cases, we find the estimated F$_{bias}$ to be high compared to the value expected for late-F stars \citep{Wood_2012,Wood_2018,Testa_2015}.  However, these values are consistent with those measured for other hot Jupiter hosting stars with suspected SPI like HD\,189733 and $\tau$\,Boo (see discussion in Section~\ref{sec:discuss}).
}
}

\subsection{Timing Analysis}\label{sec:timing}

\subsubsection{Light Curves} \label{sec:lc}

In order to explore the temporal variability in \egp, we construct light curves of counts in all the passbands of interest (see Table~\ref{table:passbands}).  We employ the Bayesian Blocks algorithm \citep{Scargle_2013} to build adaptively binned count rates, using the implementation by \citet{astropy:2022}.

\begin{figure}[!htbp]
\centering
\includegraphics[width=\columnwidth,keepaspectratio]{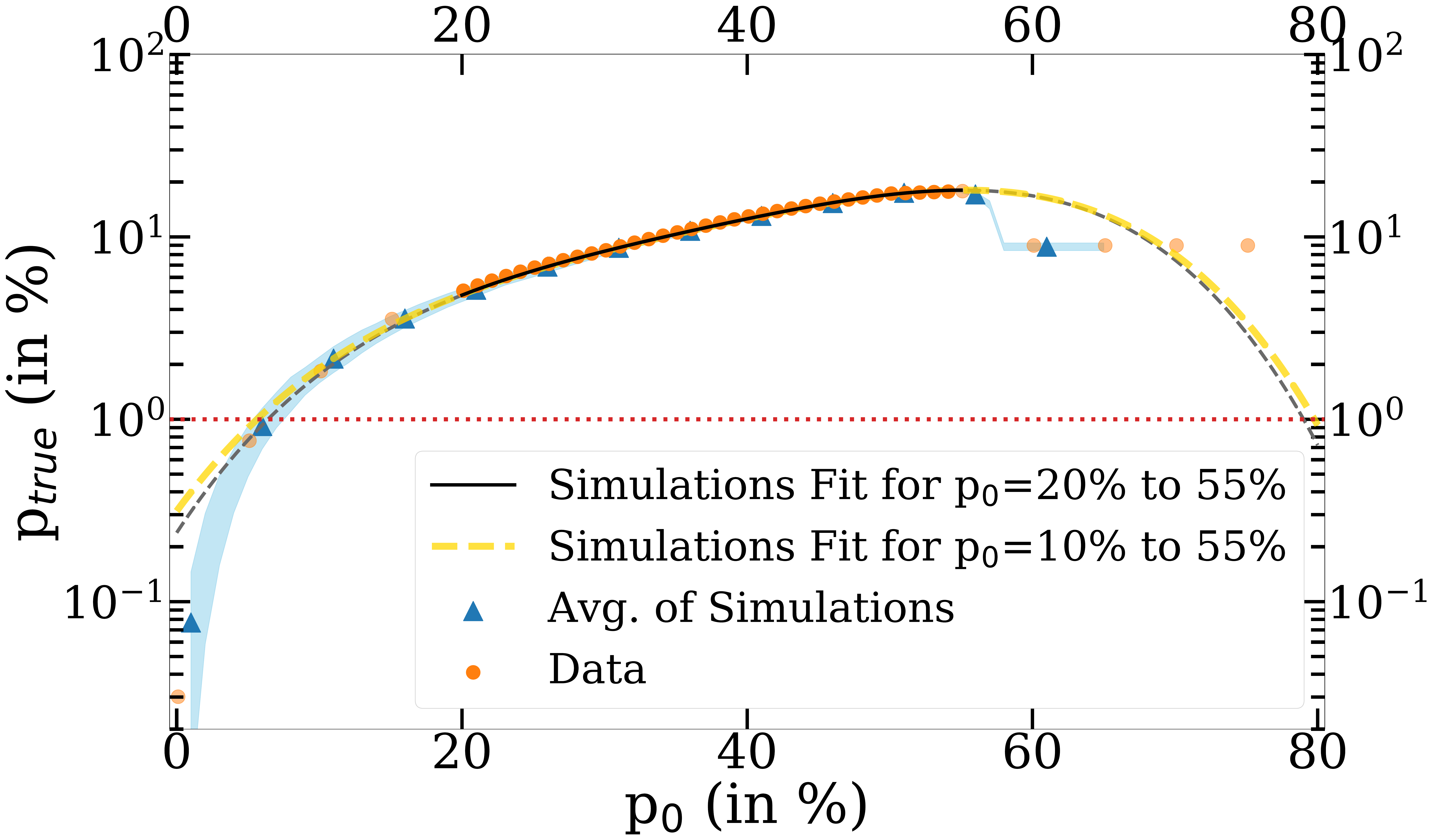}
\caption{Calibrating the correction factor for the False Alarm Probability (FAP; $p_{true}$) for the AstroPy implementation of the Bayesian Blocks algorithm. We fit the $p_0$ value obtained from simulations to the input p0 value using a quartic best-fit equation, after merging all change points separated by $<$30 sec. This provides a corrected $p_0=6$\% as the appropriate value to use for a 1\% FAP. We compute the block representation of light curves using this value (see, e.g., Figure~\ref{fig_broadlc}).}
\label{fig_calibBB}
\end{figure}

We modify the process by which the blocks are constructed in two ways: first, \updatebf{we discard any change points that occur $<10\tau$ after another as unlikely to be physically meaningful ($\tau=3.24104$~s is the CCD readout time in all of the ObsIDs)}; second, we recalibrate the parameter $p_0$ that controls the false alarm probability (FAP) to both account for the $10\tau$ filter as well as to calibrate the steepness of the prior distribution. 
We accomplish this via simulations involving the same number of photons as in the observed data, $9963$, spaced uniformly over a normalized time duration equal to the full observation duration, $152295/\tau$.  That is, each simulation represents events that would be obtained from a flat light curve that has zero true change points. Change points for several values of $p_0$ were computed for the same set of event times.  We discard all change points that follow another by $<10$, and record the number of change points $n^{cp}_{sim}(p_0)$.  Because the simulated events have zero true change points, every change point found, at any given $p_0$, represents false positives.  The fraction 
$$p^{sim}_{true}(p_0)={n^{cp}_{sim}(p_0)}/{N_{sim}}$$
thus represents the actual false alarm probability parameterized by $p_0$ for data sets of the size we are dealing with.  We repeat the simulations 50 times, and estimate ${p_{true}(p_0)}$ as the average over the simulations at each given $p_0$.  The result of these computations is shown in Figure~\ref{fig_calibBB}.  The range of $p_0 \in (10,55)\%$ is approximately linear, and we fit a quartic polynomial to this range and extrapolate it to smaller values of $p_{true}$.  We find that a value of $p_{true}=0.01$, corresponding to a 1\% false alarm probability, is obtained for $p_0{\approx}0.06$, and hence adopt $p_0=6\%$ for all the change point calculations carried out below.

\begin{figure}[!htbp]
\centering
\includegraphics[width=0.7\columnwidth,keepaspectratio]{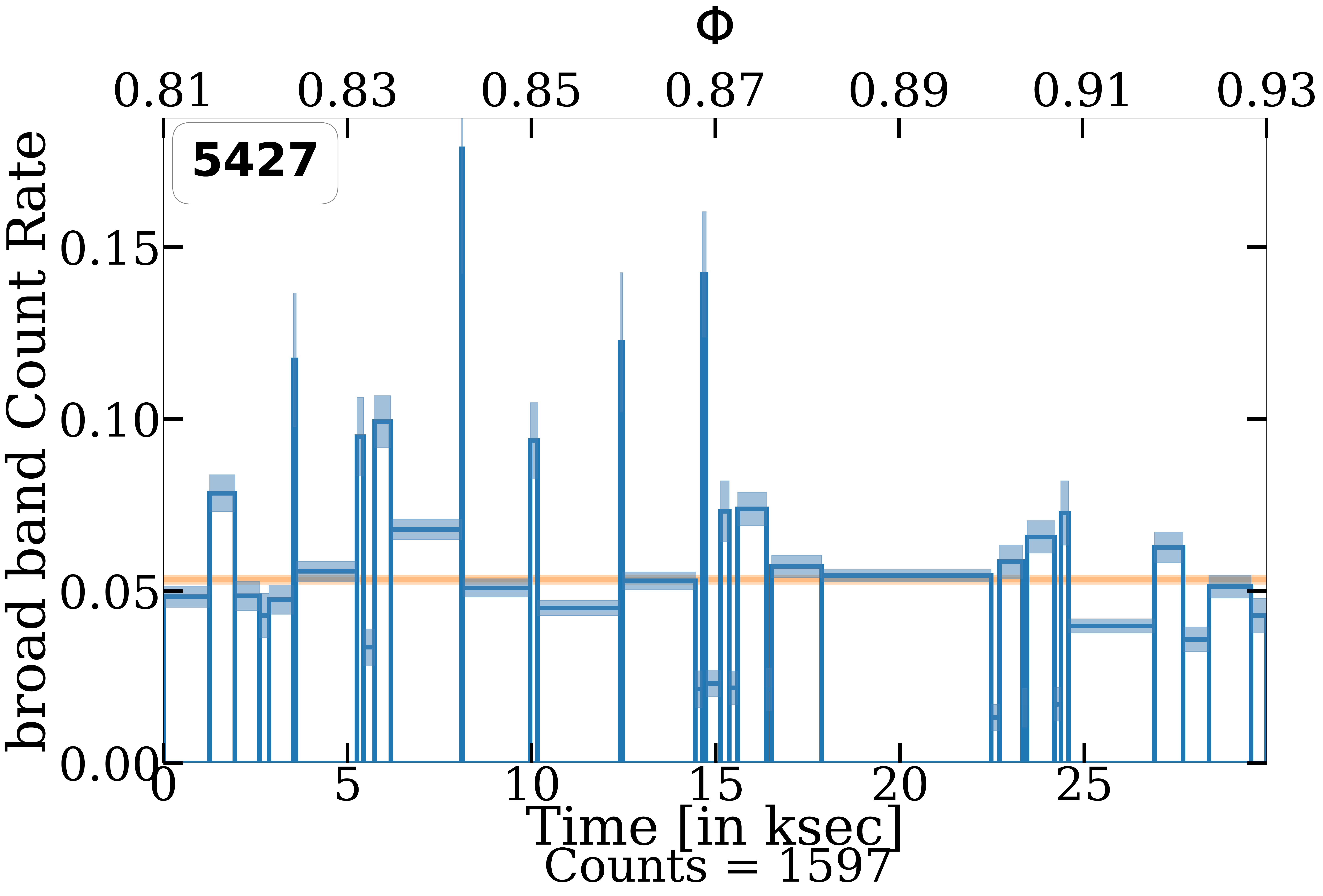}
\includegraphics[width=0.7\columnwidth,keepaspectratio]{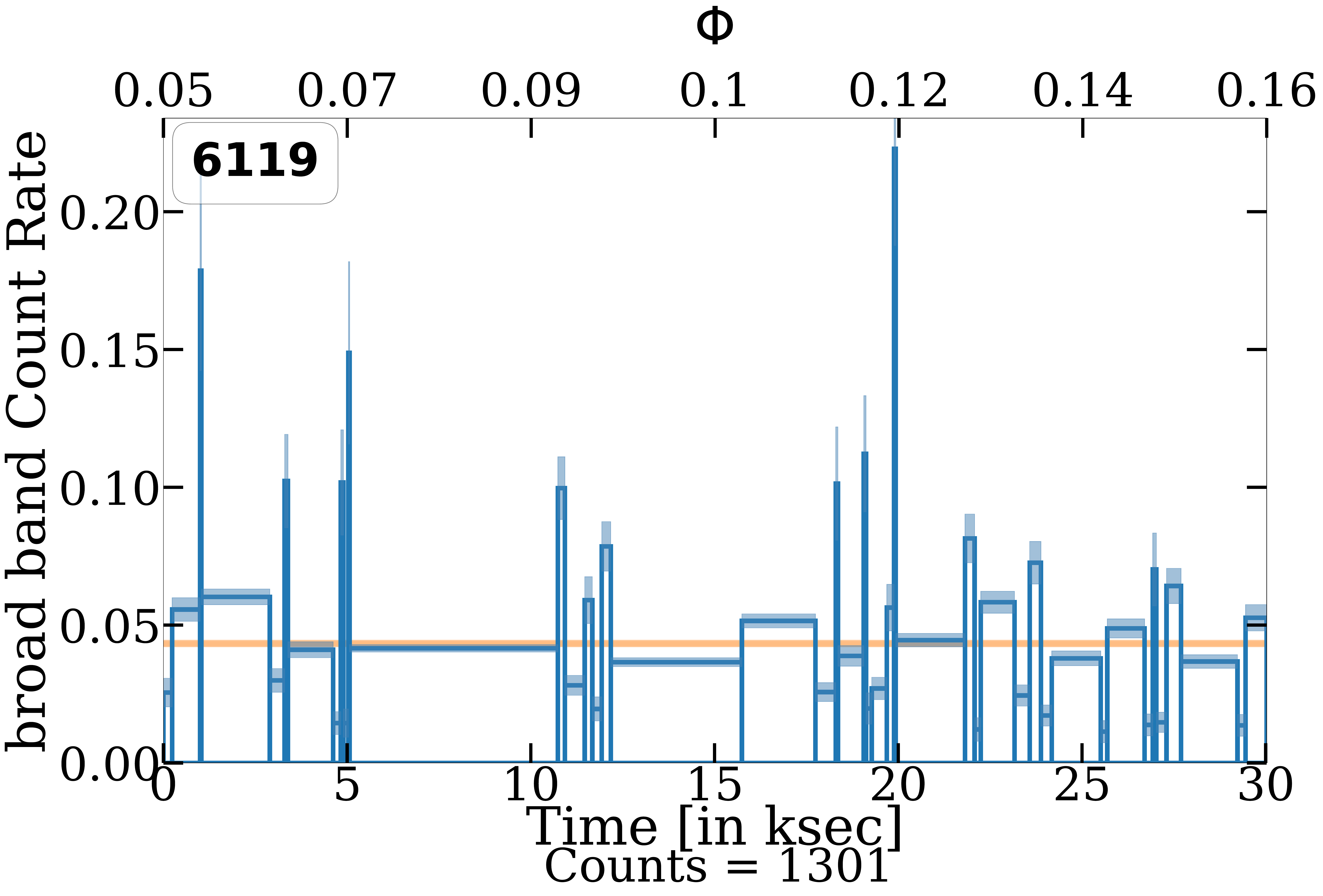}
\includegraphics[width=0.7\columnwidth,keepaspectratio]{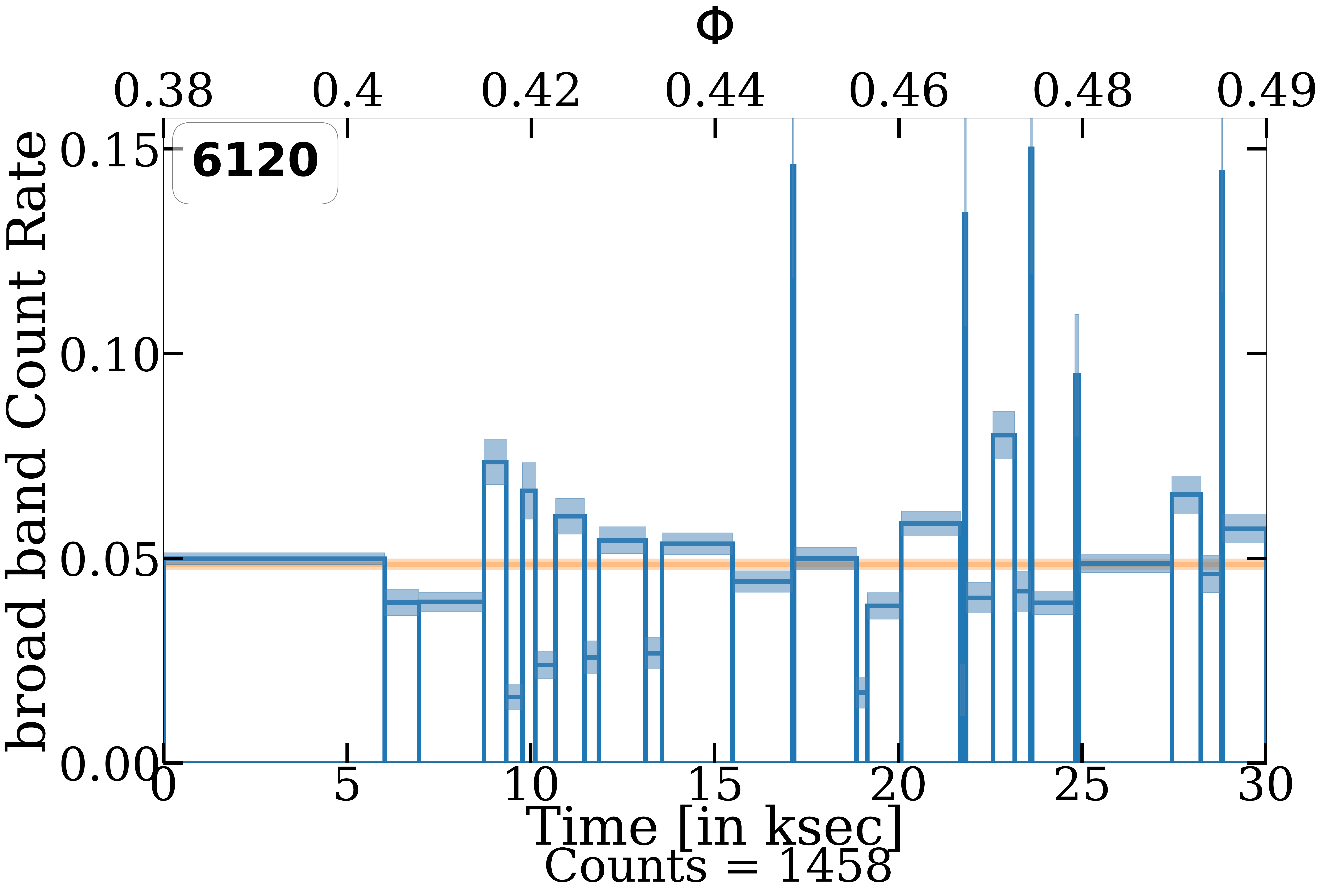}
\includegraphics[width=0.7\columnwidth,keepaspectratio]{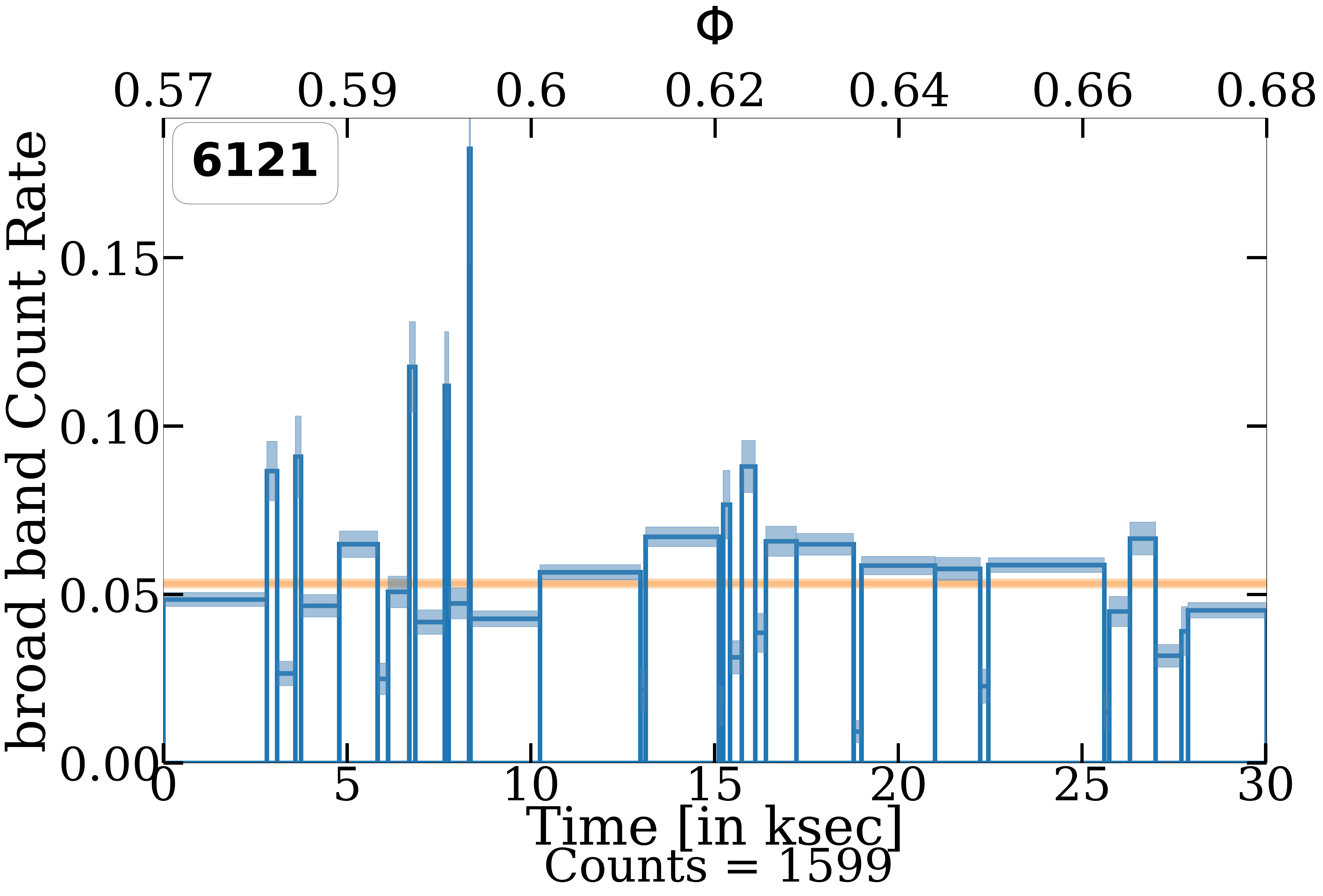}
\includegraphics[width=0.7\columnwidth,keepaspectratio]{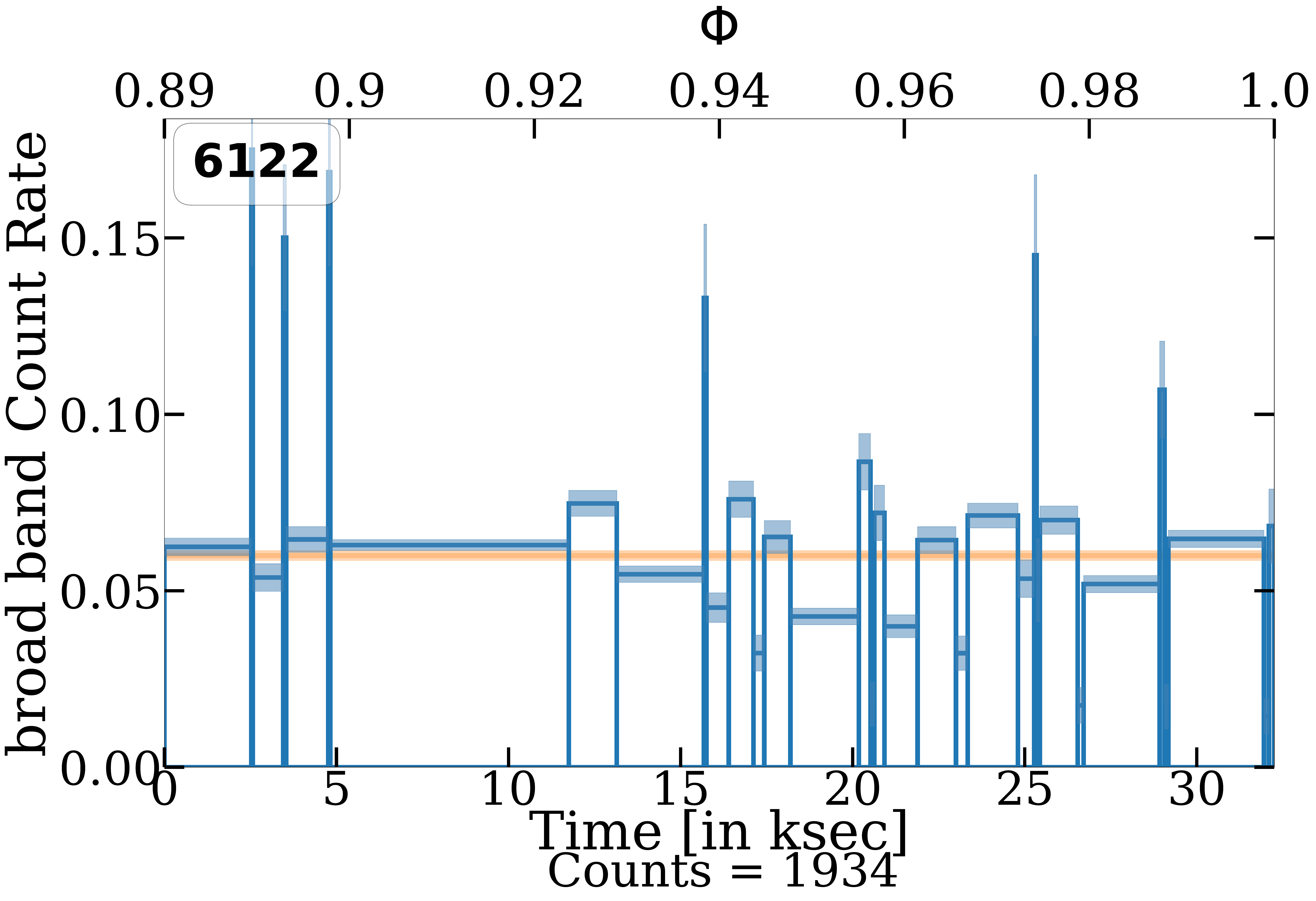}
\caption{Adaptively binned {\sl broad} band light curves for \egp\ as a function of time \updatebf{within} observation (lower \updatebf{abscissa}) and planetary phase (upper \updatebf{abscissa}).
The thickness of the horizontal bars indicates the Poisson error due to the counts in each block, \updatebf{as obtained using Bayesian Blocks \citep[][]{Scargle_1982,Scargle_2013,VanderPlas_2018}}.  The horizontal orange line represents the mean count rate over the observation.  \updatebf{No large flare-like events are present in the data.}}
\label{fig_broadlc}
\end{figure}

\begin{deluxetable}{lllllll}[!htbp]
\tabletypesize{\footnotesize}
\tablecolumns{7}
\tablewidth{0pt}
\tablecaption{Summarizing the variability in light curves \label{table:BBsummary}}
\tablehead{
\colhead{\diagbox[width=6em]{\hspace{-1em }Passband}{ObsID}} & \colhead{Parameters} &\colhead{5427} & \colhead{6119} & \colhead{6120} & \colhead{6121} & \colhead{6122}}
\startdata
Oxy & $\langle {\Delta}t_{\rm bin} \rangle$ & $257 \tau$ & $193 \tau$ & $226 \tau$ & $309 \tau$ & $237 \tau$ \\
    & N$_{\text{blocks}}$ & 36 & 48 & 41 & 30 & 42 \\
    & $\chi^2_{\rm red}$ & 15.3 & 7.7 & 14.0 & 5.1 & 9.4 \\
Fe  & $\langle {\Delta}t_{\rm bin} \rangle$ & $264 \tau$ & $206 \tau$ & $238 \tau$ & $221 \tau$ & $331 \tau$ \\
    & N$_{\text{blocks}}$ & 35 & 45 & 39 & 42 & 32 \\
    & $\chi^2_{\rm red}$ & 8.1 & 11.1 & 7.9 & 7.9 & 8.4 \\
Ne  & $\langle {\Delta}t_{\rm bin} \rangle$ & $250 \tau$ & $215 \tau$ & $215 \tau$ & $309 \tau$ & $255 \tau$ \\
    & N$_{\text{blocks}}$ & 37 & 43 & 43 & 30 & 39 \\
    & $\chi^2_{\rm red}$ & 16.1 & 13.7 & 16.3 & 6.7 & 6.2 \\
Mg  & $\langle {\Delta}t_{\rm bin}\rangle$ & $257 \tau$ & $201 \tau$ & $281 \tau$ & $238 \tau$ & $267 \tau$ \\
    & N$_{\text{blocks}}$ & 36 & 46 & 33 & 39 & 37 \\
    & $\chi^2_{\rm red}$ & 12.5 & 11.4 & 9.3 & 15.9 & 8.1 \\
u   & $\langle {\Delta}t_{\rm bin}\rangle$ & $237 \tau$ & $265 \tau$ & $290 \tau$ & $226 \tau$ & $231 \tau$ \\
    & N$_{\text{blocks}}$ & 39 & 35 & 32 & 41 & 43 \\
    & $\chi^2_{\rm red}$ & 7.6 & 9.3 & 7.9 & 14.8 & 8.8 \\
s   & $\langle {\Delta}t_{\rm bin}\rangle$ & $201 \tau$ & $226 \tau$ & $265 \tau$ & $244 \tau$ & $343 \tau$ \\
    & N$_{\text{blocks}}$ & 46 & 41 & 35 & 38 & 29 \\
    & $\chi^2_{\rm red}$ & 6.9 & 9.4 & 10.0 & 8.7 & 11.7 \\
m+h & $\langle {\Delta}t_{\rm bin} \rangle$ & $257 \tau$ & $201 \tau$ & $273 \tau$ & $238 \tau$ & $255 \tau$ \\
    & N$_{\text{blocks}}$ & 36 & 46 & 34 & 39 & 39 \\
    & $\chi^2_{\rm red}$ & 12.3 & 11.7 & 9.6 & 15.9 & 8.7 \\
broad & $\langle {\Delta}t_{\rm bin}\rangle$ & $250 \tau$ & $ 211 \tau$ & $ 299 \tau$ & $ 265 \tau$ & $ 293 \tau$ \\
    & N$_{\text{blocks}}$ & 37 & 44 & 31 & 35 & 34 \\
    & $\chi^2_{\rm red}$ & 6.9 & 10.8 & 8.9 & 5.9 & 7.7 \\
\vspace{-1em} \enddata  
\tablenotetext{}{
\updatebf{Note that the high values of $\chi^2_{\rm red}$ indicates significant variability. Further, the $\langle {\Delta}t_{\rm bin} \rangle$ values indicate a range of 27 to 48 blocks, consistent with the N$_{\text{blocks}}$ values for the typical observation durations of $\approx$30~ks.}}
\end{deluxetable}

We show the Bayesian Blocks count rate light curves for the broad band in Figure~\ref{fig_broadlc} for all five datasets \updatebf{(the light curves for the remaining passbands are available via Zenodo\footnote{\href{https://doi.org/10.5281/zenodo.7220014}{doi:10.5281/zenodo.7220014}; \citet{anshuman_acharya_2022_7181873}\label{zenodo}})}.  We compute several measures of variability:
\paragraph{Average Block size:} $\langle {\Delta}t_{\rm bin}\rangle$ is the average of the block sizes ${\Delta}t_{\rm bin}$ found with Bayesian Blocks, and represents the time scale over which the algorithm requires changes in its piece-wise constant light curve model.
\paragraph{Number of blocks:} The number of blocks, ${\rm N_{blocks}}=n^{cp}-1$, defined by $n^{cp}$ change points found by Bayesian Blocks.  This number complements $\langle{\Delta}t_{\rm bin}\rangle$ to show the range and ubiquity of variability.
\paragraph{Chi-square:} A reduced $\chi^2$, defined as
\begin{equation}
\chi^2_{\rm red} = \frac{1}{({\rm N}_{blocks}-1)} \sum_B \frac{({\rm rate}_B-{\rm rate}_{ObsID})^2}{\sigma_B^2} \,,
\label{eq:lc_chired}
\end{equation}
where ${\rm rate}_B = \frac{{\rm counts}_B}{{\Delta}t_B}$ is the observed rate in block $B$ when ${\rm counts}_B$ counts are observed over a duration ${\Delta}t_B$, ${\rm rate}_{ObsID}$ is the count rate for the full dataset (the orange horizontal line in Figure~\ref{fig_broadlc}), and $\sigma_B = \frac{\sqrt{{\rm counts}_B}}{{\Delta}t_B}$ is the uncertainty on the rate.  This represents the quality with which a model with unvarying intensity can be fit to the \updatebf{adaptively binned} light curves.

These quantities are reported in Table~\ref{table:BBsummary} for all the passbands and for all ObsIDs, and show that there is clear evidence for variability in the light curves over a broad range of time scales from $\approx$100~s to $\approx5$~ks.  We confirm that \updatebf{invariably $\chi^2_{\rm red}\gg$1, and thus} variability in \egp\ is ubiquitous across passbands and datasets.

\subsubsection{Hardness ratios}\label{sec:HR}

\begin{deluxetable*}{llllll}[!htbp]
\tabletypesize{\footnotesize}
\tablecolumns{6}
\tablewidth{0pt}
\tablecaption{Colors for different passband combinations for all ObsIDs \label{table:overallHR}}
\tablehead{
\colhead{\diagbox[width=6em]{\hspace{-1em}Color$^*$}{ObsID}} & \colhead{5427} & \colhead{6119} & \colhead{6120} & \colhead{6121} & \colhead{6122} }
\startdata
$C[\text{u/s}]$ & $-0.52^{+0.03}_{-0.01}$ & $-0.48^{+0.03}_{-0.02}$ & $-0.48^{+0.03}_{-0.02}$ & $-0.52^{+0.03}_{-0.02}$ & $-0.59^{+0.03}_{-0.01}$ \\
$C[\text{u/m+h}]$ & $+0.40^{+0.03}_{-0.04}$ & $+0.53^{+0.06}_{-0.04}$ & $+0.46^{+0.04}_{-0.04}$ & $+0.43^{+0.03}_{-0.04}$ & $+0.30^{+0.03}_{-0.04}$ \\
$C[\text{s/m+h}]$ & $+0.90^{+0.04}_{-0.03}$ & $+1.02^{+0.03}_{-0.05}$ & $+0.94^{+0.03}_{-0.05}$ & $+0.94^{+0.03}_{-0.04}$ & $+0.87^{+0.02}_{-0.04}$ \\
$C[\text{Oxy/Fe}]$ & $-0.40^{+0.03}_{-0.03}$ & $-0.37^{+0.03}_{-0.03}$ & $-0.36^{+0.03}_{-0.03}$ & $-0.38^{+0.03}_{-0.03}$ & $-0.39^{+0.04}_{-0.02}$ \\
$C[\text{Oxy/Ne}]$ & $-0.03^{+0.04}_{-0.03}$ & $+0.11^{+0.04}_{-0.04}$ & $+0.04^{+0.03}_{-0.04}$ & $+0.04^{+0.03}_{-0.04}$ & $-0.06^{+0.03}_{-0.03}$ \\
$C[\text{Oxy/Mg}]$ & $+0.25^{+0.04}_{-0.04}$ & $+0.39^{+0.05}_{-0.05}$ & $+0.31^{+0.04}_{-0.05}$ & $+0.29^{+0.04}_{-0.04}$ & $+0.21^{+0.04}_{-0.03}$ \\
$C[\text{Fe/Ne}]$ & $+0.37^{+0.03}_{-0.02}$ & $+0.47^{+0.04}_{-0.03}$ & $+0.39^{+0.03}_{-0.02}$ & $+0.34^{+0.03}_{-0.02}$ & $+0.32^{+0.03}_{-0.02}$ \\
$C[\text{Fe/Mg}]$ & $+0.65^{+0.03}_{-0.04}$ & $+0.75^{+0.05}_{-0.04}$ & $+0.67^{+0.03}_{-0.04}$ & $+0.67^{+0.03}_{-0.04}$ & $+0.59^{+0.04}_{-0.03}$ \\
$C[\text{Ne/Mg}]$ & $+0.27^{+0.04}_{-0.04}$ & $+0.30^{+0.04}_{-0.06}$ & $+0.28^{+0.03}_{-0.05}$ & $+0.33^{+0.03}_{-0.05}$ & $+0.26^{+0.04}_{-0.03}$ \\
\vspace{-1em}
\enddata 
\tablenotetext{$*$}{
\centering Showing the mode and the 68\% highest posterior density intervals \updatebf{for the overall color for different passbands given by Eq.~\ref{eq:color}}}
\end{deluxetable*}

There are insufficient counts in the blocks found for the counts light curves (Section~\ref{sec:lc}) to allow time-resolved spectroscopy.  Instead, we calculate the evolution of {\sl hardness ratios} over time and evaluate the strength of the evidence for {\sl spectral} variability (for analyses that highlight how hardness ratios can be useful, see, e.g., \citealt{Noel_2022,DiStefano_2021}).  Specifically, we compute the {\sl color}
\begin{equation}
    C[{\rm A/B}]=\log_{10}{\frac{\rm counts~in~band~A}{\rm counts~in~band~B}} \,,
    \label{eq:color}
\end{equation}
for various combinations of pairs of passbands in Table~\ref{table:passbands}.  We consider different combinations of the \updatebf{\chandra\ Source Catalog (CSC)} bands and the line dominated bands.  Variations in plasma temperatures would be reflected in changes in the CSC colors, while abundance variations could become discernible in the line dominated band ratios.  We use \behr\ \citep[Bayesian Estimation of Hardness Ratios;][]{Park_2006} to account for background and estimate the most probable values (modes) and uncertainties (68\% highest posterior density [HPD] intervals) of the color in each time interval considered.

\begin{figure}[!htbp]
\centering
\includegraphics[width=0.7\columnwidth,,keepaspectratio]{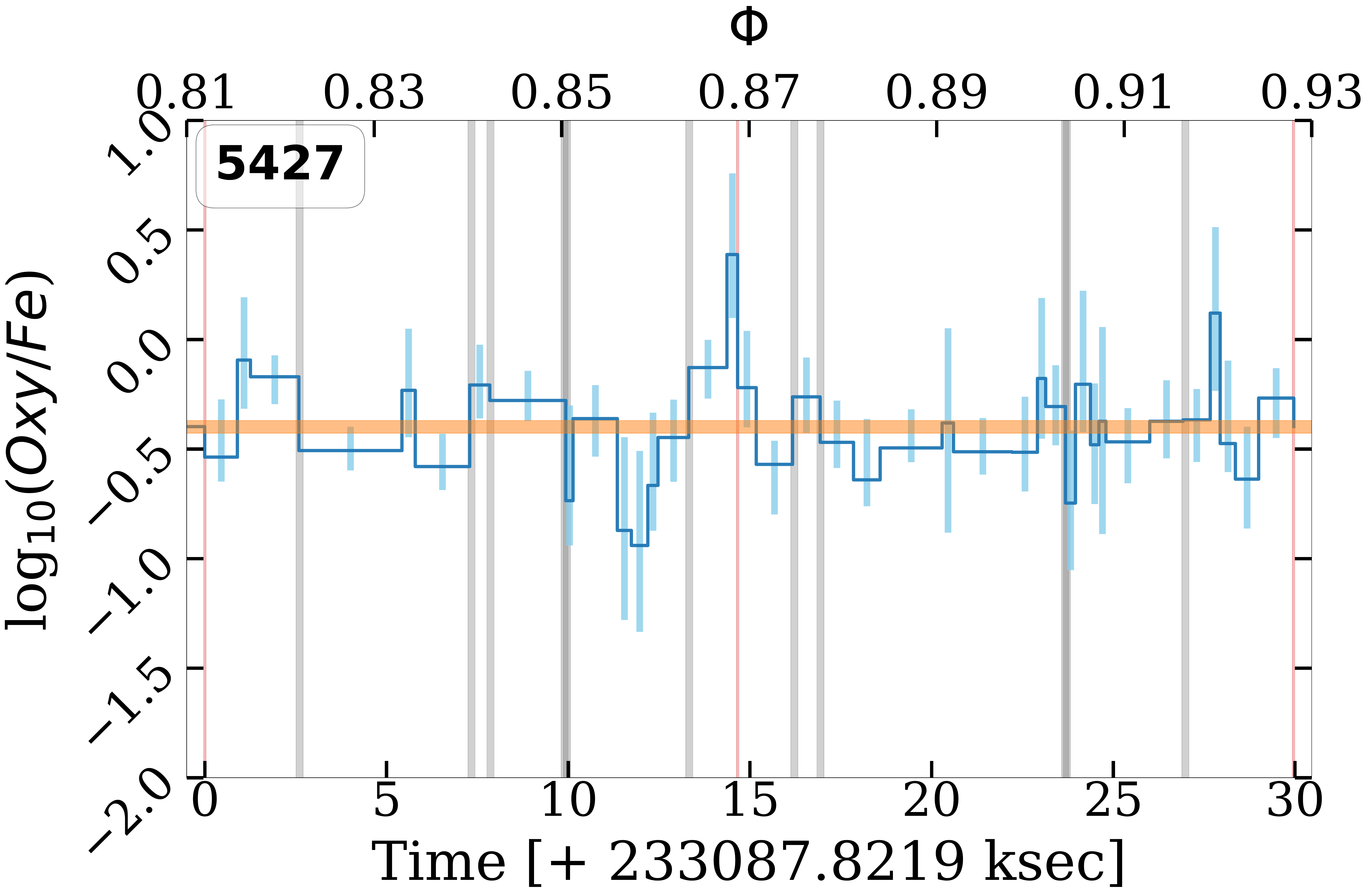}
\includegraphics[width=0.7\columnwidth,keepaspectratio]{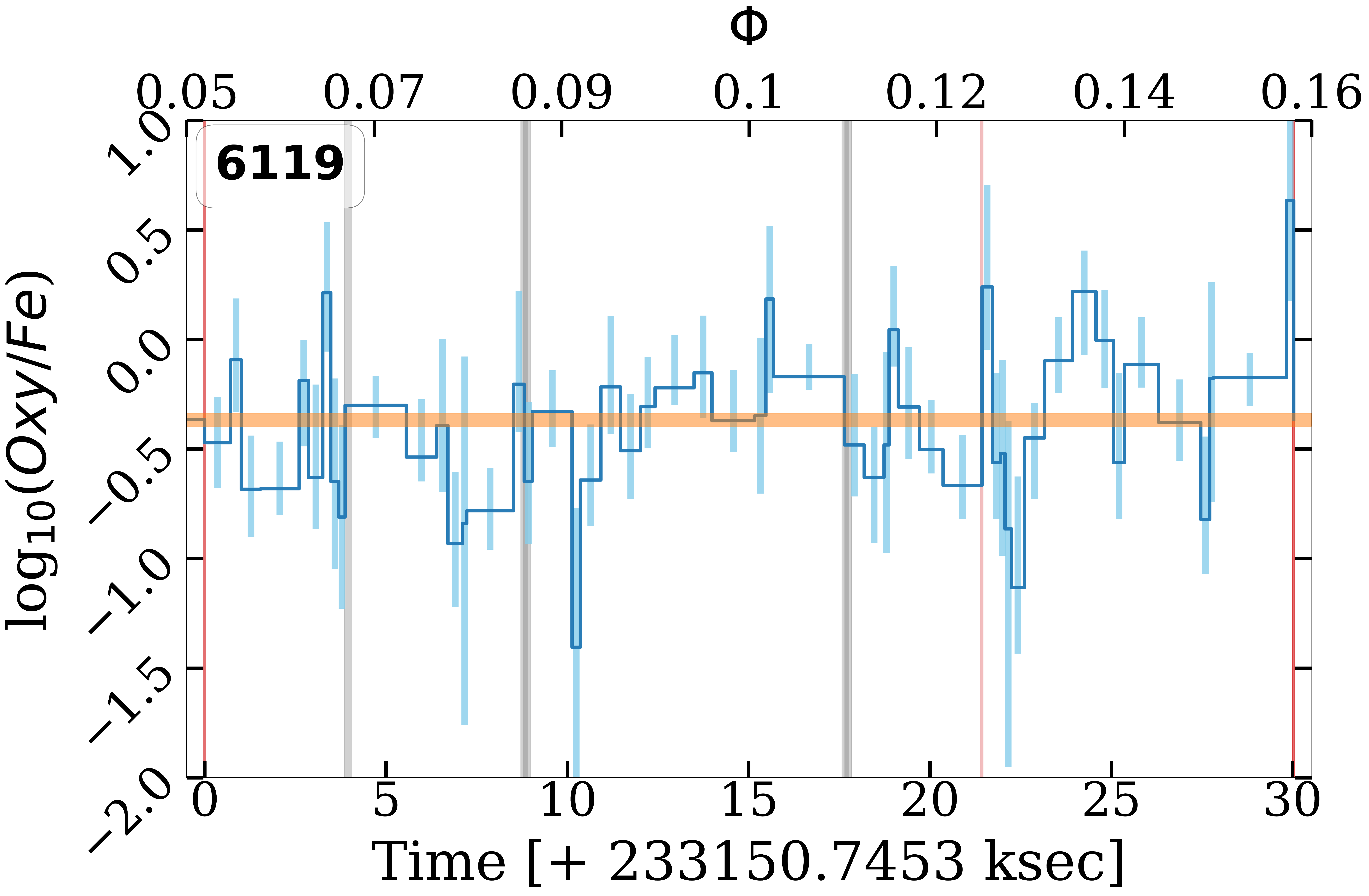}
\includegraphics[width=0.7\columnwidth,keepaspectratio]{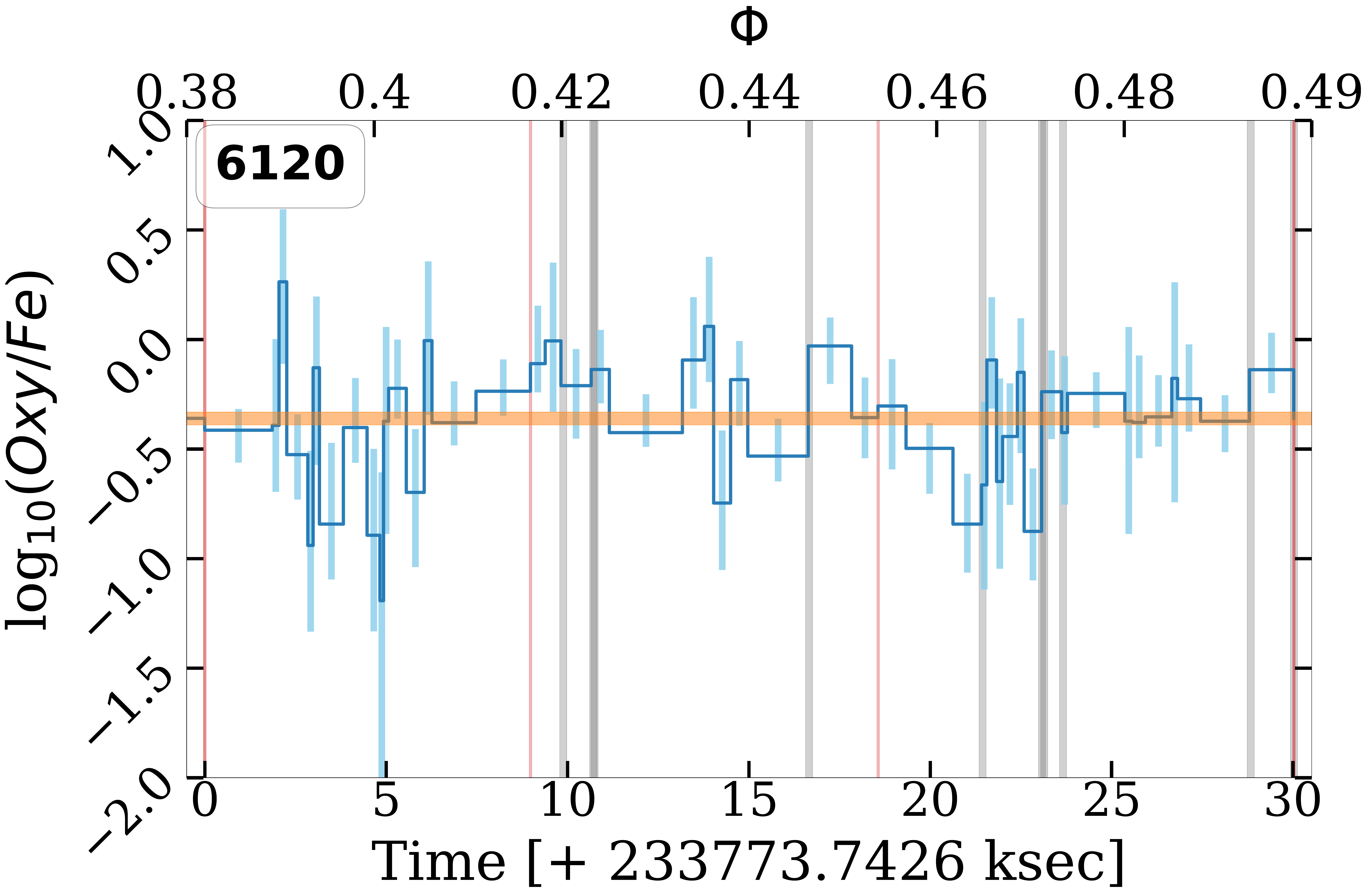}
\includegraphics[width=0.7\columnwidth,keepaspectratio]{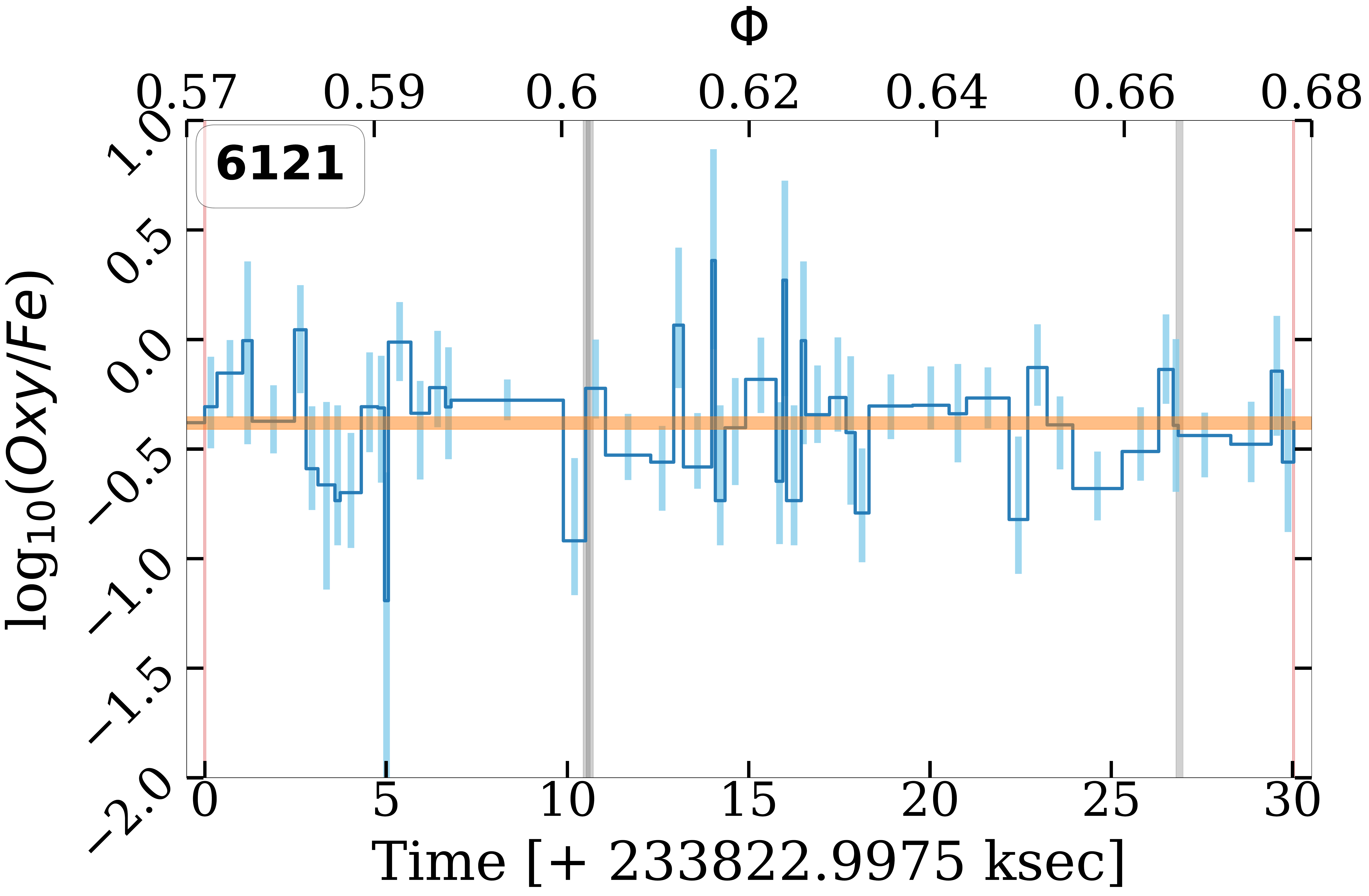}
\includegraphics[width=0.7\columnwidth,keepaspectratio]{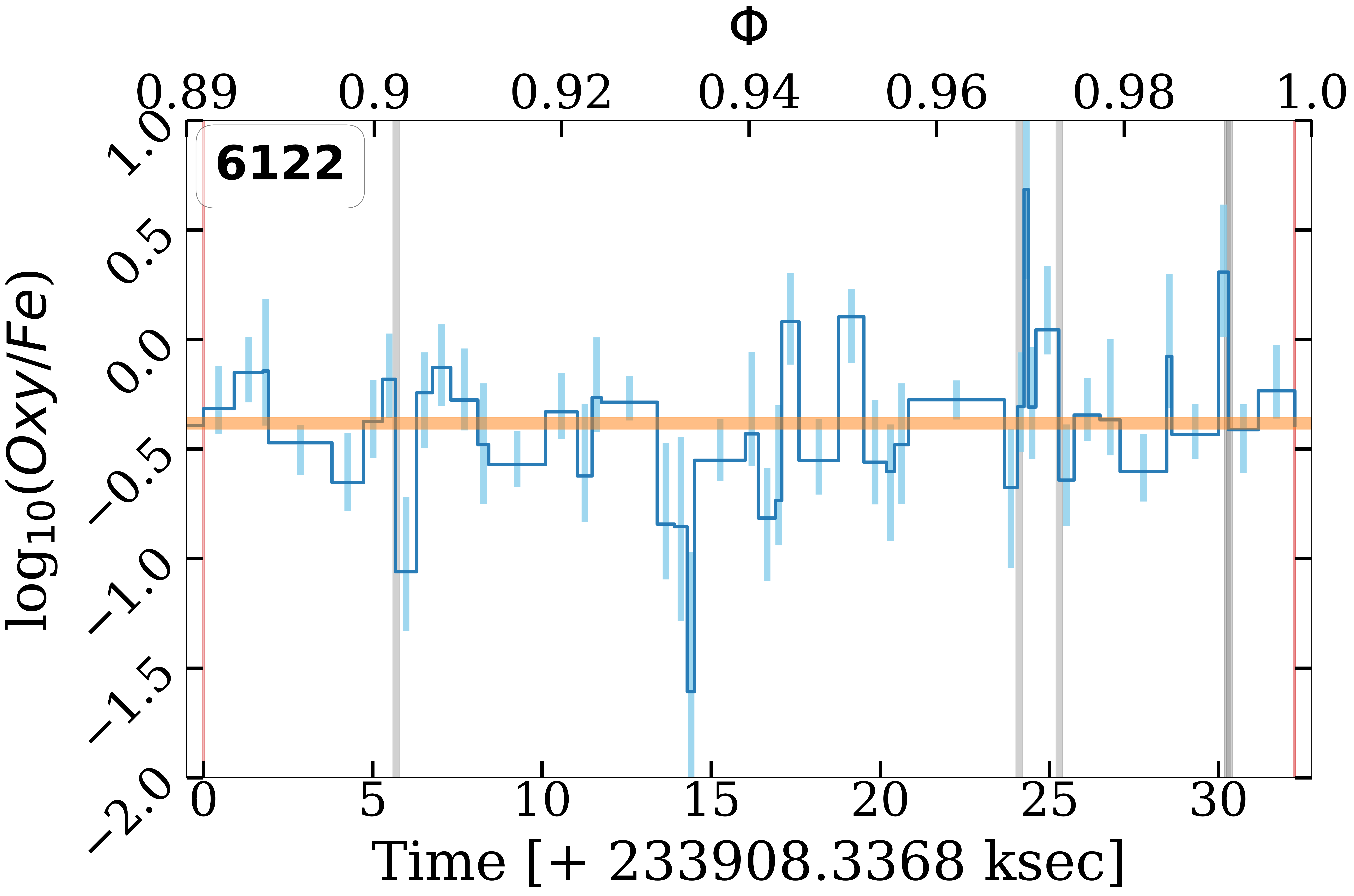}
\caption{Color Hardness Ratios for Oxy and Fe bands ($C{\rm [Oxy/Fe]}$) for the 5 datasets.  The most probable value in each time bin (blue stepped curve) and the 68\% HPD interval (blue vertical bars) are shown, along with the mode value measured for the full dataset (orange horizontal bar; the width represents the 68\% HPD interval).  The locations of {\sl close merger} change points (\updatebf{change points separated by $<10\tau\approx{32}$~s and marked by} red vertical lines) and {\sl loose merger} change points (\updatebf{change points separated by $<30\tau\approx{100}$~s and marked by} grey vertical lines) are also \updatebf{shown}.}
\label{fig_HRoxyfe}
\end{figure}

In order to define appropriate time intervals, we use the change points found in the light curves by Bayesian Blocks (Section~\ref{sec:lc}) analysis of the corresponding passbands.  These change points are then combined and pruned to form time bins, and the color is computed independently for counts collected in each bin.  The process is described in detail below.
\begin{enumerate}
    \item We begin by constructing the union of the set of change points from both passbands to form a single set of change points in temporally sorted order.
    \item All pairs of change points which are separated by $<10\tau\approx{32}$~s are then grouped together and replaced by their average, for consistency with how the change points are initially determined.  This grouping and averaging is carried out iteratively until no such pairs are found.  We call these merged change points {\sl close mergers}.
    \item We repeat the above pair-wise merging process for all remaining points which are separated by $<30\tau\approx 100$~s.  While this removes any sensitivity in our process to spectral changes at time scales shorter than 100~s, the likelihood of such quick changes being visible in low-density hot coronal plasma dominated by radiative cooling processes is low.  We name such merged points  {\sl loose mergers}.
\end{enumerate}
The time points left in the set at the end of the above merging procedure are used as the bin boundaries for events collected in both passbands.  We show an example of variations in the color ratio of the Oxy and Fe passbands in Figure~\ref{fig_HRoxyfe}.  The figure shows the mode of $C[{\rm Oxy/Fe}]$ (blue stepped curve) and associated 68\% HPD uncertainty (blue vertical bars), along with the close mergers (red vertical lines) and loose mergers (grey vertical lines).  The color estimate obtained for the full dataset is also shown (orange horizontal line), illustrating the scope of departure from constancy.  As with the count rate light curves, we again compute a reduced $\chi^2$ as a measure of variability,
\begin{equation}
    \chi^2_{red} = \frac{1}{{\rm N}_{bins}-1} \sum_k \frac{\left(\langle C{\rm [A/B]}_k \rangle - \langle C{\rm [A/B]}_{ObsID}\rangle\right)^2}{{\rm Var[}{C{\rm [A/B]}_k}{\rm ]}} \,,
    \label{eq:col_chired}
\end{equation}

where N$_{bins}$ are the number of bins in the light curve of $C{\rm [A/B]}$ after the merging process, $\langle C{\rm [A/B]}_k \rangle$ is the mean color estimated from \behr\ draws in each bin $k$, ${\rm Var[}{C{\rm [A/B]}_k}{\rm ]}$ is the corresponding variance, and $\langle C{\rm [A/B]}_{ObsID}\rangle$ is the mean color estimated for the full ObsID.  These $\chi^2_{red}$ values are reported in Table~\ref{table:chisqHR}, with values of $\chi^2_{red}>1+\sqrt{\frac{2}{{\rm N}_{bins}-1}}$ marked in italic, and $\chi^2_{red}>1+3\sqrt{\frac{2}{{\rm N}_{bins}-1}}$ marked in bold.  It is apparent that the magnitude of the spectral variations is significantly smaller than the intensity variations. This suggests that count rates in the different bands tend to rise and fall in tandem, with most of the variations attributable to emission measure changes.  Nevertheless, spectral variations over time scales of the observation duration is definitely present and are indeed detectable, as seen in several cases where significant deviations in $\chi^2_{red}$ is seen.  Furthermore, detailed examination of the color light curves can yield information on smaller time scale variations,
as, e.g., the large rise in $C{\rm [Oxy/Fe]}$ seen in ObsID 5427 between times 11-15~ks after the start of the observation (top panel of Figure~\ref{fig_HRoxyfe}).  A similar, but smaller variation is also seen in $C{\rm [Oxy/Ne]}$, but no variability is present in either the CSC band colors $C{\rm [s/m+h]}$ or in $C{\rm [Fe/Ne]}$.  This is a strong indication of a temporary surge in the abundance of Oxygen, reminiscent of variations seen in streamers and plumes around active regions on the Sun \citep[see, e.g.,][]{1998SSRv...85..283R,1999SSRv...87...55R,2015ApJ...807..145G}.  All of the color $C[{\rm A/B}]$ 
spectral variability plots are available via Zenodo\footnoteref{zenodo}. 

\begin{figure}[!htbp]
\centering
\includegraphics[width=0.7\columnwidth,,keepaspectratio]{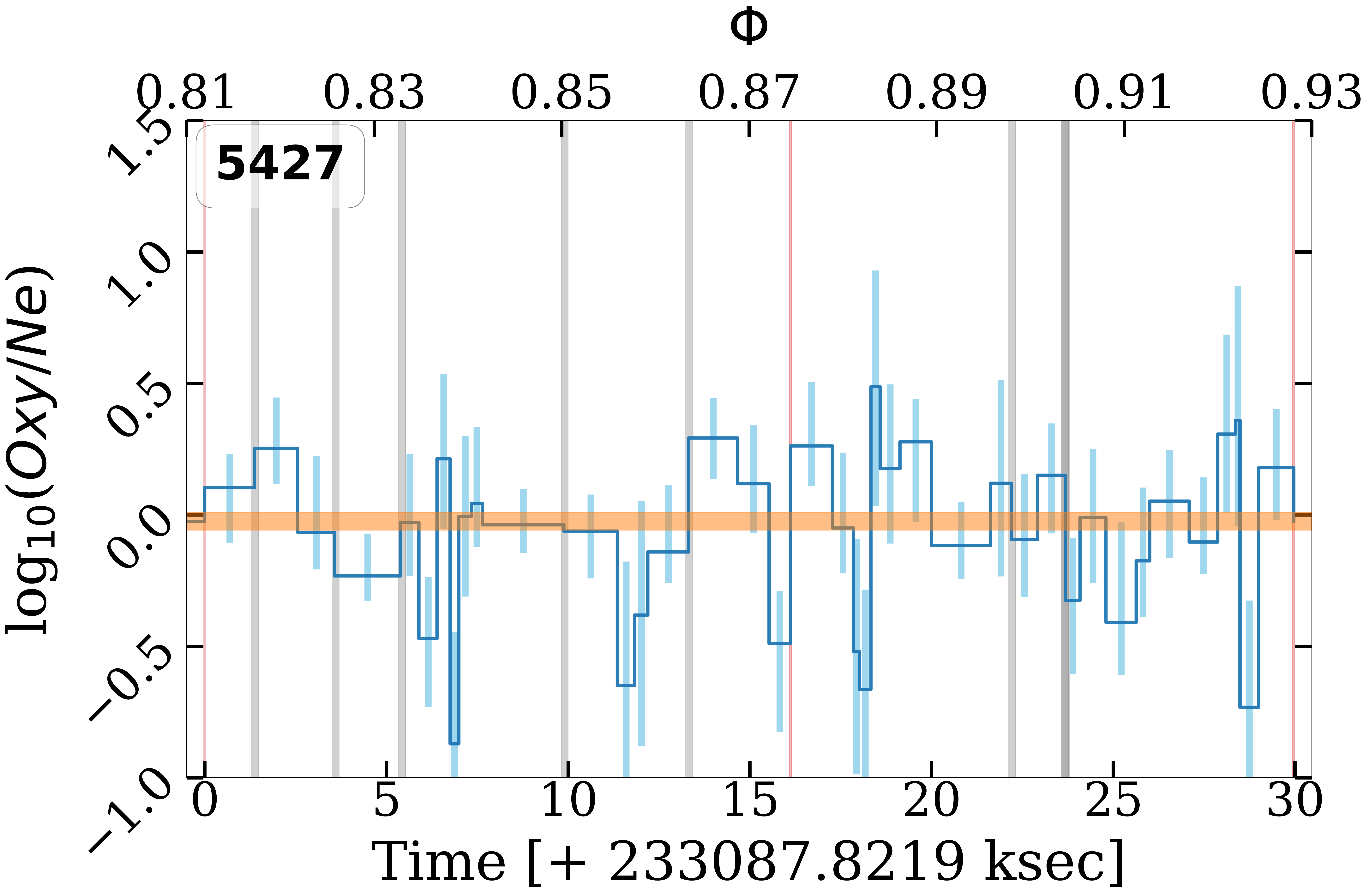}
\includegraphics[width=0.7\columnwidth,keepaspectratio]{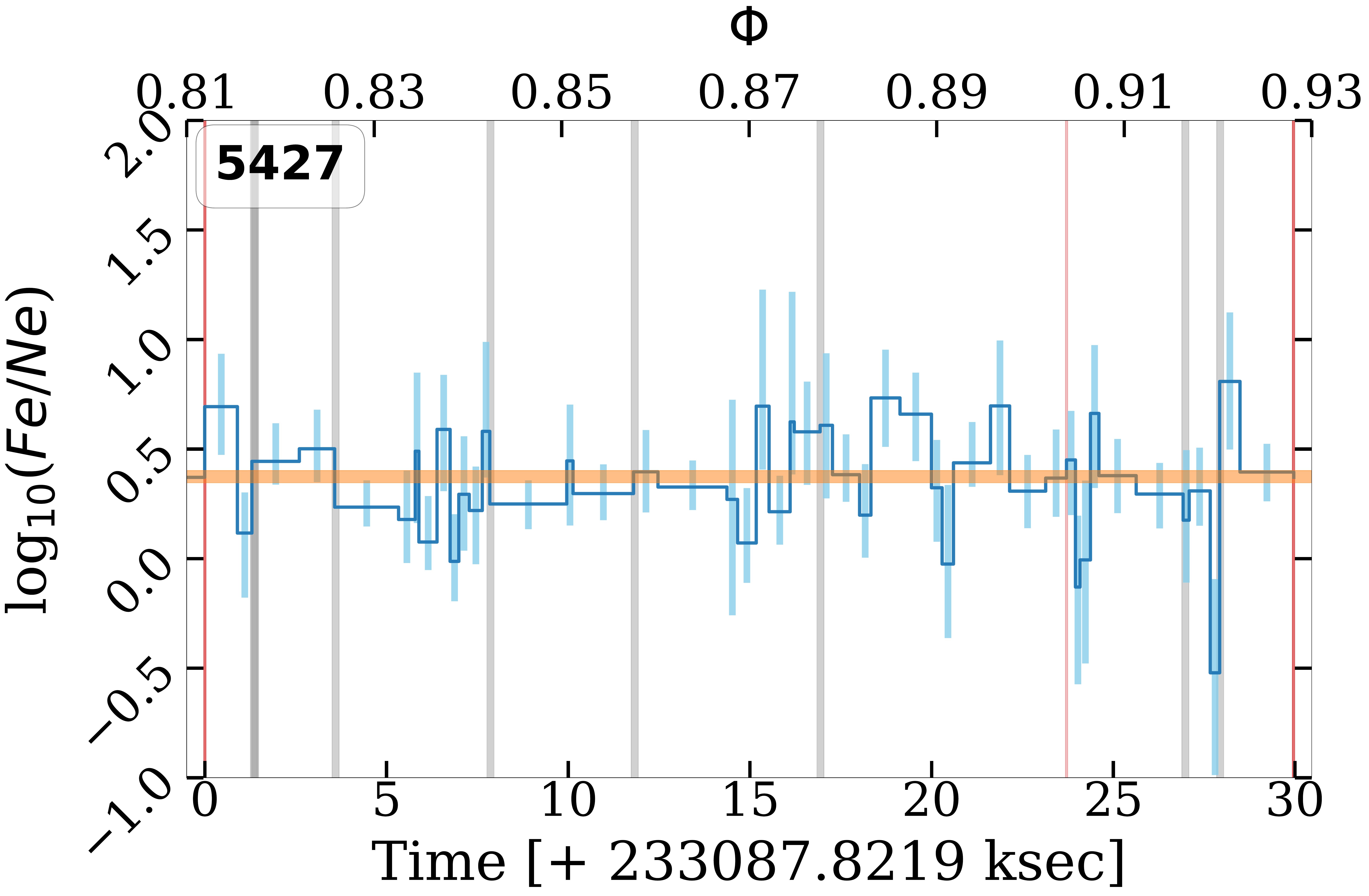}
\includegraphics[width=0.7\columnwidth,keepaspectratio]{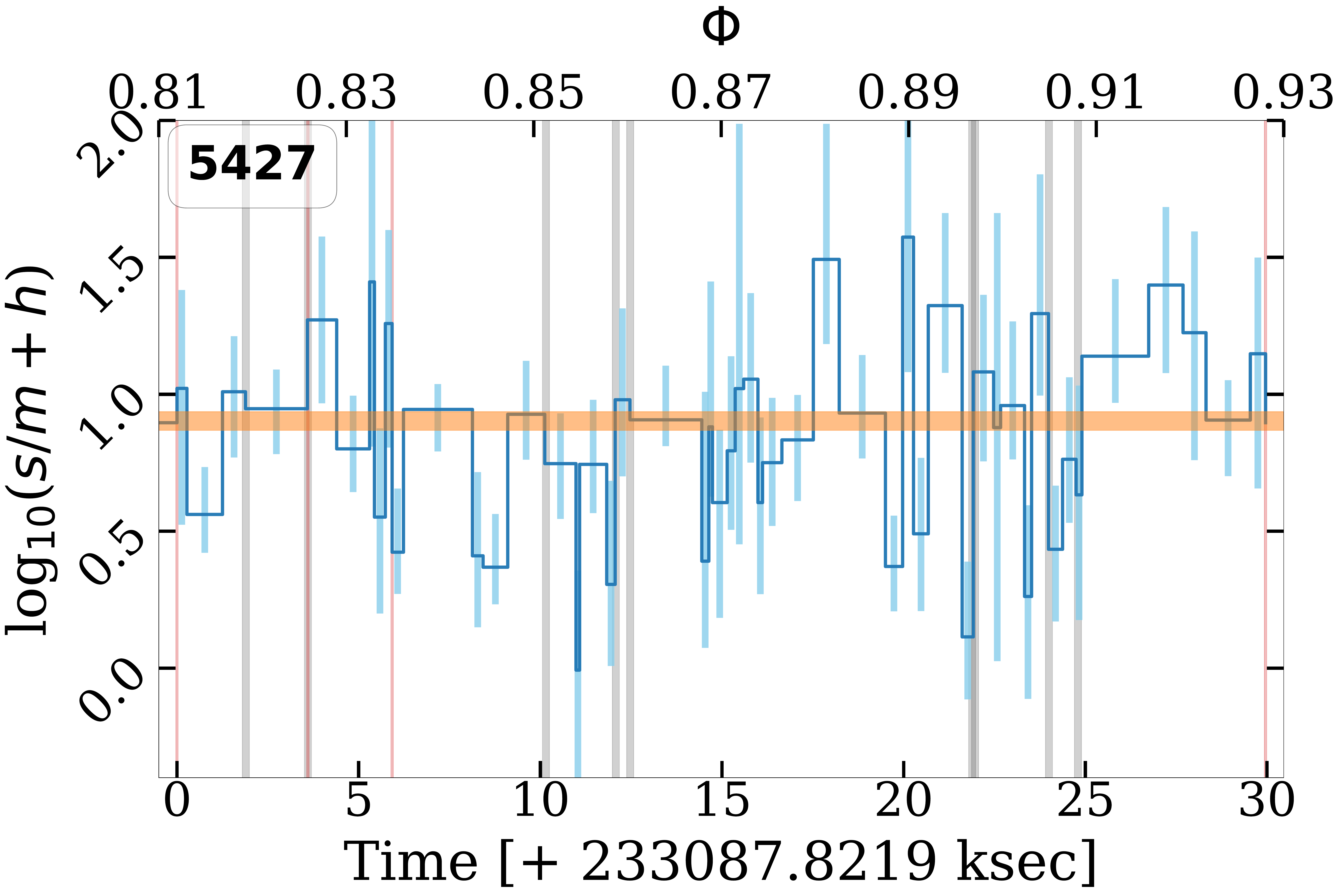}
\caption{Demonstrating the Oxygen abundance surge in ObsID 5427.  Color light curves for several passbands -- $C{\rm [Oxy/Ne]}$ (top), $C{\rm [Fe/Ne]}$ (middle), and $C{\rm [s/m+h]}$ (bottom) \updatebf{are shown}.  A similar \updatebf{increase} as in $C{\rm [Oxy/Fe]}$ (top panel of Figure~\ref{fig_HRoxyfe}) is seen in $C{\rm [Oxy/Ne]}$ \updatebf{between $\approx$12-15~ks}, but not in either $C{\rm [Fe/Ne]}$ or $C{\rm [s/m+h]}$, \updatebf{suggesting that the increase in counts in the oxygen band cannot be attributed to temperature effects}.}
\label{fig_HR5427}
\end{figure}

\begin{deluxetable}{llllll}[!htbp]
\tabletypesize{\footnotesize}
\tablecolumns{6}
\tablewidth{0pt}
\tablecaption{Summary of spectral variability$^*$
\label{table:chisqHR}}
\tablehead{ \colhead{\diagbox[width=6em]{Color}{ObsID}} & \colhead{5427} & \colhead{6119} & \colhead{6120} & \colhead{6121} & \colhead{6122} }
\startdata
$C[\text{u/s}]$ & \textit{1.4} (59) & \textit{1.6} (56) & 1.2 (48) & \textit{\textbf{1.9}} (55) & \textit{\textbf{1.8}} (53) \\
$C[\text{u/m+h}]$& 0.9 (44) & 0.8 (39) & 1.2 (38) & 1.1 (43) & 1.1 (48) \\
$C[\text{s/m+h}]$& \textit{1.6} (49) & 0.9 (41) & \textit{1.3} (46) & \textit{1.3} (45) & 0.8 (41) \\
$C[\text{Oxy/Fe}]$ & 1.1 (38) & \textit{1.6} (52) & 1.1 (47) & 1.0 (50) & \textit{\textbf{1.7}} (45) \\
$C[\text{Oxy/Ne}]$ & \textit{1.3} (39) & \textit{1.3} (57) & 1.2 (46) & 1.1 (38) & \textit{1.5} (51) \\
$C[\text{Oxy/Mg}]$ & 1.2 (35) & 0.6 (34) & 1.0 (40) & 0.9 (33) & 0.8 (44) \\
$C[\text{Fe/Ne}]$ & 1.0 (46) & 1.1 (56) & \textit{1.5} (51) & 1.0 (51) & \textit{1.4} (51) \\
$C[\text{Fe/Mg}]$ & \textit{1.4} (43) & 1.1 (41) & 1.2 (46) & 1.2 (45) & 0.8 (40) \\
$C[\text{Ne/Mg}]$ & 1.0 (41) & 0.9 (35) & 1.2 (42) & \textit{1.3} (38) & 0.9 (49) \\
\vspace{-1em} 
\enddata  
\tablenotetext{$*$}{The values of $\chi^2_{red}$ (Equation~\ref{eq:col_chired}) are shown, along with the number of bins N$_{bins}$ in parentheses.  Values with deviations $>1\sigma$ are marked in italic and deviations $>3\sigma$ are marked in bold.}
\end{deluxetable}

\subsubsection{\updatebf{Periodicity Analysis}}\label{sec:LS}

After fitting the spectral models for the 5 datasets, we calculate the resultant fluxes in the $0.15-4.0$ keV energy range. The results are shown in Table~\ref{table:fitparams} and in Figure~\ref{fig_fluxes} along with flux data from \xmm\ \citep{Scandariato_2013} and \swift\ \citep{D_Elia_2013}.  We choose to compare against \swift\ fluxes obtained assuming a plasma with temperature $\approx$5~MK as that is comparable to the plasma temperatures we find (Table~\ref{table:fitparams}.  We remove three flaring events (with flux ${\gg}7{\times}10^{-13}$~ergs~s~cm$^{-2}$) from the \swift\ dataset to focus only on periodic variability of the quiescent flux.  In addition, we normalise the \xmm\ and \swift\ flux values to the average \chandra\ flux to remove systematics like calibration uncertainties and long-term activity variations.  The resultant fluxes are shown in the bottom plot of Figure~\ref{fig_fluxes}.

\begin{figure*}[!htbp]
\centering
\includegraphics[width=\textwidth,keepaspectratio]{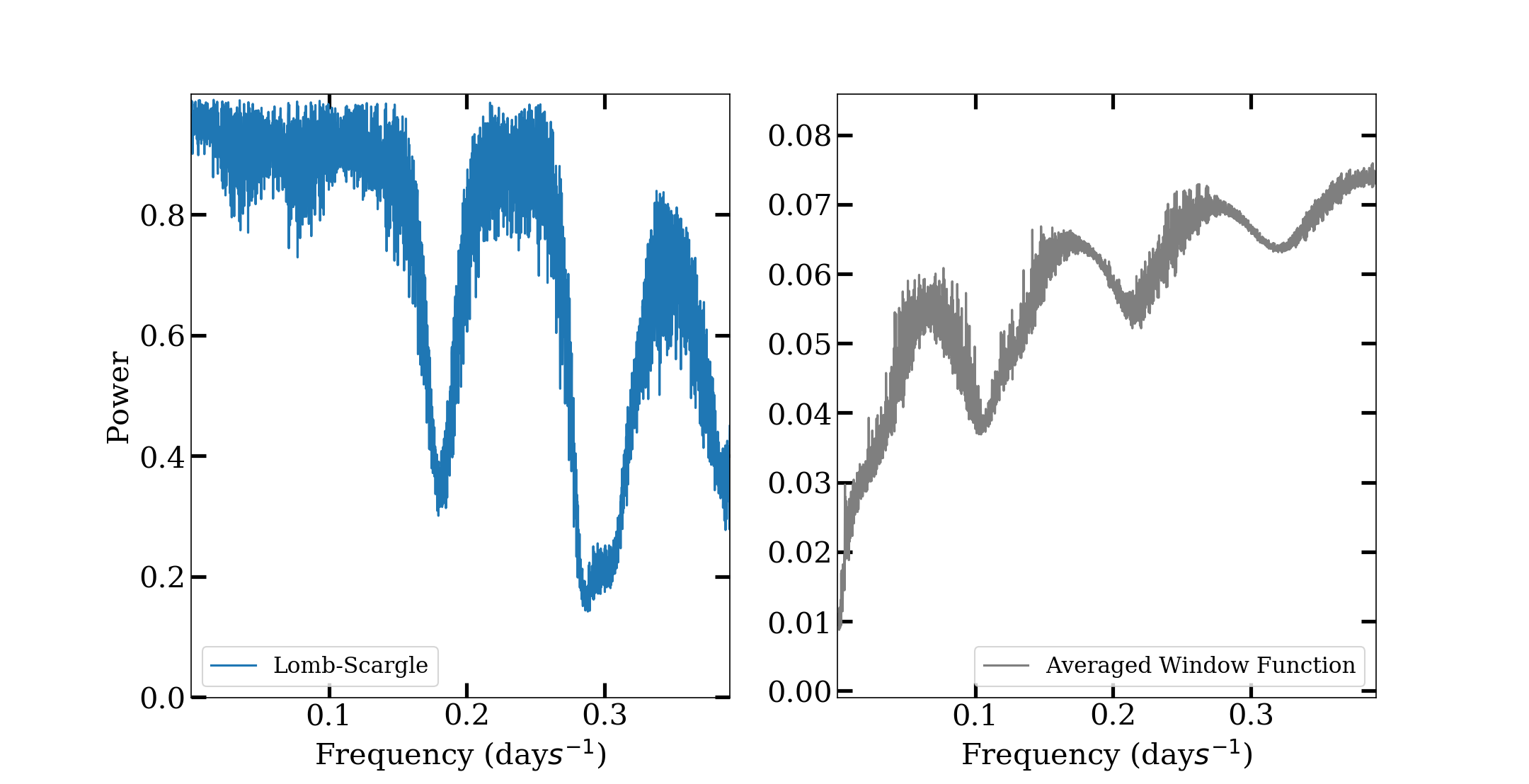}
\caption{\textbf{Left:} The Lomb-Scargle periodogram $\mathcal{LS}(f)$ for the normalized fluxes data $f$ from \chandra, \xmm, and \swift\ \updatebf{(see bottom panel of Figure~\ref{fig_fluxes}}). \textbf{Right:} The averaged window function $\langle \mathcal{LS}(W) \rangle$ \updatebf{computed for the observed cadence}.}  
\label{fig_winls}
\end{figure*}

We search for periodicity in the combined flux data $f$ using the Lomb-Scargle (L-S) periodogram $\mathcal{LS}(f)$ \citep[][]{Lomb_1976,Scargle_1982}, as implemented by \citet{VanderPlas_2018} (see left panel of Figure~\ref{fig_winls}).  In order to obtain better resolution in phase, we split the \chandra\ data further into segments of $\gtrsim$5~ks in each ObsID, converting the net counts to fluxes using a counts-to-energy conversion factor computed separately for each ObsID as the ratio of the flux measured (Table~\ref{table:fitparams}) to the net count rate (Table~\ref{table:summaryobs}) for that ObsID.

We account for the effects of statistical fluctuations and the windowing function via bootstrapping.  We first scramble the fluxes $f$ using a random permutation operator $Scr(\cdot)$ to obtain a new set of fluxes $g_k=Scr_k(f)$, where $k=1,\ldots,5000$.  We compute $\mathcal{LS}(g_k)$ for each of these permutations.  When averaged, $\langle{\mathcal{LS}(g)}\rangle$ represents a periodogram that is devoid (due to the scrambling) of an intrinsic periodic signal, but includes effects due to spacing and data level.  The scatter in the bootstrapped periodograms, ${\rm stddev}[{\mathcal{LS}(g)}]$ provides a measure of the statistical and data spacing noise.  The difference $\mathcal{LS}(f)-\langle{\mathcal{LS}(g)}\rangle$ represents the residual signal that can be attributed to intrinsic periodicity.

We then compute the window function by constructing the L-S periodograms of a unit flux dataset sampled at the same times as $f$.  In order to ameliorate numerical instabilities, we allow for a jitter in the unit flux, obtaining values in each case drawn from a Gaussian with mean $1$ and width equal to the fractional error on each flux value. We calculate the window function 5000 times and average the resulting periodograms to obtain $\langle \mathcal{LS}(W) \rangle$ (see right panel of Figure~\ref{fig_winls}).

\begin{figure*}
\centering
\includegraphics[width=1.00\textwidth,height=\textheight,keepaspectratio]{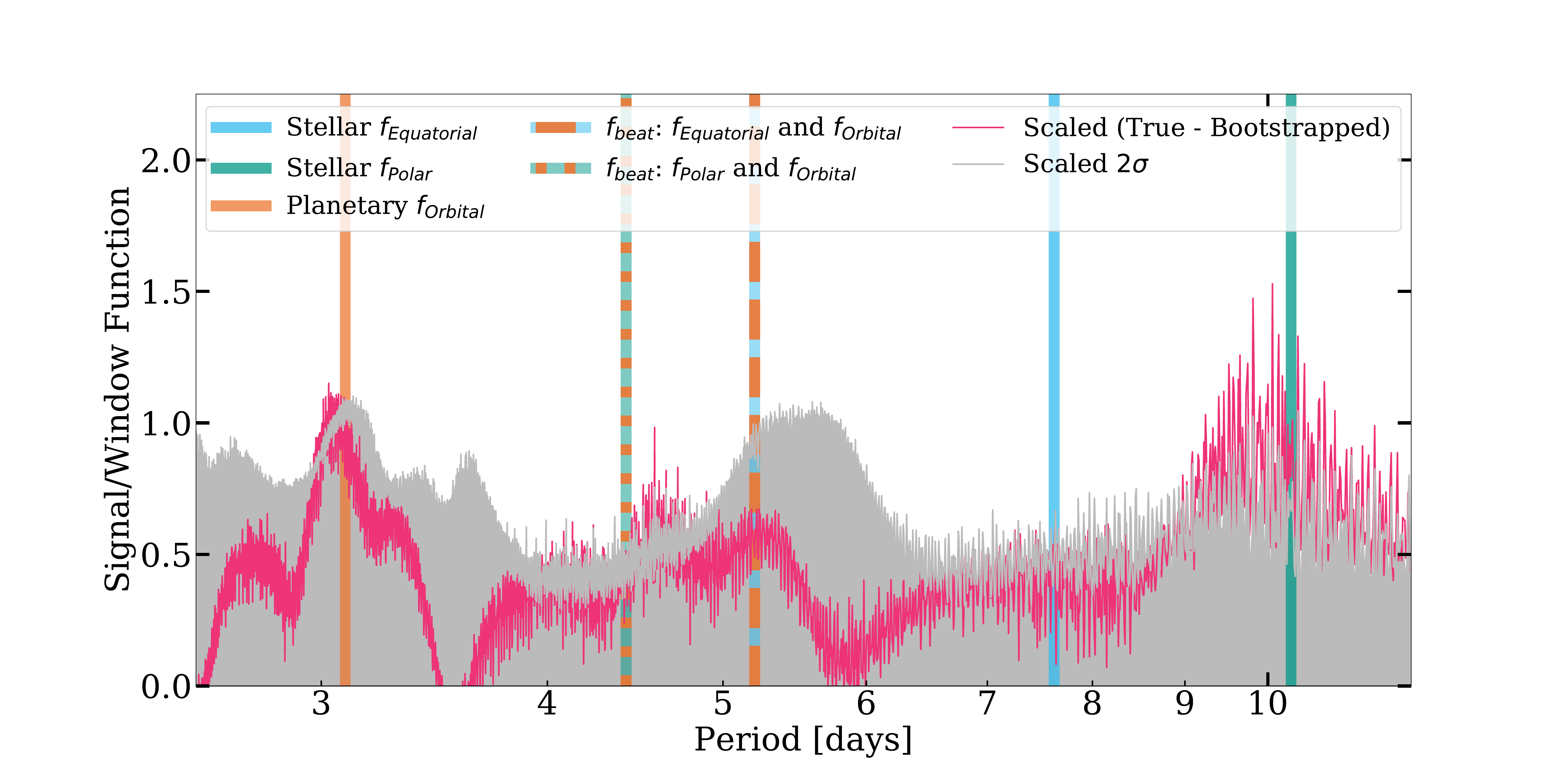}
\caption{Analysis of periodicities in \egp.  $\mathcal{P}(f) = \frac{\mathcal{LS}(f)-\langle{\mathcal{LS}(g)}\rangle}{\langle \mathcal{LS}(W) \rangle}$ which is the modified data periodogram, scaled by the averaged window function (magenta) is compared with $2\cdot\frac{{\rm stddev}[{\mathcal{LS}(g)]}}{\langle \mathcal{LS}(W) \rangle}$ which is twice the noise in the averaged time-scrambled periodogram similarly normalized (grey). Peaks with signal-to-noise ratio $>2$ \updatebf{indicate the presence of possible} periodicities of flux variation. A clear sign of variability tied to the stellar polar rotation ($f_{Polar}$) can be seen. Periodicity tied to the stellar equatorial rotation ($f_{Equatorial}$) is not detected.  \updatebf{The presence of power above noise at periods close to the planetary orbital frequency ($f_{Orbital}$), and the beat frequency ($f_{beat}$) between $f_{Polar}$ and $f_{Orbital}$ presents evidence for an SPI effect (see Sections~\ref{sec:PhaseVar},\ref{sec:SPI}).} 
}  
\label{fig_ls}
\end{figure*}

We then extract the de-aliased, window-scaled, signal of intrinsic periodicity 
\begin{equation}
    \mathcal{P}(f) = \frac{\mathcal{LS}(f)-\langle{\mathcal{LS}(g)}\rangle}{\langle \mathcal{LS}(W) \rangle} \,.
\end{equation}
This scaled periodogram is shown in Figure~\ref{fig_ls} as the magenta curve, with select periods marked with vertical bars.  We also show the expected $2\sigma$ uncertainty, constructed as $2\cdot\frac{{\rm stddev}[{\mathcal{LS}(g)]}}{\langle \mathcal{LS}(W) \rangle}$, and shown as the grey shaded region (for visibility, the $2\sigma$ curve is placed above $\mathcal{P}(f)$, covering it where it crosses).  It is clear that several peaks are present in $\mathcal{P}(f)$ at or close to periods of relevance to the \egp\ system: we detect periodicities at the stellar polar rotational period (\updatebf{$\approx$10~d}), the planetary orbital period (\updatebf{$\approx$3~d}), and the beat period of the stellar polar and planetary orbital periods (\updatebf{$\approx$4.5~d}, consistent with some Ca~II~H\&K modulations found by \citet[][]{Fares_2012}) at significances $>95$\%.  While each of these are not definitively above the usually adopted "$3\sigma$" threshold, the confluence of all three inter-related signals suggests that these periodicities do exist in the \egp\ system.

\section{Discussion} \label{sec:discuss}

\subsection{Stellar Context}\label{sec:context}

\begin{figure}[bth]
\centering
\includegraphics[width=\columnwidth,keepaspectratio]{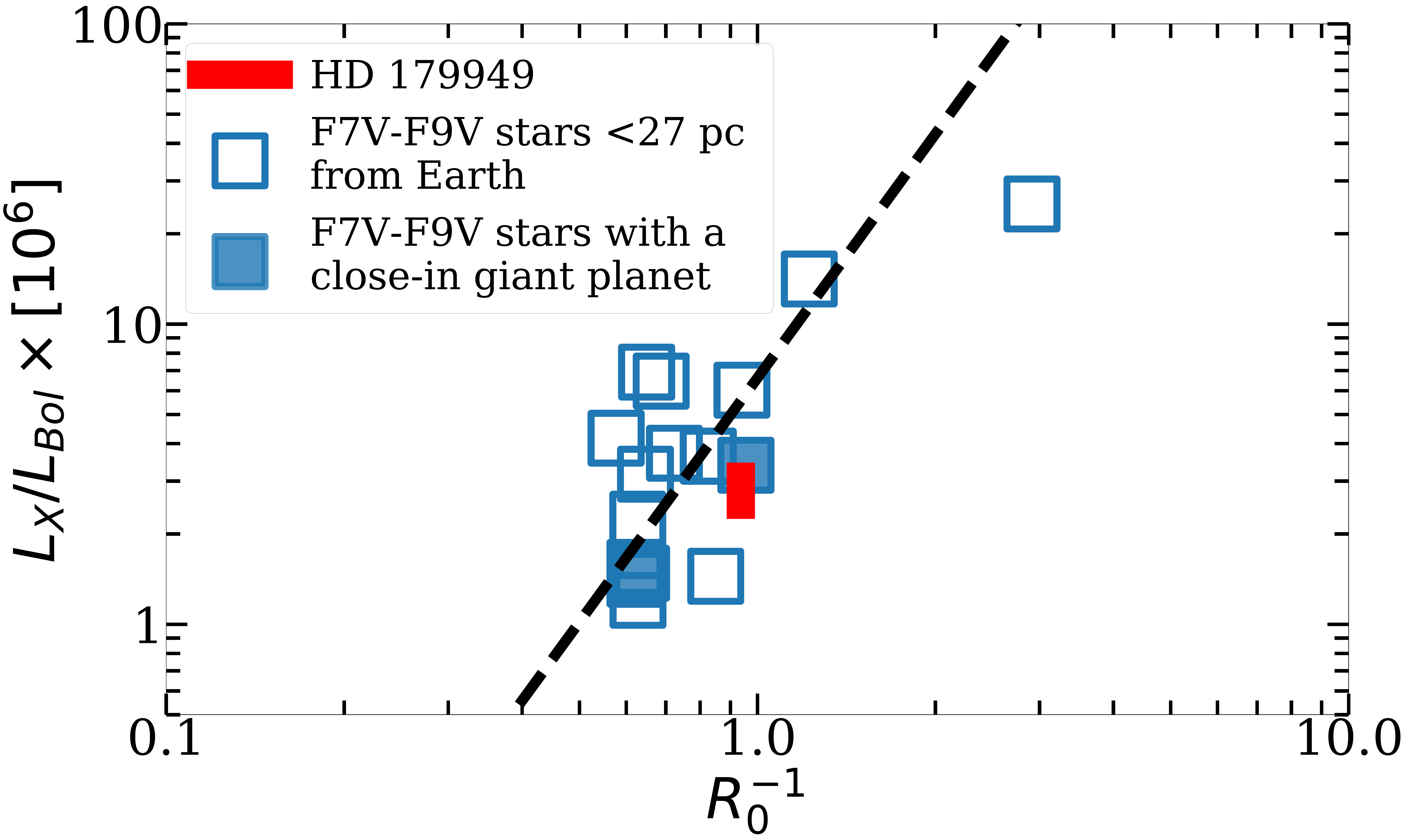}
\caption{Comparing the $L_X/L_{Bol}$ ratio of F7-9V stars within 27 pc (listed in Table~\ref{table:LXLBol_compare}) versus their Rossby number $R_{0}^{-1} = \tau_C/\rm P_{\rm Rot}$ (where convective turnover time scale $\tau_C$ is calculated using \citet{Gunn_1998} for the appropriate {\sl B-V} values) with that of \egp. The red line for HD 179949 showcases the range of $L_X/L_{Bol}$ values for the 5 datasets. The best-fit line is given by $L_X/L_{Bol} = A R_{0}^{-2.7}$, where the factor of $-2.7$ is taken from \cite{Wright_2011}. From this, we conclude that \egp\ is a typical F-type star in terms of \updatebf{its rotation and activity.}
}  
\label{fig_lxlbol_vs_r0inv}
\end{figure}

We further explore the properties of \egp\ in the context of similar stars (see Table~\ref{table:LXLBol_compare}), by comparing the relative X-ray luminosity $\frac{L_X}{L_{bol}}$, effective temperature, rotation period and metallicity.  In Figure~\ref{fig_lxlbol_vs_r0inv}, we show the variation in $\frac{L_X}{L_{bol}}$ with the Rossby number 
$R_{0}$ (the ratio of the equatorial rotational period P$_{\rm Rot}$ and the convective turnover time scale $\tau_C$), where \egp\ is represented with a red block whose height represents the variation in measured $L_X$, and for $R_0$ calculated using $\tau_C$ from \citet{Gunn_1998}. \egp\ is unremarkable in this space, with a slightly lower activity than expected by the trend line, but consistent with other F dwarfs of similar milieus. Other physical parameters like effective temperature are also consistent with other F7V-F9V stars.

In contrast, we find that the photospheric abundances are higher than similar stars. Typically, [Fe/H]$_{\phot}\lesssim{-0.1}$ for dF stars of similar activity and distance, but is $\gtrsim$0 for \egp\ \citeauthor{Gonzalez_2007}\ (2007; corrected to \citealt{Anders_1989}).  Note however that this appears to be a characteristic of the presence of close-in giant planets \citep{Fischer_2005,Wang_2015}; indeed, for the stars listed in Table~\ref{table:LXLBol_compare}, we find that for those without confirmed exoplanets (14 stars), $\langle \rm [Fe/H]_{\phot} \rangle = -0.22 \pm 0.18$ while for stars with confirmed exoplanets (3, including \egp), $\langle \rm [Fe/H]_{\phot} \rangle = -0.06 \pm 0.10$. 

\begin{deluxetable*}{llllllll}[!htbp]
\tabletypesize{\footnotesize}
\tablecolumns{8}
\tablewidth{0pt}
\tablecaption{Sample of nearby F7V-F9V stars \label{table:LXLBol_compare}}
\tablehead{ \colhead{Name} & \colhead{Spectral type} & \colhead{$B-V$} & \colhead{T$_{eff}$ [K]} & \colhead{[Fe/H]$_{\phot}$} & \colhead{log($L_X/L_{Bol}$)} & \colhead{$R_{0}^{-1}$} & \colhead{Reference} \vspace{-1em}\\ }
\startdata
$\beta$ Vir  & F8IV-F8V & 0.550 & 6071 & +0.03$^{\rm A09}$ & -5.65 & 0.628 & (1,2,3,4,11)\\
$\theta$ Per & F7V & 0.514 & 6196 & -0.16$^{\rm G07}$ & -5.81 & 0.621 & (1,2,3,4,11)\\ 
$\zeta$ dor & F7V-F8V & 0.526 & 6153 & -0.40$^{\rm A09}$ & -4.60 & 2.913 & (1,2,3,4,11)\\
36 UMa & F8V & 0.515 & 6172 & -0.25$^{\rm A09}$ & -5.43 & 0.724 & (1,2,3,4,11)\\
$\upsilon$ And$^{\dag}$ & F8V & 0.540 & 6106 & -0.07$^{\rm A09}$ & -5.85 & 0.621 & (1,2,4,5,11)\\
$\iota$ Psc & F7V & 0.500 & 6241 & -0.24$^{\rm A09}$ & -5.92 & 0.628 & (1,2,3,4,11)\\
99 Her  & F7V & 0.520 & 5947 & -0.73$^{\rm A09}$ & -5.83 & 0.637 & (2,3,4,6)\\
HD~10647$^{\dag}$ & F8V-F9V & 0.551 & 6151 & -0.18$^{\rm A09}$ & -5.47 & 0.956 & (2,3,4,7)\\
$o$ Aql  & F8V & 0.560 & 6120 & -0.05$^{\rm A09}$ & -5.84 & 0.850 & (2,4,8,9,11)\\
HD~154417 & F8.5V & 0.580 & 6042 & -0.14$^{\rm A09}$ & -4.85 & 1.224 & (2,4,9,10,11)\\
HD~16673 & F6V-F8V & 0.524 & 6301 & -0.12$^{\rm A09}$ & -5.22 & 0.941 & (2,8,4,9,11)\\
HR~4767 &  F8V-G0V & 0.557 & 6038 & -0.23$^{\rm A09}$ & -5.50 & 0.647 & (2,3,4,7,11)\\  
$\psi$1 Dra B & F8V-G0V & 0.510$^*$ & 6223 & -0.15$^{\rm G07}$ & -5.16 & 0.649 & (2,3,4,7,12)\\
HR~5581 & F7V-F8V & 0.530 & 6156 & -0.06$^{\rm G07}$ & -5.38 & 0.577 & (2,3,4,7,11)\\
HR~7793 & F7V & 0.510$^{**}$ & 6261 & -0.26$^{\rm S15}$ & -5.44 & 0.826 & (2,4,8,9)\\
HD~33632 & F8V & 0.542 & 6074 & -0.36$^{\rm A09}$ & -5.19 & 0.687 & (2,3,4,7,11)\\
\vspace{-1em} 
\enddata  
\tablenotetext{$\dag$}{ \centering Star with confirmed close-in giant planet \citep{Stassun_2019}.}
\tablenotetext{\hfil}{References:-- \\
$-$ for spectral types and $L_X/L_{Bol}$: 
[1] \cite{Schmitt_2004}, 
[6] \cite{Suchkov_2003}, 
[7] \cite{Hinkel_2017}, and 
[9] \cite{Freund_2022}; \\
$-$ for T$_{eff}$ and [Fe/H]$_{\phot}$: 
[2] \cite{Stassun_2019,2019yCat..51560102S}; \\
$-$ for $B-V$:
[11] \cite{BoroSaikia_2018}, $**$[8]~\citet[][]{Baliunas_1996}, $*$[12]~\citet[][]{Fabricius_2002}; \\
$-$ P$_{\rm Rot}$ is estimated from B-V and S$_{HK}$ relation from [3]~\cite[][]{Noyes_1984}, or directly obtained from [5]~\cite{Shkolnik_2008}, [8]~\cite{Baliunas_1996}, or [10]~\cite{Donahue_1996}; \\
$-$ $\tau_C$ values from 
[4]~\citet{Gunn_1998} calculated using $B-V$ values is used $-$ with P$_{\rm Rot}$ to obtain the Rossby number $R_0$; \\
$-$ ${\rm [Fe/H]_{ph}}$ measurements converted to
\cite{Anders_1989} from 
[A09]~\citet{Asplund_2009},
[G07]~\citet{Grevesse_2007}, and 
[S15]~\citet{Scott_2015}. }
\end{deluxetable*}

\subsection{Spectral and Temporal Variations}\label{sec:disc_specvar}

As we show in Section~\ref{sec:timing}, considerable variability is present in both intensity and color over a range of time scales.  We first note that the existence of close and loose mergers of change points from several bands point to there being substantial energy release incidents that reinforce the presence of ubiquitous variability in the corona of \egp.  Several passband color combinations show the presence of variability over the durations of the observations (see Table~\ref{table:chisqHR}, and small time scale changes (see Figures~\ref{fig_HRoxyfe},\ref{fig_HR5427}).

While the source is not bright enough to allow us to carry out time-resolved spectroscopy, we can infer the presence of spectral variability through hardness ratio light curves.  We find that pervasive variability at time scales $\sim$ksec is present throughout the \chandra\ observations (see Section~\ref{sec:HR}).  Table~\ref{table:chisqHR} lists the estimated goodness of a constant hardness ratio model for all passband combination for all passband combinations and observations; we find that $\chi^{2}_{\rm red} \gtrsim 3$ for all cases.  This variability suggests that the corona of \egp\ is dynamic, with intermittent impulsive energy releases occurring continuously.

This small time scale variability also translates to the corona in its gross characteristics over the $\gtrsim$8-hour observation durations.  Variations of factors of 2$\times$ is present in all fitted parameters (Table~\ref{table:fitparams}).  Nevertheless, it is notable that large flares are not present.  We speculate that this is due to the inhibition of large stresses from being built up in the magnetic loops in the corona because periodic interference of the planetary magnetic field which would act to dissipate the stresses continuously.
The estimated \updatebf{abundances show that Fe is} \textit{depleted} in \egp\ \updatebf{(see Section~\ref{sec:abun})}. Further, due to the under-abundance of Neon, following from \citet{Laming_2015}, we conclude that the star is not active. A comparison of $\frac{L_X}{L_{bol}}$ which is found to be [-5.6, -5.5] for \egp\ with other F-type stars like HD~17156 with $\frac{L_X}{L_{bol}}$ = [-5.8, -6.0] further supports the state of quiescence \citep{Maggio_2015}.

\subsection{Phased variations} \label{sec:PhaseVar}

From \updatebf{the Lomb-Scargle periodogram analysis (Section~\ref{sec:LS}, Figure~\ref{fig_ls}),} we note a definitive correlation with stellar polar rotation, in agreement with \citet{Scandariato_2013}. Apart from this, the periodogram also suggests likely  variability with the planetary orbital period, thus confirming past results of chromospheric observations \citep{Shkolnik_2003,Shkolnik_2005,Shkolnik_2008}. Lastly, it is also likely that there is variability tied to the beat frequency between the stellar polar rotation and the planetary orbital period, which was suggested by \citet{Fares_2012}. These results can further be improved with more observations to improve phase coverage.

Note that \citet{Shkolnik_2008} suggested an on/off nature of star-planet interactions, and during 2 out of 6 observation epochs, they found a periodicity of \updatebf{$\approx$7 days}.  We have not detected this period in our results, but non-detection of SPI in some observations is understandable, as the field stretching effect discussed earlier should occasionally cease.  This could be due to decay of the active polar region on the star, or because the connection to the star is broken by excessive ``field wrapping" due to the differing $P_{\rm orb}$ and $P_{\rm polar}$. In either case, some time would be needed to reestablish a sturdy star-planet connection. 
In \citet{Scandariato_2013}, their X-ray light curves also suggest variability tied to a period of \updatebf{$\approx$4 days}, apart from variability with the stellar polar rotation. Due to limitations of the data used, and due to non-detection of variability tied to the planetary orbital period, they did not conclude the presence of SPI. However, we believe our result of possible variability tied to the beat period of the stellar polar rotation and the planetary orbital period agrees with their measured period of \updatebf{$\approx$4 days}, and could be  indicative of the presence of SPI.  Alternatively, the 4 day signal could be due to two active longitudes on the star at a latitude a bit removed from the equator.

\subsection{Coronal Abundances} \label{sec:abun}

We find consistent coronal abundance measurements for $A({\rm O})/A({\rm Fe})$ and $A({\rm Ne})/A({\rm Fe})$ across all datasets, with $A({\rm O})/A({\rm Fe})\sim 0.7$ and $A({\rm Ne})/A({\rm Fe})\ll 0.1$. In contrast, $A({\rm N})/A({\rm Fe})$ is found to be $\gg 1.0$ for some datasets (ObsIDs 5427, 6119, and 6120). Further, the average $Z_{\text{Fe}} \sim 0.2$ for all datasets except 6119, where it rises to $\approx$0.45. Note that for ObsID 6119 (phase $\phi\approx{0.1}$ \updatebf{occurring} just after planetary conjunction), the abundances of other metals are boosted as well, though within the uncertainty bounds from other ObsIDs. Overall, we conclude that the corona of \egp\ is significantly deficient in Fe relative to the Sun.
We also find indications of abundance variations at small time scales in the color light curves.  For instance, we find an extended duration ($\gtrsim$3~ks) where $C{\rm [Oxy/Fe]}$ and $C{\rm [Oxy/Ne]}$ increase (see Section~\ref{sec:HR}) without an accompanying change in the temperature sensitive $C{\rm [s/m+h]}$ or in $C{\rm [Fe/Ne]}$, suggesting a short duration surge in O abundance.  However, detailed spectral analyses, exploring whether the abundances in specific elements can be estimated, are statistically unsupportable.  None of the models {\bf 2v/Z} were found to be {\sl required} to explain the spectra, though some instances of models of {\bf 2v/Al} suggest that Al/Fe $>$ solar.

\updatebf{We further estimate the abundances of high FIP elements like C, N, O, and Ne relative to Fe in the corona in order to evaluate the FIP bias in \egp.  We find that typically the ranges in ${\rm [C,O/Fe]_{\coro|\angr}} \approx -0.37 {\rm~to~} +0.04$ and ${\rm [Ne/Fe]_{\coro|\angr}} \approx -0.87 {\rm~to~} -0.56$ are sub-solar, and the range in ${\rm [N/Fe]_{\coro|\angr}} \approx 0 {\rm~to~} +1$ is super-solar (Table~\ref{table:Fbiassummary}).  We compute the FIP bias by comparing our measurements with those compiled in the literature.  In particular, we use the measurements of \citet{Ecuvillon_2004,2016ApJS..225...32B,Luck_2018} together with an assumed photospheric correction for ${\rm [Ne/O]}_{\testa|\angr}=+0.48$ (see Appendix~\ref{sec:appendix_abun}) to compute F$_{bias}$ (see Equation~\ref{eq:fipbias}).  F$_{bias}<0$ indicates that stars have a solar-like FIP effect, and F$_{bias}>0$ incidates an inverse-FIP effect.  For stars of type late-F and relatively low activity like \egp\ ($L_X/L_{Bol}{\approx}3{\times}10^{-6}$, $R_0^{-1}{\approx}1$; see Figure~\ref{fig_lxlbol_vs_r0inv}), F$_{bias}\lesssim{-0.5}$ are expected \citep[][]{Wood_2012,Testa_2015,Wood_2018}.  Here, we find estimates (see Table~\ref{table:Fbiassummary}) based on \citet{2016ApJS..225...32B} to be  ${\rm F}_{bias}|_\brwr = -0.09 \pm 0.02~{\rm scatter} \pm 0.29~{\rm statistical}$ and based on \citet[][]{Ecuvillon_2004} and \citet[][]{Luck_2018} to be ${\rm F}_{bias}|_{\ecu,\luck} = -0.33 \pm 0.03~{\rm scatter} \pm 0.32~{\rm statistical}$. These two estimates differ from each other, though the statistical uncertainties are large enough that they are not statistically inconsistent.  Both estimates are however larger than the nominal expected value.

However, \citet{Wood_2018} suggest that F$_{bias}$ is larger than the nominal trend for stars with close-in Jupiter-mass planets: $\tau$\,Boo\,A has an observed F$_{bias}=-0.21{\pm}0.9$ when the expected value is ${\approx}-0.50$, and HD\,189733 is observed to be at $0.24{\pm}0.22$ instead of the expected ${\approx}-0.2$.  In both these cases, as well as for \egp, the observed F$_{bias}$ is less solar-like than expected and consistent with a reduced magnitude of FIP effect.  We speculate that this is due to the FIP effect being diluted due to mechanical mixing (see Section~\ref{sec:SPI})
}

\subsection{Magnetic SPI}\label{sec:SPI}

The star's large-scale magnetic geometry which has been mapped using ZDI (Zeeman-Doppler Imaging; \citealt{Semel_1989}, \citealt{Donati_1997}) and near contemporaneously with \chandra\ observations during 2007 and 2009 \citep[][]{Fares_2012}.  The well-observed 2009 data yield a strongly dipolar configuration, but with the magnetic pole inclined 70$^{\circ}$ to the rotational axis.  Our detection of periodicity corresponding to the polar rotational period thus implies magnetic activity around the magnetic equator but confined to the rotational pole.  The emission measures we estimate from the spectral analysis (see Table~\ref{table:fitparams}) suggest that the active area ranges from 1 - 20\% of the surface area depending on the assumed coronal plasma density,\footnote{We compute the fraction of the surface area occupied by the coronal plasma at temperature $T$ as 
$$f_{\rm fill} \equiv \frac{V(\text{EM},n_e)}{h_p(g_*,T)} \frac{1}{A_*} = \frac{\rm EM }{n_e^2} \frac{\mu m_H g_*}{k_B T} \frac{1}{4 \pi R_*^2}~,$$
where $V$ is the coronal volume, $n_e$ is the plasma electron density, $h_p$ is the pressure scale height at temperature $T$ for a surface gravity $g_*$, EM is the emission measure of the temperature component, $R_*$ is the radius of \egp, and $\mu \approx 1$ is the fractional atomic mass relative to $m_H$, the mass of a H atom.
Thus, 
$$f_{\rm fill} = 0.064 \frac{\rm EM}{10^{51}~{\rm cm}^{-3}} \left(\frac{n_e}{5 \times 10^{9}~{\rm cm}^{-3}}\right)^{-2} \left(\frac{T}{2 \times 10^6~{\rm K}}\right)^{-1} \,.$$
A circular region that corresponds to this fraction of the surface area would extend from the rotational pole down to a latitude of $$b = 90 - \frac{180}{\pi} \cos^{-1}(1 - 2 f_{\rm fill})~{\rm degrees}.$$
} and is thus consistent with this scenario.

\citet{Cauley_2019} study SPI in the context of several models of potential interactions, specifically testing which models can generate sufficient power to generate the observed enhancement in Ca II H\&K emission of several systems, including \egp.  They found reconnection (flaring) between the star and planet was insufficient to heat H\&K in many cases; instead, a model where heating is derived from field-line stretching was found more adequate to the task.  Here we employ the same  models to test whether the (weaker) coronal enhancement  might be reconnection heated. The two main models are heating by reconnection \citep[][]{Lanza_2009,Lanza_2012}, and heating by field line stretching \citep[][]{Lanza_2013}. In the reconnection case \citep[i.e., essentially flaring between star and planet as proposed by][]{Cuntz_2000}, the expected power \updatebf{due to reconnection,
$$P_r = \gamma (\pi/\mu) R_p^2 B_p^{2/3} B_{*}(a_p)^{4/3} v_{\rm rel} \,,$$
%
}
where $0 \leq \gamma \leq 1$ is a constant related to the relative magnetic geometry between star and planet, $\mu$ is the magnetic permeability of the vacuum, $R_p$ and $B_p$ are the planetary radius and dipolar magnetic field strength, respectively, $B_{*}(a_p)$ is the stellar field strength at the location of the planet, and $v_{\rm rel}$ is the relative velocity between the orbiting planet and the rotating magnetic footpoint on the star.  
\updatetwo{We assume that the stellar dipole field strength scales as the cube of the distance, i.e., $B_*(a_p) = B_*(R_*)\cdot\left({a_p}/{R_*}\right)^{-3} \approx 0.005$~G.  For $\gamma\approx\mu\approx{1}$ and $v_{rel}=150$~km~s$^{-1}$ the Keplerian velocity of the planet in its orbit, we find (see Table~\ref{table:baseparams}) $P_r{\approx}6{\times}10^{25}$~ergs~s$^{-1}$, consistent with the estimate by \citet{Lanza_2012}.}

In the field stretching case, the 
stellar field power \updatebf{available due to field stretching,
$$P_s = \frac{2 \pi}{\mu} f_{B_p} R_p^2 B_p^2 v_{\rm rel} \,.$$
where} $f_{B_p}$ is the fractional area of the planet
\updatetwo{covered by the magnetic footprint.  Using the values from Table~\ref{table:baseparams}, we find $P_s{\approx}5{\times}10^{30} \frac{f_{B_p}}{0.1}$~ergs~s$^{-1}$.  Note that the average X-ray luminosity of \egp\ is $L_{\rm X}{\approx}2{\times}10^{28}$~ergs~s$^{-1}$, with variations ${\Delta}L_{\rm X}{\gtrsim}10^{27}$~ergs~s$^{-1}$ between observations.  Thus, $P_s{\gg}L_{\rm X}>{\Delta}L_{\rm X}{\gg}P_r$, and we conclude that only the field stretch model of \citet{Lanza_2013} is capable of producing the required power for \egp. 
}
Thus, while the SPI appears magnetic in origin \citep[viz.][]{Saar_2004}, the power requirements point to a non-flare like origin.

\updatebf{Additional justification for this picture comes from the estimate of the FIP bias, which measures the relative abundances of high-FIP and low-FIP elements between the corona and the photosphere.  The estimated value of the FIP bias,
F$_{bias} = (-0.33 - -0.09)$
for \egp\ (see Section~\ref{sec:abun}), suggests that FIP mechanisms operating in solar-like situations, such as ponderomotive forces \citep{Laming_2015} or proton drag in the chromosphere \citep{1996ApJ...464L..91W}, could be diluted.  We speculate that the driver for this dilution is the direct mechanical mixing of the metals in the corona leading to the homogenization of abundances between the photosphere and corona.  That is, the stretching and kneading of flux tubes \updatetwo{due to magnetic SPI} could set up flows that move metals from the photosphere up to the corona, leading to a FIP bias closer to zero.}

\section{Summary} \label{sec:summary}

We have carried out a comprehensive spectral and temporal analysis of the \chandra\ observations of \egp.

We first carry out spectral fits (Section~\ref{sec:spec}) using two-temperature APEC spectral models, for which we also estimate abundances by carefully analyzing and minimizing structures in residuals using a novel technique.   \updatebf{The estimated coronal $\frac{A({\rm Fe})}{A({\rm H})}\ll$1, while $\frac{A({\rm O}),A({\rm N})}{A({\rm Fe})}\,\gtrsim\,$1, $\frac{A({\rm Ne})}{A({\rm Fe})}\,\ll\,$1, relative to the solar baseline of $\angr$ \citep[][]{Anders_1989}.}  The abundances relative to the photosphere (see Appendix~\ref{sec:appendix_abun}) demonstrate that \egp\ has a larger FIP bias than is expected for stars of similar types, similar to other F stars with close-in planets.

We find that light curves in all passbands are characterized by significant variability over time scales $10^2-10^4$~s.  No \updatebf{large scale} flares are detected, suggesting a low-activity but dynamic environment.  Hardness ratio variations are also seen, albeit at a lower significance than that of intensity, suggesting that spectral variations also occur in the corona at time scales $\lesssim$10$^4$~s.

Period analysis using the Lomb-Scargle periodogram method yields several congruous peaks that exceed $2\sigma$ significance.  We find power peaks at the stellar polar rotational period (\updatebf{$\approx$10~d}), the planetary orbital period (\updatetwo{$\approx$3~d}), and the beat period between the two (\updatebf{$\approx$4.5~d}).
The presence of peaks at the planetary orbital time period are in agreement with \citet{Shkolnik_2003,Shkolnik_2005,Shkolnik_2008}. The presence of peaks at the beat frequency between the planetary orbital time period and the stellar polar rotation period suggest agreement with \citet{Fares_2012} and both together are indicative of SPI, with a magnetic connection between the planetary magnetic field and the stellar magnetic equatorial field. Mechanisms of field stretching given by \citet{Lanza_2013} are explored as the possible mechanism leading to the observed SPI, as it would not always lead to chromospheric variability, thus agreeing with the on/off nature noted in \citet{Shkolnik_2008}, and the lack of detection of SPI in \citet{Scandariato_2013}. We also observe variability tied to the stellar polar rotation in agreement with \cite{Scandariato_2013}. The calculated area fractions provide further support for the field stretching phenomenon expected to be powering the SPI.

We also note that while \egp\ is quite similar to other F stars for most parameters, its photospheric metallicity [Fe/H] is higher than most of them, consistent with it having a planetary system \citep[viz.,][]{Fischer_2005}. 

Further, the calculated $F_{\rm bias}$ from the coronal abundances of different metals indicates that \egp\ shows a \updatebf{tendency of a shift towards an} inverse FIP effect instead of the solar-like FIP effect expected according to \citet{Wood_2018} and \citet{Testa_2015}. \updatebf{Indeed, the measured FIP bias is consistent with there being a dilution of the expected solar-like FIP effect, similar to the FIP biases exhibited by some other planet-hosting stars like $\tau$\,Boo\,A and HD\,189733A. \updatetwo{Both of these stars have shown evidence for magnetic SPI, as was highlighted in \citet{Wood_2018, Pillitteri_2014a}}. Thus, our measured abundances lend credence to the possibility of SPI operating to affect the coronal composition of \egp\ as well.}

More data are necessary to improve phase coverage and the significance of the results found here. Given the large amounts of small-scale variability in the color-color ratios, it is important to cover both soft and hard passbands using a monitoring X-ray telescope. Augmenting this with Ca II observations would be useful to test the possible physical phenomena discussed in this paper.

\vspace{-2em}
\acknowledgements
We thank the anonymous referee for useful comments that helped to significantly clarify and improve the paper. 
AA acknowledges support from \chandra\ Grant AR6-17003X, SAO fund 040208IH00VK, KVPY Fellowship by DST-India, and administrative support from IISER-Mohali. 
VLK acknowledges support from NASA Contract to \chandra\ X-ray Center NAS8-03060 and \chandra\ Grant AR6-17003X.
SHS acknowledges support from NASA XRP grant 80NSSC21K0607 and \chandra\ grant GO5-6019A.
KPS thanks the Indian National Science Academy for support under the INSA Senior Scientist Programme.  This research has made use of data obtained from the Chandra Data Archive and software provided by the Chandra X-ray Center (CXC) in the application packages CIAO and Sherpa.
This research has made use of the SIMBAD database (DOI : 10.1051/aas:2000332), and the VizieR catalogue access tool, operated at CDS, Strasbourg, France (DOI : 10.26093/cds/vizier).

\facilities{\chandra, XMM-{\sl Newton}, {\sl Swift}}

\software{\ciao\ \citep[][]{Fruscione_2006}, PINTofALE \citep[][]{2000BASI...28..475K}, astropy \citep{astropy:2022}, VizieR \citep{2000A&AS..143...23O}}

\clearpage
\appendix

\section{Abundance Comparisons}\label{sec:appendix_abun}

\updatebf{Since most abundance values that are reported in the literature are relative to some baseline (set usually to solar photospheric), it is imperative that the comparisons are made with a well-established baseline when comparing abundance measurements.  Here we describe in detail how the \egp\ photospheric measurements we compile are converted to \angr\  \citep[][]{Anders_1989}, for which 
$${\rm [X=\{C,N,O,Ne,Fe\}]}_\angr \equiv 12 + \log_{10} \frac{A({\rm X})}{A({\rm H})} = \{8.56,8.05,8.93,8.09,7.67\}$$
respectively, relative to H=12.  We begin by noting how the generic number abundance of an element X relative to Y in baseline system {\tt a} relative to baseline system {\tt b} can be transformed to baseline system {\tt c},
\begin{eqnarray}
    {\rm [X/Y]}_{\tt a | b} &\equiv& \log_{10} \frac{A({\rm X})_{\tt a}}{A({\rm Y})_{\tt a}} - \log_{10} \frac{A({\rm X})_{\tt b}}{A({\rm Y})_{\tt b}} \nonumber \\
    &=& \log_{10} \frac{A({\rm X})_{\tt a}}{A({\rm Y})_{\tt a}}
    - \log_{10} \frac{A({\rm X})_{\tt c}}{A({\rm Y})_{\tt c}}
    + \log_{10} \frac{A({\rm X})_{\tt c}}{A({\rm Y})_{\tt c}}
    - \log_{10} \frac{A({\rm X})_{\tt b}}{A({\rm Y})_{\tt b}} \nonumber \\
    &\equiv&  {\rm [X/Y]}_{\tt a | c} + {\rm [X/Y]}_{\tt c | b} \,.
\end{eqnarray}
Notice that F$_{bias}$ (Equation~\ref{eq:fipbias}) is invariant to the chosen baseline, since each term in the sum is computed as a relative difference between the corona ($\coro$) and the photosphere ($\phot$).  Thus, abundance estimates relative to $\angr$ are exactly equivalent to estimates in an alternate system {\tt b},
\begin{eqnarray}
    {\rm [X/Fe]}_{\coro|\angr} - {\rm [X/Fe]}_{\tt \phot|\angr} &=& \left( {\rm [X/Fe]}_{\coro|b} + {\rm [X/Fe]}_{\tt b|\angr} \right) - \left( {\rm [X/Fe]}_{\tt \phot|b} + {\rm [X/Fe]}_{\tt b|\angr} \right) \nonumber \\
    &\equiv& {\rm [X/Fe]}_{\coro|b} - {\rm [X/Fe]}_{\tt \phot|b} \,,
\end{eqnarray}
provided that all the abundances are reported for the same baseline.

Consider, as an example case, the abundance of Ne relative to O.  \citet[][]{Drake_2005}, based on an extensive analysis of active stars, suggested that the Ne abundance in the solar photosphere must be increased by a factor $\approx2.5\times$ relative to the \aspl\ \citep[][]{Asplund_2009} values.  Thus, 
$${\rm [Ne/O]_{\testa|\angr}} = {\rm [Ne/O]_{\testa|\aspl} + {\rm [Ne/O]_{\aspl|\angr}}} = \log_{10}(2.5) + {\rm [Ne/O]}_{\aspl|\angr} = +0.48 \,,$$ or an increase by a factor $\approx3\times$ relative to \angr.  Since ${\rm [Ne/O]}_\angr=-0.84$, in absolute terms, $${\rm [Ne/O]}_{\testa|{\tt abs}} = {\rm [Ne/O]}_{\testa|\angr} + {\rm [Ne/O]}_{\angr|{\tt abs}} = -0.36 \,.$$
Photospheric Ne abundances are difficult to measure, and are unavailable for \egp.  Therefore, in order to compute F$_{bias}$, we use
\begin{eqnarray}
    {\rm [Ne/Fe]}_{\phot|\angr} &=& {\rm [Ne/O]}_{\phot|\angr} + {\rm [O/Fe]}_{\phot|\angr} \nonumber \\
    &=& +0.48 + {\rm [O/Fe]}_{\phot|\luck} + {\rm [O/Fe]}_{\luck|\angr} = +0.64 \nonumber \\ 
    &=& +0.48 + {\rm [O/Fe]}_{\phot|\brwr} + {\rm [O/Fe]}_{\luck|\brwr} = +0.42 \,, 
    \label{eq:NeFeangr}
\end{eqnarray}
where $\luck$ and $\brwr$ refer to the photospheric abundance measurements of \egp\ by \citet[][]{Luck_2018} and \citet[][]{2016ApJS..225...32B} respectively.  Similarly, we list various abundance measurements obtained from the literature, corrected to the $\angr$ baseline, in Table~\ref{tab:phabun}.
}

\begin{deluxetable}{llll}[!htb]
\tablecaption{Photospheric abundances of \egp}\label{tab:phabun}
\tablehead{ \colhead{Abundance} & \colhead{Reference} & \colhead{Baseline} & \colhead{corrected to $\angr$}}
\startdata
${\rm [C/H]}_{\phot|\luck}=+0.16 \pm 0.10$ & \citet[][]{Luck_2018} & C=8.50$^\dag$ & ${\rm [C/H]}_{\phot|\angr}=+0.10 \pm 0.10$ \\
${\rm [C/H]}_{\phot|\brwr}=+0.14 \pm 0.026$ & \citet[][]{2016ApJS..225...32B} & C=8.39 & ${\rm [C/H]}_{\phot|\angr}=-0.03 \pm 0.026$ \\
\hline
${\rm [N/H]}_{\phot|\ecu}=+0.33 \pm 0.12$ & \citet[][]{Ecuvillon_2004} & N=8.05 & ${\rm [N/H]}_{\phot|\angr}=+0.33 \pm 0.12$ \\
${\rm [N/H]}_{\phot|\brwr}=+0.25 \pm 0.042$ & \citet[][]{2016ApJS..225...32B} & N=7.78 & ${\rm [N/H]}_{\phot|\angr}=-0.02 \pm 0.042$ \\
\hline
${\rm [O/H]}_{\phot|\luck}=+0.35 \pm 0.10$ & \citet[][]{Luck_2018} & O=8.76$^\ddag$ & ${\rm [O/H]}_{\phot|\angr}=+0.18 \pm 0.10$ \\
${\rm [O/H]}_{\phot|\brwr}=+0.22 \pm 0.036$ & \citet[][]{2016ApJS..225...32B} & O=8.66 & ${\rm [O/H]}_{\phot|\angr}=-0.05 \pm 0.036$ \\
\hline
${\rm [Fe/H]}_{\phot|{\tt GL07}}=+0.22 \pm 0.055$ & \citet[][]{Gonzalez_2007} & Fe=7.50$^*$ & ${\rm [Fe/H]}_{\phot|\angr}=0.056 \pm 0.055$ \\
${\rm [Fe/H]}_{\phot|\luck}=+0.22 \pm 0.06$ & \citet[][]{Luck_2018} & Fe=7.46$^{\dag\dag}$ & ${\rm [Fe/H]}_{\phot|\angr}=+0.01 \pm 0.06$ \\ 
${\rm [Fe/H]}_{\phot|\ecu}=+0.22 \pm 0.05$ & \citet[][]{Ecuvillon_2004} & Fe=7.47$^{\ddag\ddag}$ & ${\rm [Fe/H]}_{\phot|\angr}=+0.02 \pm 0.05$ \\
${\rm [Fe/H]}_{\phot|\brwr}=+0.23 \pm 0.01$ & \citet[][]{2016ApJS..225...32B} & Fe=7.45 & ${\rm [Fe/H]}_{\phot|\angr}=+0.01 \pm 0.01$ \\
\hline
${\rm [C/Fe]}_{\phot|\luck}=-0.06 \pm 0.12$ & \citet[][]{Luck_2018} & \hfil & ${\rm [C/Fe]}_{\phot|\angr}=+0.09 \pm 0.12$ \\
${\rm [C/Fe]}_{\phot|\brwr}=-0.09 \pm 0.028$ & \citet[][]{2016ApJS..225...32B} & \hfil & ${\rm [C/Fe]}_{\phot|\angr}=-0.04 \pm 0.028$ \\
${\rm [N/Fe]}_{\phot|\ecu}=+0.11 \pm 0.13$ & \citet[][]{Ecuvillon_2004} & \hfil & ${\rm [N/Fe]}_{\phot|\angr}=+0.31 \pm 0.13$ \\
${\rm [N/Fe]}_{\phot|\brwr}=+0.02 \pm 0.043$ & \citet[][]{2016ApJS..225...32B} & \hfil & ${\rm [N/Fe]}_{\phot|\angr}=-0.03 \pm 0.043$ \\
${\rm [O/Fe]}_{\phot|\luck}=+0.12 \pm 0.12$ & \citet[][]{Luck_2018} & \hfil & ${\rm [O/Fe]}_{\phot|\angr}=+0.16 \pm 0.12$ \\
${\rm [O/Fe]}_{\phot|\brwr}=-0.01 \pm 0.037$ & \citet[][]{2016ApJS..225...32B} & \hfil & ${\rm [O/Fe]}_{\phot|\angr}=-0.06 \pm 0.037$ \\
${\rm [Ne/Fe]}_{\phot|\luck}={\rm N/A}^{**}$ & \citet[][]{Luck_2018} & \hfil & ${\rm [Ne/Fe]}_{\phot|\angr}=+0.64 \pm 0.12$ \\
${\rm [Ne/Fe]}_{\phot|\brwr}={\rm N/A}^{**}$ & \citet[][]{2016ApJS..225...32B} & \hfil & ${\rm [Ne/Fe]}_{\phot|\angr}=+0.42 \pm 0.037$ \\
\enddata
\tablenotetext{$\dag$}{~~Based on \citealt{SuarezAndres2017}. }
\tablenotetext{$\ddag$}{~~Based on \citealt{BdeLis+2015} and \citealt{Caffau+2008}. }
\tablenotetext{$*$}{~~Based on $\grsa$, \citet[][]{1998SSRv...85..161G}. }
\tablenotetext{$\dag\dag$}{~~Based on \citealt{Grevesse+2015}. }
\tablenotetext{$\ddag\ddag$}{~~Based on \citealt{Santos+2004}. }
\tablenotetext{$**$}{~~Computed as ${\rm [Ne/Fe]}_{\phot|\angr}={\rm [Ne/O]}_{\phot|\angr}+{\rm [O/Fe]}_{\phot|\angr}$ (see Equation~\ref{eq:NeFeangr}). }
\end{deluxetable}

\section{CuSum plots} \label{sec:cusum}

\updatebf{Here we show the Cumulative Sum (CuSum) plots of the best fit models for all datasets.}
The count rate light curves for all the passbands for all ObsIDs, hardness ratio light curves for all relevant passband pairs for all ObsIDs, cumulative sum residuals plots for both models {\bf 2m} and {\bf 2v} for all ObsIDs, \updatebf{and corner plots for all spectral fits}  are available on Zenodo \citep[][]{anshuman_acharya_2022_7181873}\footnote{\href{https://doi.org/10.5281/zenodo.7220014}{10.5281/zenodo.7220014}}.

\begin{figure*}[!htbp]
\centering
\begin{tabular}{cc}
\includegraphics[width=0.35\textwidth,height=\textheight,keepaspectratio]{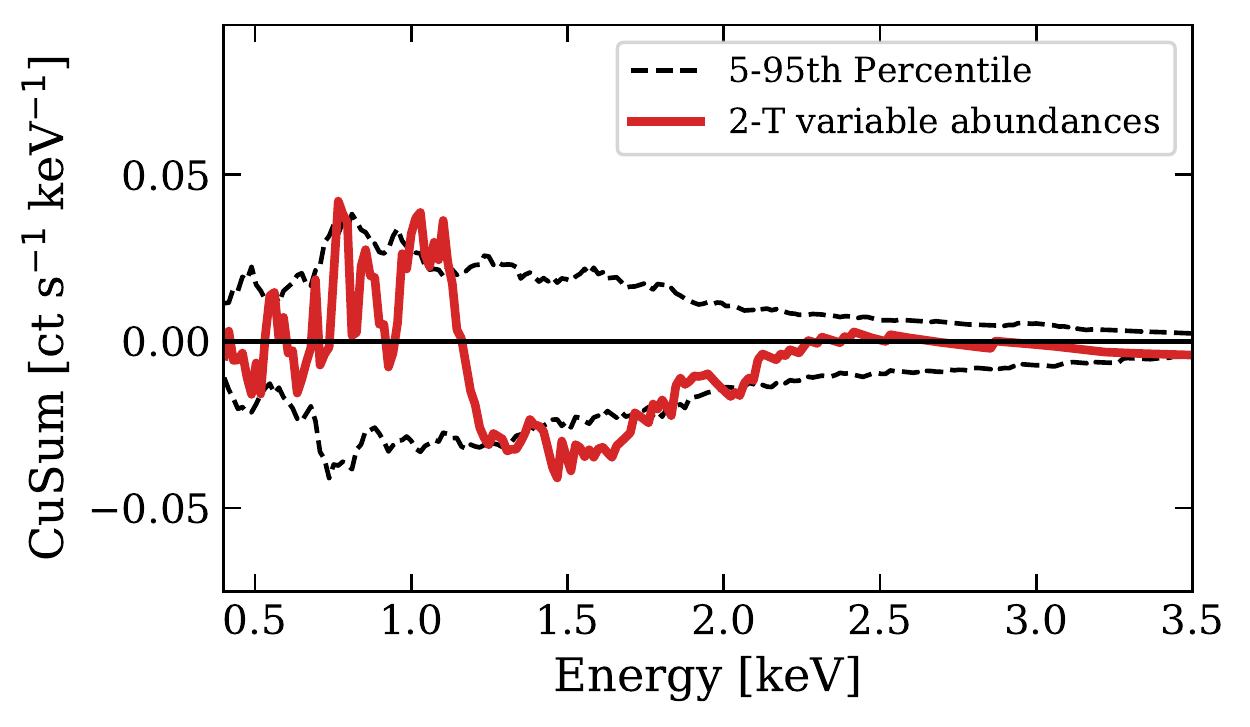} &
\includegraphics[width=0.32\textwidth,height=\textheight,keepaspectratio]{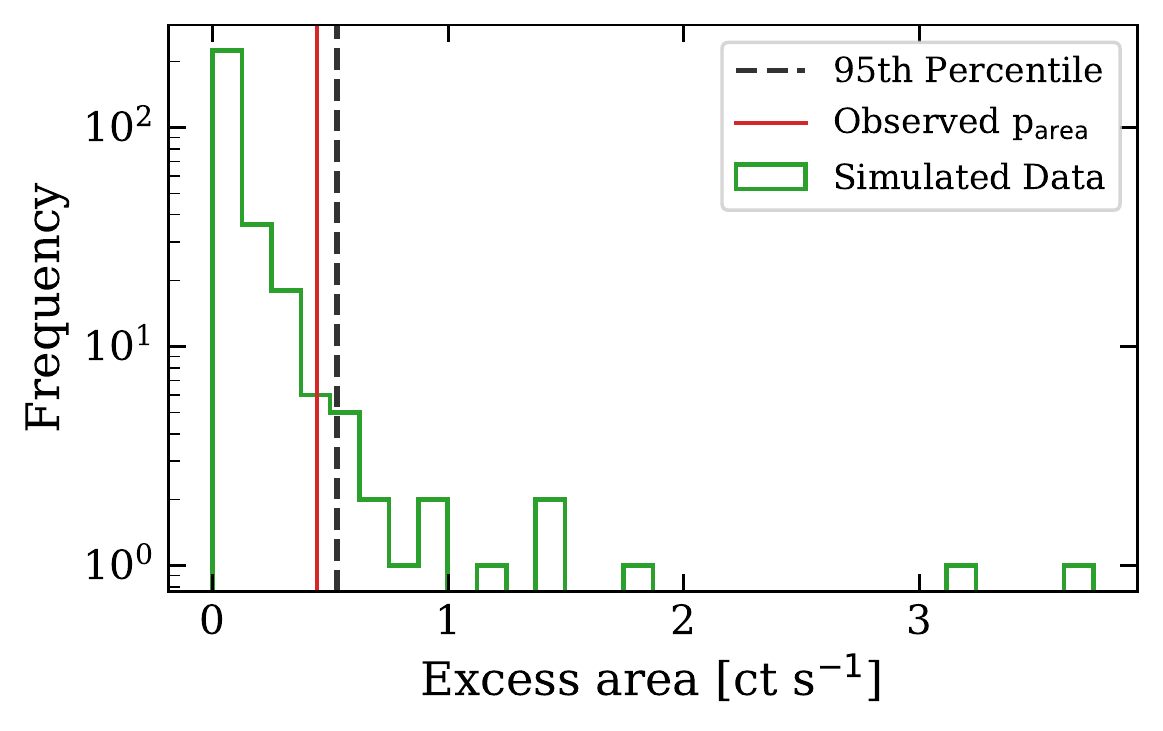} \\
\includegraphics[width=0.35\textwidth,height=\textheight,keepaspectratio]{figures/Model2_6119_residuals_for_CuSum.pdf} &
\includegraphics[width=0.32\textwidth,height=\textheight,keepaspectratio]{figures/Model2_6119_p-value_for_CuSum.pdf} \\
\includegraphics[width=0.35\textwidth,height=\textheight,keepaspectratio]{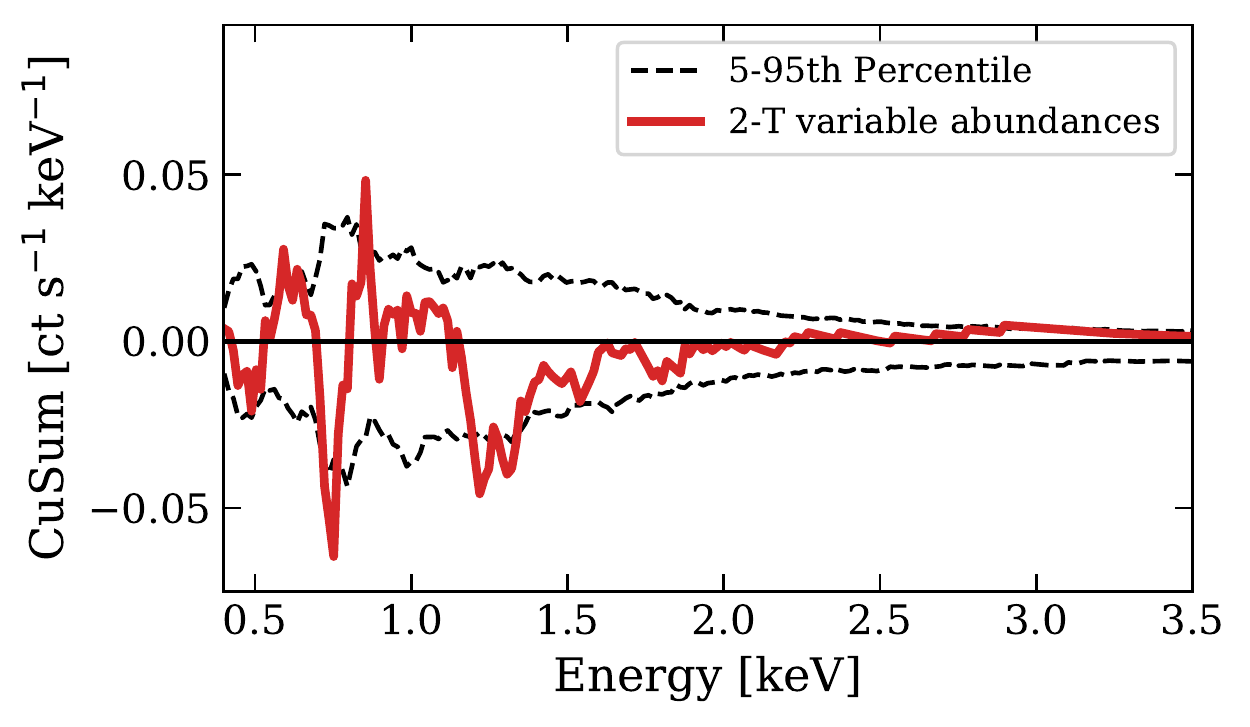} &
\includegraphics[width=0.32\textwidth,height=\textheight,keepaspectratio]{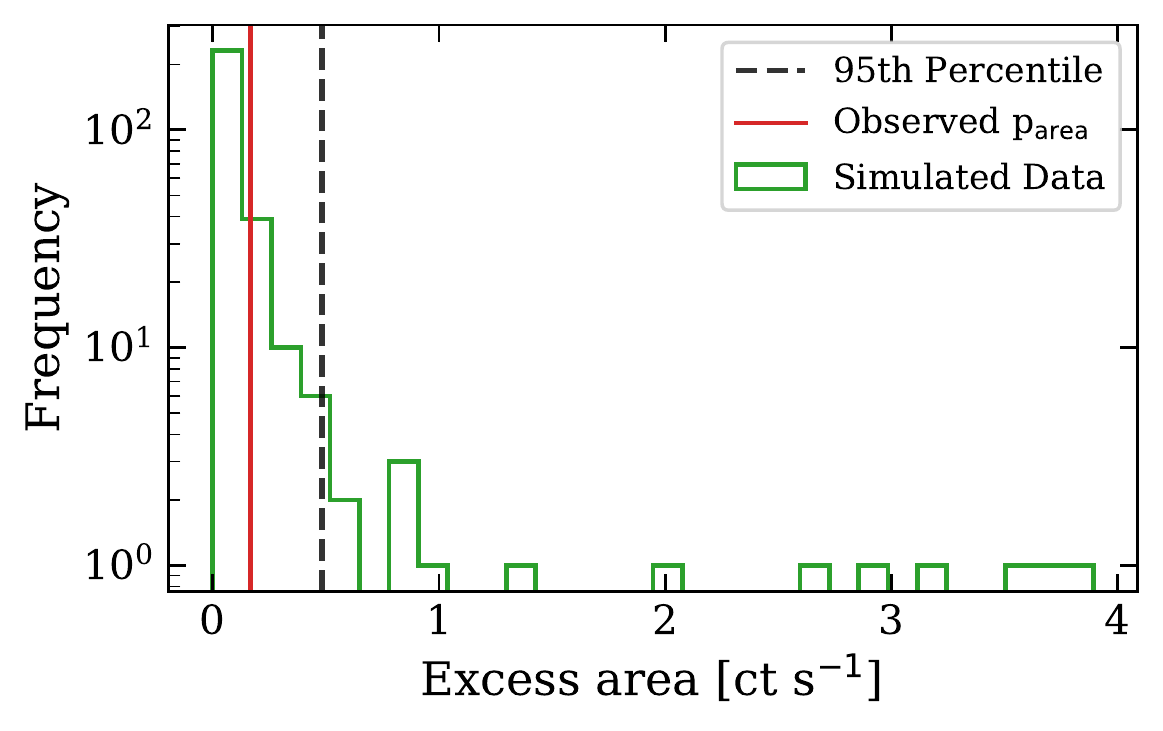} \\
\includegraphics[width=0.35\textwidth,height=\textheight,keepaspectratio]{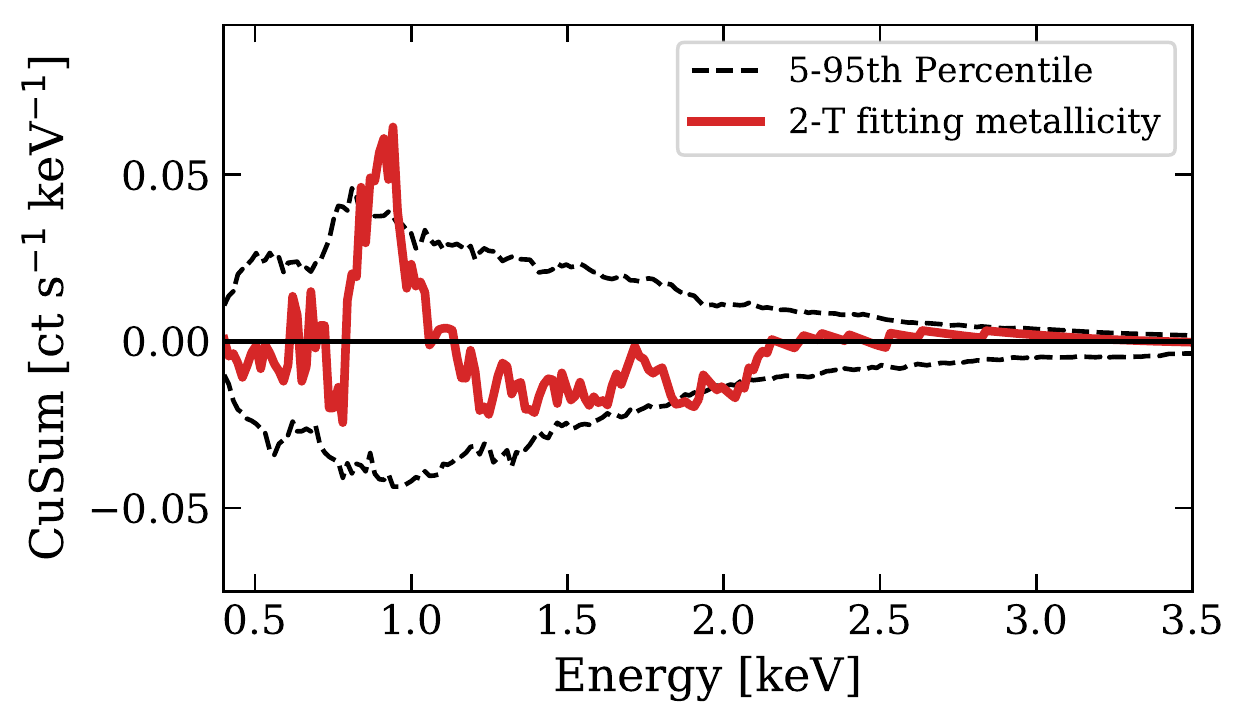} &
\includegraphics[width=0.32\textwidth,height=\textheight,keepaspectratio]{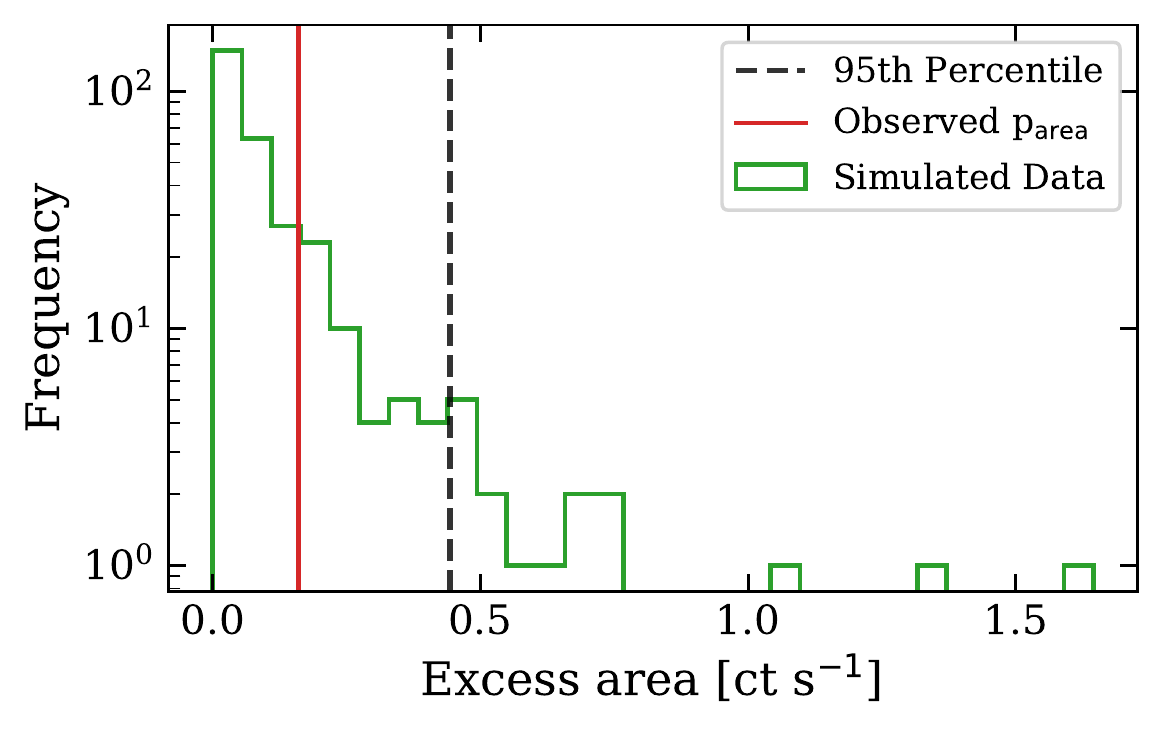} \\
\includegraphics[width=0.35\textwidth,height=\textheight,keepaspectratio]{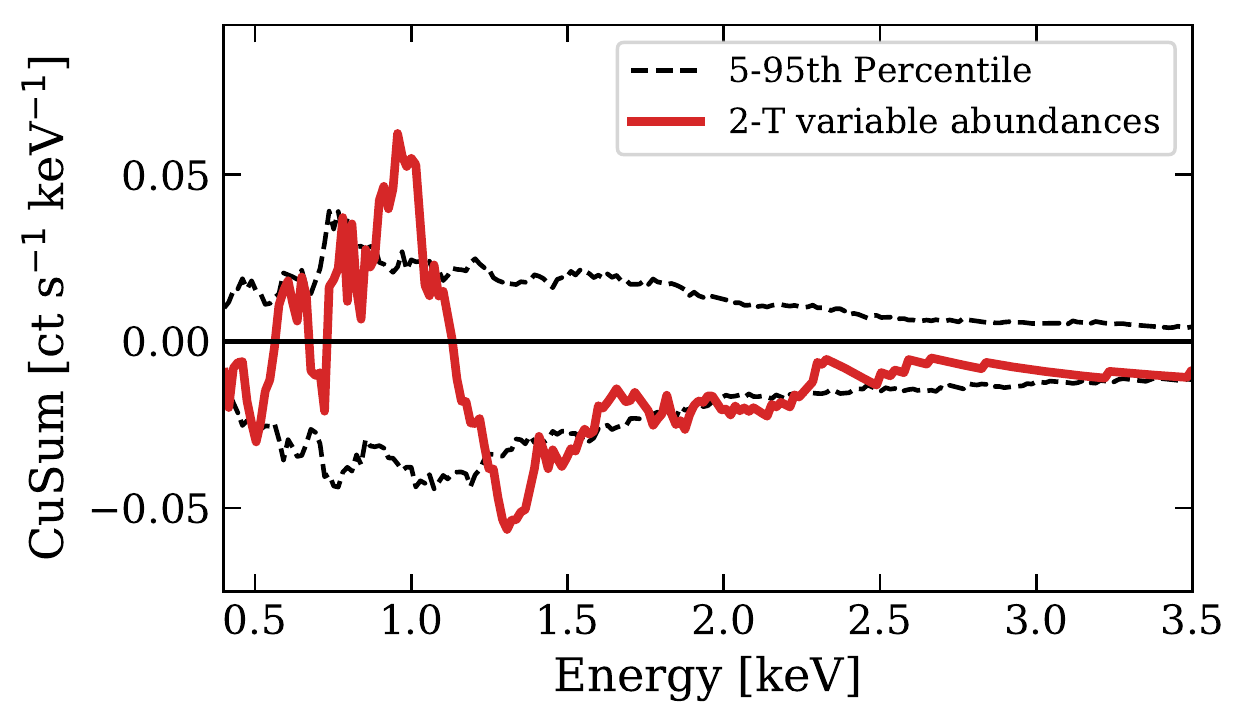} &
\includegraphics[width=0.32\textwidth,height=\textheight,keepaspectratio]{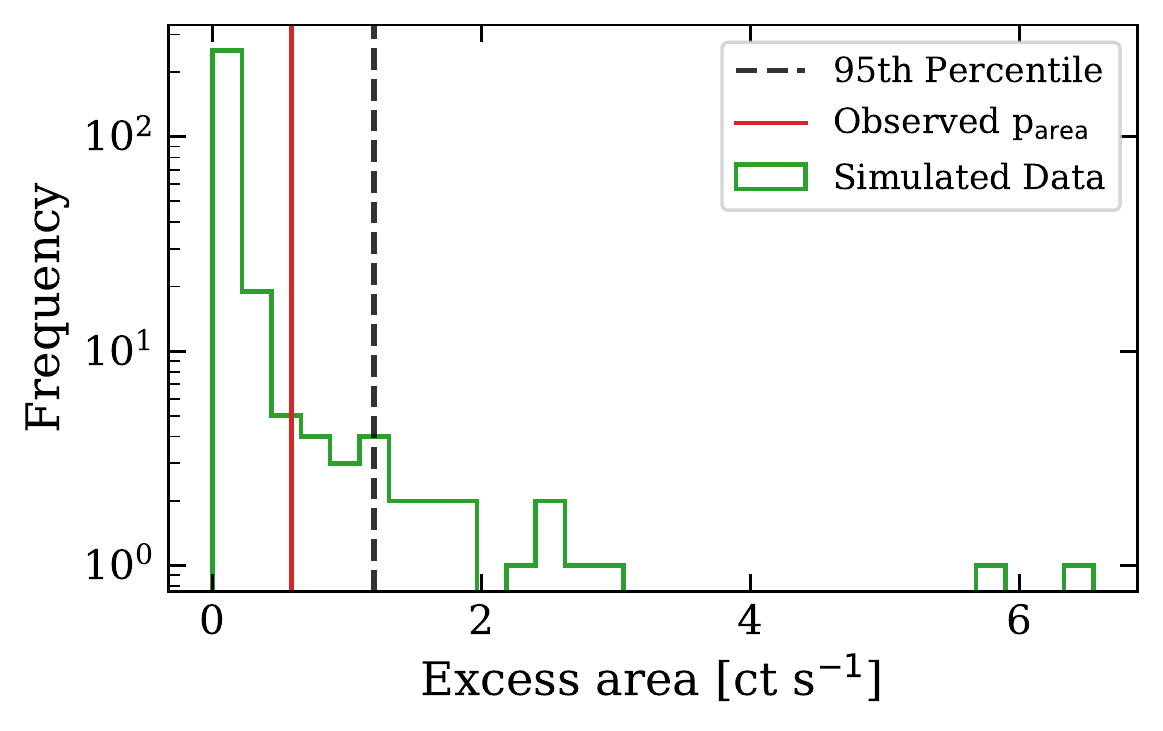}
\end{tabular}
\caption{As in Figure~\ref{fig_cusumcompare}, showing the Cumulative Sum (CuSum) deviations and the excess CuSum for the best-fit model for each ObsID, ranging in chronological order from 5427 (top) to 6122 (bottom).}
\label{fig_cusum}
\end{figure*}
\clearpage

\bibliography{allrefs}{}
\bibliographystyle{aasjournal}
\label{lastpage}
\end{document}